\documentclass[journal,twoside,web]{ieeecolor}
\usepackage{generic}
\usepackage{cite}
\usepackage{amsmath,amssymb,amsfonts}
\usepackage{algorithmic}
\usepackage{graphicx}
\usepackage{algorithm,algorithmic}
\usepackage{hyperref}
\hypersetup{hidelinks=true}
\usepackage{textcomp}
\def\BibTeX{{\rm B\kern-.05em{\sc i\kern-.025em b}\kern-.08em
  T\kern-.1667em\lower.7ex\hbox{E}\kern-.125emX}}
\usepackage{booktabs}
\usepackage{mathtools,mathrsfs,optidef,bm}
\mathtoolsset{showonlyrefs=true}

\newtheorem{definition}{Definition}
\newtheorem{assumption}{Assumption}
\newtheorem{theorem}{Theorem}
\newtheorem{proposition}{Proposition}
\newtheorem{lemma}{Lemma}
\newtheorem{corollary}{Corollary}

\newtheorem{remark}{Remark}

\newcommand{\C}{\mathbb{C}}
\newcommand{\R}{\mathbb{R}}
\newcommand{\Z}{\mathbb{Z}}
\newcommand{\N}{\mathbb{N}}

\DeclareMathOperator{\argmin}{arg\,min}

\DeclareMathOperator{\rank}{rank}
\DeclareMathOperator{\tr}{tr}

\newcommand{\T}{\top}
\newcommand{\E}{\mathbb{E}}
\newcommand{\Prob}{\mathbb{P}}

\newcommand{\Aset}{\mathcal{A}}
\newcommand{\Bset}{\mathcal{B}}
\newcommand{\Cset}{\mathcal{C}}

\newcommand{\Iset}{\mathcal{I}}
\newcommand{\Kset}{\mathcal{K}}

\newcommand{\Mset}{\mathcal{M}}
\newcommand{\Oset}{\mathcal{O}}
\newcommand{\Sset}{\mathcal{S}}
\newcommand{\Tset}{\mathcal{T}}
\newcommand{\Borel}{\mathscr{B}}

\newcommand{\midd}{\,\middle|\,}

\usepackage{tikz}
\usetikzlibrary{arrows.meta,calc,positioning,shapes}
\tikzset{
  every node/.style = {outer sep=0.12cm, inner sep=0},
  arrow/.style = {-{Triangle[length=0.25cm]}, thick},
  line/.style = {thick},
  block/.style = {rectangle, draw, minimum height=0.8cm, minimum width=2cm, thick, outer sep=0},
  sum/.style = {thick, circle, draw, inner sep=0, minimum size=6pt, outer sep=0},
  point/.style = {radius=2pt}
}

\makeatletter
\let\originaltitlepagestyle\ps@titlepagestyle
\def\ps@titlepagestyle{%
  \originaltitlepagestyle
  \def\@oddfoot{%
    \hfil
    \parbox{0.95\textwidth}{%
      \centering\fontsize{7}{8}\selectfont
      This work has been submitted to the IEEE for possible
      publication. Copyright may be transferred without notice,
      after which this version may no longer be accessible.
    }%
    \hfil
  }%
}
\makeatother

\begin{document}
% \title{Event-Based Control via Sparsity-Promoting Regularization With Performance Guarantees}
% \title{Suboptimal Event-Based Sparse Control With Performance Guarantees}
% \title{Event-Based Intermittent Control With Performance Guarantees}
% \title{Event-Based Control With Sparsity-Induced Penalty}
% \title{Event-Based Sparse Control With Performance Guarantees: A Rollout Approach}
% \title{Event-Based Sparse Control for Linear Quadratic Problems With Performance Guarantees}
\title{Linear Quadratic Control With Event-Based Control Actions: Performance Beyond Periodic Control}
% \title{LQ Control With Event-Based Control Actions: Performance Beyond Periodic Control}
\author{Shumpei Nishida, \IEEEmembership{Student Member, IEEE} and Kunihisa Okano, \IEEEmembership{Member, IEEE}
  \thanks{This work was supported by JST SPRING Grant Number JPMJSP2101 and JSPS KAKENHI Grant Number JP26K07565.
  }
  \thanks{The authors are with the Graduate School of Science and Engineering,
    Ritsumeikan University,
    Shiga 525-8577 Japan.
    E-mails: {\footnotesize \{\texttt{re0158ff@ed.}, \texttt{kokano@fc.}\}%
    \texttt{ritsumei.ac.jp}}.
  }
}

\maketitle

\begin{abstract}
  % This paper proposes event-based sparse control for stochastic linear systems to balance control performance and actuation rate.
  This paper investigates linear quadratic (LQ) control for discrete-time stochastic systems to balance control performance and actuation rate.
  In our framework, the control input is sparse in time, and the actuation timings are determined in an event-based manner.
  We examine two types of problems: A sparsity-promoting regularization problem and a hard-constraint problem with a prescribed average actuation rate.
  Due to the combinatorial nature of these problems, finding optimal solutions is generally hard.
  Inspired by rollout algorithms in the context of dynamic programming, we develop a rollout-based approach to obtain tractable suboptimal policies.
  We show closed-loop stability and establish theoretical performance guarantees with respect to classical periodic control.
  Moreover, under additional conditions, we quantify the performance improvement over periodic control and prove that this improvement is strictly positive.
  A numerical example demonstrates the effectiveness of the proposed method in balancing control performance and actuation rate.
\end{abstract}

\section{Introduction}
\label{sec:introduction}

\PARstart{T}he efficient use of limited physical and cyber resources has become an important issue in control systems.
For example, reducing the duration of actuator operation in electric vehicles can decrease power consumption~\cite{chan2007state}.
In networked control systems, reducing unnecessary data transmissions can save communication load and prolong the battery life of wireless devices~\cite{sadi2014minimum,park2018wireless}.
However, reducing resource usage can degrade closed-loop performance or even compromise stability.
% This trade-off has motivated research on sparse control~\cite{nagahara2014sparse,nagahara2016maximum} and event-based control~\cite{heemels2012introduction,hertneck2025current}.
% This background has motivated control strategies that utilize actuation and communication resources only when needed.
% However, reducing resource usage for control may degrade closed-loop performance or compromise stability.
To balance control performance and resource efficiency, various resource-aware control strategies have been studied in recent decades.
In particular, sparse control~\cite{nagahara2014sparse,nagahara2016maximum} and event-based control~\cite{heemels2012introduction,hertneck2025current} have gained significant attention.
% Consequently, systematic methods that balance control performance and resource usage have received considerable research attention~\cite{nagahara2014sparse}.

% Motivated by this background, liner quadratic control with intermittent actuation, which aims to reduce the number of control actions while maintaining control performance, has been studied.
Sparse control seeks to achieve control objectives with sparse input signals, which allows efficient use of actuation resources.
Within this framework, LQ control with intermittent actuation has been examined to maintain control performance while limiting the number of control actions.
One line of research considers optimal control with explicit constraints on the number of control actions.
For example, Imer and Ba\c{s}ar~\cite{imer2006optimal} studied this problem for scalar systems, and subsequent studies considered higher-order linear systems~\cite{bommannavar2008optimal,gao2011cardinality,shi2013finite,nishida2024sparsity}.
% Gao and Li~\cite{gao2011cardinality} developed a branch-and-bound method to compute a suboptimal solution.
% In~\cite{nishida2024sparsity}, a greedy approach was employed for a sparsity-constrained linear quadratic problem to find a suboptimal solution efficiently.
Another line of research promotes sparsity in control signals by adding a sparsity-promoting penalty to the objective function~\cite{zhang2021linear}.
% For instance, Zhang and Nagahara~\cite{zhang2021linear} considered an LQR problem with sparsity-promoting regularization and proposed a convex relaxation method.
% These studies provide several methods for constructing sparse control inputs while accounting for control performance.
% However, most of these methods focus on noise-free systems, which limits their applicability to systems subject to disturbances.
% For such systems, an event-based approach, which determines actuation timings and control inputs online in response to system behavior, offers a promising way to achieve preferable closed-loop performance against disturbances while reducing resource usage.

% In parallel, performance-oriented event-based control, which seeks to balance closed-loop performance and communication usage, has developed along several directions.
In parallel, event-based control has been developed in the field of networked control systems to achieve control objectives while limiting the use of cyber resources.
% Since event-based control triggers transmissions of information, it offers a promising way to achieve preferable closed-loop performance against disturbances.
A substantial body of research focuses on sensor-to-controller communication, in which triggering policies determine when sensor measurements are transmitted.
From an optimal control perspective, such triggering policies can be designed by solving an optimization problem that accounts for both closed-loop performance and resource usage~\cite{molin2014price,soleymani2021value}.
Following this direction, consistent event-triggered control, which aims to achieve better control performance than periodic control at the same average transmission rate, has been proposed.
This concept was developed in the context of LQ control~\cite{antunes2018consistent,khashooei2018consistent} and has been extended to other settings~\cite{brunner2018stochastic,antunes2020consistent,yu2022stochastic,meister2026improving}.
% This concept was developed for LQ control~\cite{antunes2018consistent,khashooei2018consistent} and extended to output-feedback control~\cite{antunes2020consistent}, stochastic event-triggering~\cite{brunner2018stochastic,yu2022stochastic}, and multi-agent systems~\cite{meister2026improving}.
% This concept was developed for linear quadratic control~\cite{antunes2018consistent,khashooei2018consistent} and subsequently extended to several other settings~\cite{antunes2020consistent,brunner2018stochastic,yu2022stochastic,meister2026improving}.
% In these approaches, however, the control law is fixed, and only the triggering policy is designed.
% Moreover, these studies mainly consider intermittent communication from sensors to controllers, where the triggering decision determines what sensing information is made available to the controller.
% This setting differs from sparse actuation, in which the controller must determine whether a control input should be transmitted and applied by the actuator.

% This motivates an event-based control problem in which the controller determines both when a control input is transmitted to the actuator and what input value is used.
% For controller-to-actuator communication settings, performance-oriented event-based control has been studied to balance quadratic performance and resource usage.
% In controller-to-actuator communication settings, the controller determines both when to transmit a control input and what input value to transmit.
Another area of research on event-based control considers controller-to-actuator communication, in which the controller determines both the control inputs and their update timings.
As in sensor-to-controller communication, one of the central objectives is to design the control and triggering policies to achieve desirable closed-loop performance with limited resource usage.
% The objective in this framework is to design the control and triggering policies so as to achieve desirable closed-loop performance while limiting the use of communication resources.
% Since this co-design problem is generally hard to solve directly, substantial studies.
Cogill~\cite{cogill2009event} designed a triggering policy to minimize an upper bound on the cost function.
Khashooei \emph{et al.}~\cite{khashooei2017output} developed output-based event-triggered control that guarantees a prescribed level of quadratic performance.
Although these studies established suboptimal event-based methods with performance guarantees, the triggering policy is optimized for a predesigned state-feedback control law, rather than jointly optimizing the control and triggering policies.
% In addition, other studies jointly optimize the control and triggering policies.

% To jointly design the control and triggering policies, several studies have employed \emph{rollout algorithms}, which are sequential optimization techniques based on dynamic programming~\cite{bertsekas2017dynamic-i,bertsekas2012dynamic-ii}.
% Using a rollout algorithm, the control and triggering policies over a finite horizon are optimized, with the cost after the horizon approximated using a given base policy.
% % This optimization is repeated in a receding-horizon manner.
% A key property of the rollout is that it facilitates performance improvements over the base policy.
% This property makes the rollout approach attractive for event-based control, where the objective is to reduce resource usage without excessively degrading closed-loop performance.
To jointly design the control and triggering policies, several studies have employed \emph{rollout algorithms}, which are sequential optimization techniques based on dynamic programming~\cite{bertsekas2017dynamic-i,bertsekas2012dynamic-ii}.
These algorithms optimize policies over a finite horizon while approximating the remaining cost using a given base policy.
A key property of the rollout is that it facilitates performance improvements over the base policy.
% This property makes the rollout approach attractive for event-based control, where the objective is to reduce resource usage without excessively degrading closed-loop performance.
% By exploiting the possibility of performance improvement over the base policy, rollout provides an attractive approach to resource-aware control.
% In this context, Gommans \emph{et al.}~\cite{gommans2014self} proposed a self-triggered control method that jointly determines the control input and the next update time subject to a prescribed performance constraint.
% Khashooei \emph{et al.}~\cite{khashooei2017output} developed an event-triggered control method using a one-step rollout procedure to design the triggering policy under a fixed control policy while guaranteeing a prescribed level of quadratic performance.
% Whereas their method treats the desired performance as a constraint, Antunes and Heemels~\cite{antunes2014rollout} considered minimizing a quadratic control cost subject to an average transmission rate constraint.
% In this context, Antunes and Heemels~\cite{antunes2014rollout} considered minimizing a quadratic control cost subject to an average transmission rate constraint.
% They developed rollout event-triggered control and showed closed-loop stability and performance guarantees with respect to traditional periodic control.
% It is worth noting that they established strict performance improvement over periodic control at the same average transmission rate, which is generally hard to prove for rollout methods.
In this context, Antunes and Heemels~\cite{antunes2014rollout} considered minimizing a quadratic control cost subject to an average transmission rate constraint.
They developed rollout event-triggered control and established closed-loop stability and, under additional conditions, strict performance improvement over periodic control at the same average transmission rate.
% Subsequently, Gommans \emph{et al.}~\cite{gommans2017resource} applied rollout-based co-design to resource-aware MPC for constrained linear systems.
% Further studies extended rollout-based event-triggered control to token-bucket networks~\cite{wildhagen2019predictive,wildhagen2023uncertainties} and time-slotted networks~\cite{linsenmayer2021controller}.
Further studies extended rollout-based event-triggered control to nonlinear systems~\cite{gommans2015resource}, constrained linear systems~\cite{gommans2017resource}, and networked systems with traffic specifications~\cite{wildhagen2019predictive,linsenmayer2021controller,wildhagen2023uncertainties}.
Despite these developments, studies on sparse control frameworks that balance quadratic performance and actuation resource usage for stochastic systems are limited.
Existing studies on sparse control mainly consider noise-free systems~\cite{shi2013finite,gao2011cardinality,nishida2024sparsity,zhang2021linear,gommans2017resource}, which limits their applicability to systems subject to disturbances.
Although several studies investigate stochastic systems~\cite{imer2006optimal,bommannavar2008optimal,cogill2009event}, they exclude control energy from their objective functions.
Existing rollout-based event-triggered control methods have typically been developed under zero-order-hold actuation, under which the previously transmitted control input remains applied when no new transmission occurs.
In contrast, in sparse control settings, the control input is set to zero whenever no actuation occurs.
This difference changes the closed-loop dynamics, and hence frameworks established for zero-order-hold actuation cannot be directly applied to sparse control.

In addition, it is important to clarify whether event-based control yields a performance advantage over classical periodic control to assess its potential for resource-aware control.
% A nonstrict performance guarantee only provides that event-based control performs no worse than periodic control and does not exclude the possibility that both methods achieve the same performance.
% A nonstrict performance guarantee cannot establish such an advantage, as both methods may achieve the same performance.
Specifically, strict performance improvement results are fundamental to characterizing the performance-resource trade-off, since they reveal whether event-based control improves this trade-off beyond periodic control.
% Therefore, strict improvement results are needed to establish that adapting the triggering decisions yields a strictly better trade-off between control performance and resource usage.
Although such results have been established for sensor-to-controller communication settings~\cite{antunes2018consistent,khashooei2018consistent,antunes2020consistent,yu2022stochastic}, they cannot be directly applied to controller-to-actuator communication since these two settings have different information structures.
Moreover, strict improvement results for controller-to-actuator communication have been established only for specific problem settings, including deterministic control over a token bucket network~\cite{wildhagen2019predictive} and LQ control with full-state information under a rate constraint~\cite{antunes2014rollout}.
Consequently, it remains unclear whether strict performance improvement results can be established for event-based schemes beyond these settings.
% Consequently, whether event-based sparse control can achieve and quantify a strict performance improvement over periodic control remains unresolved.

% In this paper, we propose \emph{event-based sparse control}, in which the control signal is sparse in time and the actuation timings are determined in an event-based manner.
\subsubsection*{Contributions}
% \subsection{Contributions}
In this paper, we develop an LQ control strategy that balances control performance and the frequency of control actions.
In our framework, the control signal is sparse in time and the actuation timings are determined in an event-based manner.
Building on our previous work~\cite{nishida2024feedback}, which investigated a finite-horizon LQ problem, we formulate an infinite-horizon problem that incorporates quadratic control performance and a penalty on resource usage, and propose a rollout-based algorithm that jointly determines the control and triggering policies.
% We formulate an infinite-horizon average-cost problem that incorporates quadratic control performance and a penalty on resource usage, and develop a rollout-based algorithm that jointly determines the control and triggering policies.

The contributions of this paper are summarized as follows:
\begin{enumerate}
  \item For the sparsity-promoting regularization problem, we establish the mean-square stability of the closed-loop system and prove that our method guarantees no worse performance than periodic control.
  \item Under additional conditions, we quantify the explicit performance improvement over periodic control and show that this improvement is strictly positive.
  \item We also investigate a LQ problem subject to a hard constraint on the average actuation rate and develop a corresponding rollout-based algorithm with stability and performance guarantees.
\end{enumerate}

A preliminary version of this work was presented in~\cite{nishida2026event}, where we developed a rollout-based algorithm for the sparsity-promoting regularization problem and established stability and performance guarantees.
The present article additionally addresses the hard-constraint problem and, more importantly, quantifies the performance improvement over periodic control and proves its strict positivity under additional conditions.
% % In~\cite{nishida2024feedback}, we proposed a one-step rollout algorithm that sequentially determines the control input and actuation timing by minimizing a one-step cost and established a performance guarantee for the resulting policy.
% % However, this method did not provide stability guarantee and strict performance improvement results.
% In that work, we proposed a rollout-based algorithm for the sparsity-promoting regularization problem and establied performance and stability guarantees.
% However, that work considered only the sparsity-promoting formulation, while this article also addresses the hard-constraint problem.
% More importantly, the present article quantifies the performance improvement over periodic control and proves that this improvement is strictly positive under additional conditions, neither of which was established in the preliminary work.

\subsubsection*{Organization}
% \subsection{Organization}
This paper is organized as follows.
% In Section~\ref{sec:preliminaries}, we provide the necessary preliminaries on the theory of Markov chains used throughout this paper.
In Section~\ref{sec:problem_formulation}, we formulate the optimal control problem via sparsity-promoting regularization.
Section~\ref{sec:rollout} presents the proposed controller design method, followed by the stability and performance analysis in Section~\ref{sec:performance_and_stability_guarantees}.
We next investigate the hard-constraint formulation in Section~\ref{sec:hard_constraint}.
A numerical example is demonstrated in Section~\ref{sec:simulation_example}, and Section~\ref{sec:conclusion} concludes the paper.
Additional technical results and proofs are provided in the Appendix.

\subsubsection*{Notation}
% \subsection{Notation}
We denote by $\N$ the set of positive integers and by $\Z_{\geq 0}$ the set of nonnegative integers.
% The $n \times n$ identity matrix is represented by $I_n$.
% For a positive-semidefinite matrix $P$, its square root is denoted by $P^{1/2}$.
For a set $\Aset$, the indicator function is denoted by $\mathbb{I}_\Aset(\cdot)$.
% \begin{equation}
%   \mathbb{I}_\Aset(x) \coloneqq 
%   \begin{cases}
%     1 & \text{if } x \in \Aset, \\
%     0 & \text{otherwise}.
%   \end{cases}
% \end{equation}
% We denote by $\sigma(\cdot)$ the $\sigma$-algebra generated by a collection of sets or random variables.
The Borel field generated by all open sets in $\R^{n}$ is represented by $\Borel(\R^{n})$.
% The probability and conditional probability are denoted by $\Prob(\cdot)$ and  $\Prob(\cdot \mid \cdot)$, respectively.
The probability measure on the underlying probability space is denoted by $\Prob$.
For a random variable $x$, $\E[x]$ and $\E[x \mid \cdot]$ denote its expectation and conditional expectation, respectively.
For integers $p$ and $q \neq 0$, the notation $p \equiv 0 \pmod{q}$ indicates that $p$ is an integer multiple of $q$.

\section{Problem Formulation}
\label{sec:problem_formulation}

% \subsection{System Model and Optimal Control Problem}
% \label{subsec:model_problem}

We consider the discrete-time linear system
\begin{align}
  x_{k+1} &= Ax_k + Bu_k + w_k, \label{eq:plant} \\
  y_k &= Cx_k + v_k, \label{eq:measurement}
\end{align}
where $x_k \in \R^{n_x}$, $u_k \in \R^{n_u}$, and $y_k \in \R^{n_y}$ denote the state, the control input, and the output, respectively.
The noise processes $\{w_k\}_{k\in\Z_{\geq 0}}$ and $\{v_k\}_{k\in\Z_{\geq 0}}$ are assumed to be independent, zero-mean, i.i.d. Gaussian sequences with covariance matrices $\Omega_w \succ 0$ and $\Omega_v \succ 0$, respectively.
The initial state $x_0$ also follows a Gaussian distribution with mean $\bar{x}_0$ and covariance matrix $\Omega_{x_0} \succeq 0$, and is independent of both $w_k$ and $v_k$ for all $k \in \Z_{\geq 0}$.
We assume that the pairs $(A,B)$ and $(A,\Omega_w^{1/2})$ are controllable, and that $(A,C)$ is observable.

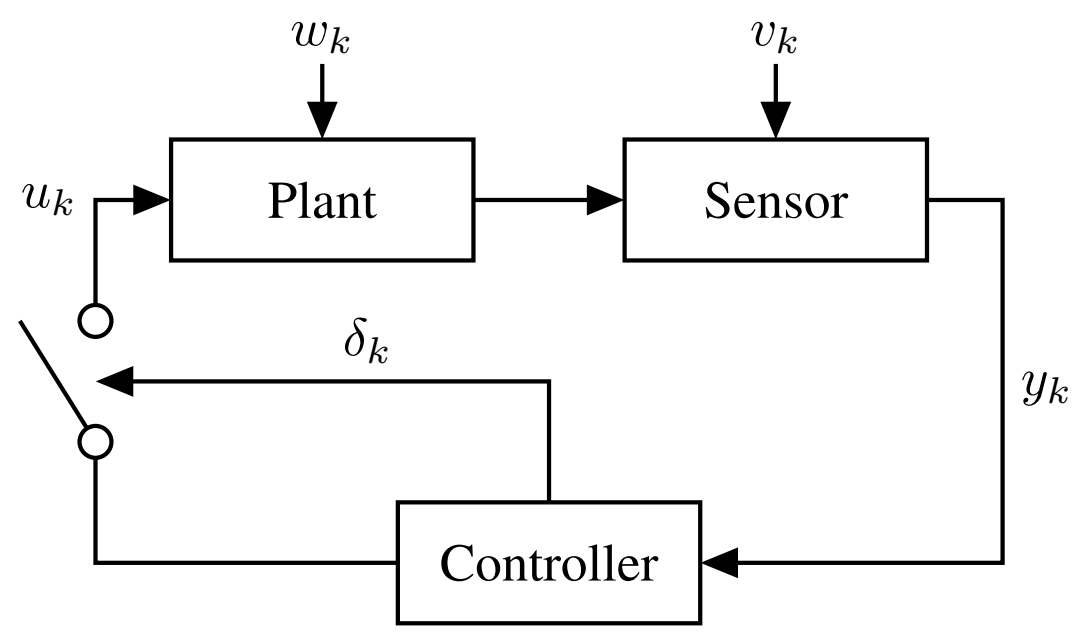
\begin{figure}[t]
  \centering
  \vspace{1em}
  \begin{tikzpicture}
    \node [block] (P){Plant};
    \node [block, right=1cm of P] (S){Sensor};
    \coordinate (mid1) at ($(P)!0.5!(S)$);
    \node [block] (C) at ($(mid1) + (0,-2.4cm)$) {Controller};
    \coordinate (mid2) at ($(mid1)!0.5!(C)$);
    \coordinate (Seast) at ($(S.east) + (0.5cm,0)$);
    \coordinate (Pwest) at ($(P.west) + (-0.5cm,0)$);
    \node [sum] at ($(Pwest)+(0,-0.8cm)$) (sum1){};
    \node [sum] at ($(sum1)+(0,-0.8cm)$) (sum2){};
    \coordinate (w) at ($(P.north)+(0,0.5cm)$);
    \coordinate (v) at ($(S.north)+(0,0.5cm)$);

    \draw [arrow] (w) -- node[above, yshift=2mm] {$w_k$} (P.north);
    \draw [arrow] (v) -- node[above, yshift=2mm] {$v_k$} (S.north);
    \draw [arrow] (P.east) -- (S.west);
    \draw [arrow] (S.east) -- (Seast) |- node[pos=0.26, right] {$y_k$} (C.east);
    \draw [line] (C.west) -| (sum2);
    \draw [arrow] (sum1) |- node [midway,left] {$u_k$} (P.west);
    \draw [line] (sum2) -- +(-0.5cm, 0.8cm);
    \draw [arrow] (C.north) |- node[pos=0.7, above] {$\delta_k$} ($(sum1)!0.5!(sum2)$);
    \end{tikzpicture}
  \caption{Feedback system considered in this paper.}
  \label{fig:feedback_system}
\end{figure}

The feedback system we consider is illustrated in Fig.~\ref{fig:feedback_system}.
The controller transmits control signals to the actuators only at selected time instants, whereas the control input is set to zero at all other time steps.
This intermittent actuation mechanism is represented by a binary variable $\delta_k \in \{0,1\}$, where $u_k = 0$ if $\delta_k = 0$,
% and $u_k \in \R^{n_u}$ otherwise.
while $u_k$ can be nonzero in $\R^{n_u}$ otherwise.
At each time $k$, the controller determines both $u_k$ and $\delta_k$ based on the available information up to time $k$.
The information set is denoted by
\begin{align}
  \Iset_k &\coloneqq \{y_i, u_j, \delta_j \colon 0\leq i\leq k,\ 0\leq j\leq k-1 \}, \quad k\geq 1\\
  \Iset_0 &\coloneqq \{y_0\}.
\end{align}
Let $\mu^u_k \colon \Iset_k \to \R^{n_u}$ and $\mu^\delta_k \colon \Iset_k \to \{0,1\}$ be the control policy and the triggering policy, respectively, both of which are measurable.
Then $u_k$ and $\delta_k$ are given by
\begin{equation}  \label{eq:control_triggering_policy}
  u_k = \mu^u_k(\Iset_k),\quad \delta_k = \mu^{\delta}_k(\Iset_k).
\end{equation}
The sequences of policies are denoted by $\mu^u \coloneqq \{\mu^u_k\}_{k\in\Z_{\geq 0}}$ and $\mu^{\delta} \coloneqq \{\mu^{\delta}_k\}_{k\in\Z_{\geq 0}}$.

At each time step, the controller computes a state estimate from the available information.
The estimate $\hat{x}_k \coloneqq \E[x_k \mid \Iset_k]$
% \footnote{Throughout this paper, $\E[\cdot\mid\Iset_k]$ is used as shorthand for $\E[\cdot\mid\sigma(\Iset_k)]$.}
is updated according to
\begin{equation} \label{eq:kalman_state_estimate_update}
  \hat{x}_{k+1} = A\hat{x}_{k} + Bu_k + G_{k+1}(y_{k+1} - C(A\hat{x}_k + Bu_k))
\end{equation}
with the initial condition $\hat{x}_0 = \bar{x}_0 + G_0(y_0 - C\bar{x}_0)$, where $G_k = \Sigma^-_kC^{\T}(C\Sigma^-_kC^{\T} + \Omega_v)^{-1}$ is the Kalman gain.
Let $\Sigma^-_k$ and $\Sigma_k$ be a priori and a posteriori error covariance matrices, respectively.
% The a priori and a posteriori error covariance matrices are given by
These matrices are calculated by
\begin{align}
  \Sigma^-_k &= A\Sigma_{k-1}A^{\T} + \Omega_w, \label{eq:kalman_prior_error_covariance} \\
  \Sigma_k &= \Sigma^-_k - \Sigma^-_kC^{\T}\left(C\Sigma^-_kC^{\T} + \Omega_v\right)^{-1}C\Sigma^-_k, \label{eq:kalman_posterior_error_covariance}
\end{align}
where $\Sigma^-_0 = \Omega_{x_0}$.
% and $\Sigma_k = \E[(x_k - \hat{x}_k)(x_k - \hat{x}_k)^{\T} \mid \Iset_k]$.
Under the controllability and observability assumptions, there exist positive-definite matrices $\tilde{\Sigma}$ and $\Sigma$ such that $\lim_{k \to \infty} \Sigma^-_k = \tilde{\Sigma}$ and $\lim_{k \to \infty} \Sigma_k = \Sigma$, and the steady-state Kalman gain is $G = \tilde{\Sigma}C^{\T}(C\tilde{\Sigma}C^{\T} + \Omega_v)^{-1}$.

The objective of this paper is to design $\mu^u$ and $\mu^{\delta}$ so as to balance control performance and actuation frequency.
The \emph{average control performance} is measured by the following quadratic cost.
\begin{equation} \label{eq:average_control_cost}
  J_c(\mu^u,\mu^{\delta}) = \limsup_{N \to \infty} \frac{1}{N} \E\left[ \sum_{k=0}^{N-1} x_k^{\T}Qx_k + u_k^{\T}Ru_k \right],
\end{equation}
where $Q \succeq 0$ and $R \succ 0$.
We assume that the pair $(A,Q^{1/2})$ is observable.
Moreover, the \emph{average actuation rate} is quantified by
\begin{equation} \label{eq:average_actuation_rate}
  J_r(\mu^{\delta}) = \limsup_{N \to \infty} \frac{1}{N} \E\left[ \sum_{k=0}^{N-1} \delta_k \right].
\end{equation}
% Using \eqref{eq:average_control_cost} and \eqref{eq:average_actuation_rate}, the we formulate the optimal control problem as
Therefore, we formulate the optimal control problem as
\begin{mini}|s|
  {\mu^u,\mu^{\delta}}
  {J(\mu^u,\mu^{\delta})}
  {\label{prob:sparsity_promoting_regularization}}
  {}
  \addConstraint{u_k = 0\quad}{\text{if}\ \delta_k = 0,}
\end{mini}
where 
\begin{equation}
\label{eq:sparsity_regularized_objective_function}
  J(\mu^u,\mu^{\delta}) = \limsup_{N \to \infty} \frac{1}{N} \E\left[ \sum_{k=0}^{N-1} x_k^{\T}Qx_k + u_k^{\T}Ru_k + \theta\delta_k\right].
\end{equation}
The parameter $\theta > 0$ adjusts the trade-off between the LQ performance and the actuation rate.
% Since $\mu^u_k$ takes values in $\R^{n_u}$ whereas $\mu^{\delta}_k$ takes values in the binary set $\{0,1\}$, Problem~\eqref{prob:sparsity_promoting_regularization} is generally hard to solve directly.
The problem~\eqref{prob:sparsity_promoting_regularization} involves the joint optimization of control and triggering policies, resulting in a nonconvex optimal control problem due to the coupling between the control input $u_k$ and the binary triggering variable $\delta_k$.
Consequently, the problem cannot be reduced to a standard LQ problem, making it challenging to solve directly.
% In addition, $\mu^u$ and $\mu^\delta$ are coupled through the constraint
% Therefore, we propose a tractable algorithm to obtain a pair of policies with both performance and stability guarantees.

Instead of solving \eqref{prob:sparsity_promoting_regularization} directly, our focus is to design admissible control and triggering policies $(\mu^u,\mu^\delta)$ that achieve better average performance than periodic control, under which control inputs are applied periodically.
% Specifically, we aim to establish non-strict and strict performance guarantees for the proposed method.
% Let $(\mu^{u,\text{per}},\mu^{\delta,\text{per}})$ be the periodic control and triggering policies with a fixed control period $p \in \N$.
Let $\mu^{\delta,\mathrm{per}}$ denote the periodic triggering policy with a fixed control period \(p\in\mathbb N\) and $\mu^{u,\mathrm{per}}$ denote the corresponding control policy that minimizes \eqref{eq:average_control_cost} for this fixed triggering policy.
% Then our design problem is formally stated as follows.
Then we aim to find a pair of policies $(\mu^u,\mu^{\delta})$ such that
\begin{equation}
  \label{eq:target_nonstrict_periodic_performance_bound}
  J(\mu^u,\mu^{\delta}) \leq J(\mu^{u,\text{per}},\mu^{\delta,\text{per}}).
\end{equation}
Futhermore, under mild conditions, we seek to establish
\begin{equation}
  \label{eq:target_strict_periodic_performance_improvement}
  J(\mu^u,\mu^{\delta}) < J(\mu^{u,\text{per}},\mu^{\delta,\text{per}}).
\end{equation}

\begin{remark}
    The control period $p$ in \eqref{eq:target_nonstrict_periodic_performance_bound} and \eqref{eq:target_strict_periodic_performance_improvement} is a given parameter.
    Our objective is to establish the performance guarantees with respect to the optimal periodic policy for this prescribed period.
    In particular, $p$ need not be optimal among all periodic control periods.
\end{remark}

% \blue{
% \begin{problem}
% \label{prob:performance_guarantee}
%   Given a parameter $\theta>0$ and a control period $p\in\N$, find a pair of policies $(\mu^u,\mu^{\delta})$ such that
%   \begin{equation}
%     \label{eq:target_nonstrict_periodic_performance_bound}
%     J(\mu^u,\mu^{\delta}) \leq J(\mu^{u,\text{per}},\mu^{\delta,\text{per}}).
%   \end{equation}
%   Furthermore, under additional conditions, establish
%   \begin{equation}
%     \label{eq:target_strict_periodic_performance_improvement}
%     J(\mu^u,\mu^{\delta}) < J(\mu^{u,\text{per}},\mu^{\delta,\text{per}}).
%   \end{equation}
% \end{problem}
% }
% \blue{
% \begin{problem}
% \label{prob:strict}
%   Given a parameter $\theta>0$ and a control period $p\in\N$, find a pair of policies $(\mu^u,\mu^{\delta})$ such that
%   \begin{equation}
%     \label{eq:target_strict_periodic_performance_improvement}
%     J(\mu^u,\mu^{\delta}) < J(\mu^{u,\text{per}},\mu^{\delta,\text{per}}).
%   \end{equation}
% \end{problem}
% }

In Section~\ref{sec:rollout}, we propose a design method to obtain an admissible $(\mu^u,\mu^{\delta})$ for the problem~\eqref{prob:sparsity_promoting_regularization}.
In Section~\ref{subsec:performance_guarantee} and Section~\ref{subsec:strict_performance_guarantee}, the performance guarantees in \eqref{eq:target_nonstrict_periodic_performance_bound} and \eqref{eq:target_strict_periodic_performance_improvement} are established, respectively.

\section{Rollout Approach to Event-Based Control}
\label{sec:rollout}

% To obtain tractable suboptimal policies for Problem \eqref{prob:sparsity_promoting_regularization}, we employ a rollout algorithm~\cite{bertsekas2017dynamic-i,bertsekas2012dynamic-ii}, which is a sequential optimization technique based on dynamic programming.
% The idea is as follows:
% Given a lookahead horizon $h \in \N$, the rollout algorithm optimizes the policies over the next $h$ steps while approximating the cost-to-go beyond the horizon using a given base policy.
% Since directly applying the rollout idea to the infinite-horizon average cost problem is analytically challenging, we consider the lookahead minimization problem with a tractable terminal cost function.

% Problem~\eqref{prob:sparsity_promoting_regularization} is generally hard to solve directly since $\mu^u_k$ takes values in $\R^{n_u}$ whereas $\mu^{\delta}_k$ takes values in the binary set $\{0,1\}$.
% Problem~\eqref{prob:sparsity_promoting_regularization} involves the joint optimization of control and triggering policies, resulting in a nonconvex optimal control problem due to the coupling between the control input $u_k$ and the binary triggering variable $\delta_k$.
% Consequently, the problem cannot be reduced to a standard LQ problem, making it challenging to solve directly.

To address \eqref{prob:sparsity_promoting_regularization}, we introduce an finite-horizon lookahead problem.
Specifically, we employ a rollout algorithm~\cite{bertsekas2017dynamic-i,bertsekas2012dynamic-ii}, which is a sequential optimization technique based on dynamic programming.
The idea is as follows:
Given a lookahead horizon $h \in \N$, the rollout algorithm optimizes the policies over the next $h$ steps while approximating a cost function beyond the horizon using a given policy.
% Since directly applying the rollout idea to the infinite-horizon average cost problem is analytically challenging, we consider the lookahead minimization problem with a tractable terminal cost function.
By repeatedly solving the $h$-step lookahead problem, we construct control and triggering policies over the infinite horizon.

Formally, given $\Iset_k$, we consider the following $h$-step lookahead minimization problem:
\begin{mini}|s|
  {\substack{
    u_k,\dots,u_{k+h-1},\\
    \delta_k,\dots,\delta_{k+h-1}}
  }
  {
    \E\left[\sum_{i=k}^{k+h-1} c(x_i, u_i, \delta_i) + \tilde{V}_{k+h}(\Iset_{k+h}) \,\middle|\, \Iset_k \right]
  }
  {\label{prob:h_step_lookahead}}
  {}
  \addConstraint{u_i = 0\ \text{if}\ \delta_i = 0,\quad i \in \{k,\dots,k+h-1\},}
\end{mini}
% \begin{align}
%   &\min_{\substack{\mu^u_k,\dots,\mu^u_{k+h-1},\\ \mu^{\delta}_k,\dots,\mu^{\delta}_{k+h-1}}}
%   \quad
%   \E\Bigg[
%     \sum_{i=k}^{k+h-1} c(x_i, \mu^u_i(\Iset_i), \mu^\delta_i(\Iset_i)) \\[-2ex]
%   &\hspace{12em} + \tilde{V}_{k+h}(\Iset_{k+h}) \,\Bigm|\, \Iset_k
%   \Bigg]
%   \label{prob:h_step_lookahead}
%   \\
%   &\hspace{2.2em} \text{s.t.}\quad u_i = 0\quad \text{if}\ \delta_i = 0,\quad i \in \{k,\dots,k+h-1\}.
%   \nonumber
% \end{align}
% \begin{align}
%     \hat{V}_k(\Iset_k)
%     =
%     \min_{\substack{u_k,\dots,u_{k+h-1},\\\delta_k,\dots,\delta_{k+h-1}}}
%     &\E\left[\sum_{i=k}^{k+h-1} c(x_i, u_i, \delta_i) \right. \notag \\
%     &\quad \left. + \tilde{V}_{k+h}(\Iset_{k+h}) \,\middle|\, \Iset_k \right], \label{eq:bellman_equation_multistep_lookahead}
% \end{align}
where
\begin{equation}
    c(x,u,\delta) \coloneqq x^{\T}Qx + u^{\T}Ru + \theta\delta
\end{equation}
and $\tilde{V}_{k+h}(\Iset_{k+h})$ is
% a terminal cost function induced by some fixed base policy, which is given in the following.
a cost function induced by a fixed base policy, which is given in the following.
At each time $k=\ell h$, $\ell\in\mathbb Z_{\ge0}$, the policies over the next $h$ steps are obtained by solving \eqref{prob:h_step_lookahead}.
These policies are applied over $[k,k+h-1]$, and the optimization procedure is repeated at the next update time.
Hence, the control and triggering policies over the next $h$ steps are determined at $k=\ell h$ and implemented over the entire window.

% The remaining issue in \eqref{prob:h_step_lookahead} is the choice of $\tilde V_{k+h}$.
In order to solve \eqref{prob:h_step_lookahead}, we proceed in two steps.
First, we construct $\tilde V_{k+h}(\Iset_{k+h})$ with a periodic policy as the base policy.
% from the periodic control policy, under which
Under this policy, actuation is allowed only at periodic time instants.
This choice is motivated by the following reasons.
Since the periodic control policy is a simple admissible policy of the form \eqref{eq:control_triggering_policy}, it can serve as a benchmark within the same class of policies considered in this paper.
In addition, to solve the lookahead minimization problem \eqref{prob:h_step_lookahead}, it is useful to have the terminal cost in explicit form.
Periodic control is preferable for this purpose, since it provides explicit expressions for the optimal control law and the corresponding cost, from which we can construct a computable $\tilde V_{k+h}(\Iset_{k+h})$ without introducing a simulation-based approximation.
Second, given $\Iset_k$ and the constructed $\tilde V_{k+h}(\Iset_{k+h})$, we obtain the optimal policies for the problem~\eqref{prob:h_step_lookahead}.

In the next subsection, we derive the optimal periodic control policy under the fixed actuation period, together with the corresponding cost in \eqref{eq:sparsity_regularized_objective_function}.
After that, we describe the implementation of the rollout-based algorithm.

\subsection{Periodic Policy With a Fixed Control Period}
\label{subsec:opt_periodic}

Let $p \in \N$ be a fixed control period.
Then the corresponding triggering policy is described as
\begin{equation} \label{eq:periodic_triggering_policy}
  \mu^\delta_k(\Iset_k) = \begin{cases}
    1 & \text{if } k \equiv 0 \pmod{p}, \\
    0 & \text{otherwise},
  \end{cases}
\end{equation}
that is, actuation is allowed every $p$ steps, while $u_k=0$ at all other time instants.
% In what follows, the sequences of periodic control and triggering policies with the triggering mechanism \eqref{eq:periodic_triggering_policy} are denoted by $\mu^{u,\mathrm{per}}$ and $\mu^{\delta,\mathrm{per}}$, respectively.
% In the following, we assume that $h/p \in \N$.

For each $\ell \in \Z_{\geq 0}$, the state evolution from $k=\ell p$ to $k=(\ell+1)p$ is represented by the following system
\begin{equation} \label{eq:periodically_lifted_state_dynamics}
  x_{(\ell+1)p} = A^p x_{\ell p} + B_p u_{\ell p} + D_p \bar{w}_\ell^{(p)},
\end{equation}
where
\begin{align}
  &B_p \coloneqq A^{p-1}B,\quad D_p \coloneqq \begin{bmatrix}
    A^{p-1} & A^{p-2} & \cdots & I_{n_x}
  \end{bmatrix}, \\
  &\bar{w}_\ell^{(p)} \coloneqq \begin{bmatrix}
    w_{\ell p}^{\T} & w_{\ell p+1}^{\T} & \cdots & w_{(\ell+1)p-1}^{\T}
  \end{bmatrix}^{\T}.
\end{align}
The random vectors $\bar{w}_\ell^{(p)}$ are independent zero-mean Gaussian vectors with covariance matrix $\Omega_w^{(p)} = I_p \otimes \Omega_w$.

To express the average cost criterion in \eqref{prob:sparsity_promoting_regularization} using the system~\eqref{eq:periodically_lifted_state_dynamics}, define
\begin{align}
  Q_p &\coloneqq \sum_{i=0}^{p-1} (A^i)^{\T} Q A^i,\quad
  S_p \coloneqq \sum_{i=1}^{p-1} (A^i)^{\T} Q A^{i-1} B, \\
  R_p &\coloneqq R + \sum_{i=1}^{p-1} B^{\T}(A^{i-1})^{\T} Q A^{i-1} B.
\end{align}
From the definitions of $Q_p$, $S_p$, and $R_p$, we have
\begin{equation}
  \begin{bmatrix}
    Q_p & S_p \\
    S_p^\T & R_p
  \end{bmatrix} \succeq 0,\quad
  R_p \succ 0.
\end{equation}
We also define $c_p \colon \R^{n_x} \times \R^{n_u} \to \R$ by $c_p(x,u) \coloneqq x^{\T} Q_p x + 2x^{\T} S_p u + u^{\T} R_p u$.
Then using \eqref{eq:periodically_lifted_state_dynamics}, the average control cost can be written as
\begin{equation}
  J_c(\mu^u, \mu^{\delta, \mathrm{per}})
  = \limsup_{K \to \infty} \frac{1}{pK}
  \E\left[\sum_{\ell=0}^{K-1} c_p(x_{\ell p}, u_{\ell p})\right]
  + \frac{d(p)}{p},
  \label{eq:periodic_average_quadratic_control_cost}
\end{equation}
where
\begin{equation}
  d(p) \coloneqq \sum_{i=1}^{p-1} \sum_{j=0}^{i-1}
  \tr\left(Q A^j \Omega_w (A^j)^{\T}\right).
\end{equation}
Since $\mu^{\delta,\mathrm{per}}$ is fixed independently of $\mu^u$, we can explicitly compute $J_r$ by
\begin{equation}
  \label{eq:periodic_actuation_rate}
  J_r(\mu^{\delta, \mathrm{per}}) = \frac{1}{p}.
\end{equation}
Therefore, once $\mu^{\delta,\mathrm{per}}$ is fixed as the triggering policy, the problem~\eqref{prob:sparsity_promoting_regularization} reduces to minimizing $J_c$ with respect to $\mu^u$, which is a standard infinite-horizon LQ problem.
However, even if the original system $(A,B)$ is controllable, we might fail to find a stabilizing control policy, since the stabilizability of $(A^p,B_p)$ is nontrivial under the intermittent actuation.

To ensure the stabilizability of $(A^p, B_p)$ under the controllability assumption of $(A,B)$, we impose the following assumption \cite{pasand2018controllability}.
\begin{assumption}
  \label{assump:no_periodic_eigenvalue_aliasing}
  For any two distinct eigenvalues of $A$, denoted by $\lambda_1$ and $\lambda_2$, it holds that
  \begin{equation} \label{eq:no_periodic_eigenvalue_aliasing}
    \lambda_1 \neq \lambda_2 \exp\left( \frac{2\pi q \sqrt{-1}}{p} \right),
    \quad \forall q \in \mathbb{Z}.
  \end{equation}
\end{assumption}

In order to derive the optimal periodic control policy under the fixed contro period $p$, we also need the following lemma.

\begin{lemma}
  \label{lemma:observability_lifted_system}
  If $(A,Q^{1/2})$ is observable, then for any $p \in \N$, the pair $(A^p, Q_p^{1/2})$ is also observable.
\end{lemma}

\begin{proof}
  Let $\lambda_p \in \C$ be an eigenvalue of $A^p$ and $\xi_p \in \C^{n_x} \setminus \{0\}$ be a corresponding eigenvector, i.e., $A^p \xi_p = \lambda_p \xi_p$.
  Suppose that there exists such a vector $\xi_p$ satisfying $Q_p^{1/2} \xi_p = 0$.
  From the definition of $Q_p$, it follows that
  \begin{equation}
    0 = \xi_p^\ast Q_p \xi_p = \sum_{i=0}^{p-1} (A^i \xi_p)^\ast Q A^i \xi_p = \sum_{i=0}^{p-1} \| Q^{1/2} A^i \xi_p \|^2.
  \end{equation}
  Hence, $Q^{1/2} A^i \xi_p = 0$ for all $i = 0, 1, \dots, p-1$.
  Writing $s = qp + r$ with $q \in \Z_{\geq 0}$ and $r \in \{0, 1, \dots, p-1\}$, we obtain
  \begin{equation}
    \label{eq:lifted_eigenvector_power_identity}
    A^s \xi_p = A^{qp + r} \xi_p = \lambda_p^q A^r \xi_p.
  \end{equation}
  Since $Q^{1/2} A^r \xi_p = 0$ for all $r = 0, 1, \dots, p-1$, Equation~\eqref{eq:lifted_eigenvector_power_identity} implies that $Q^{1/2} A^s \xi_p = 0$ for all $s \in \Z_{\geq 0}$.
  Thus, we have
  \begin{equation}
    \xi_p \in \{x \in \C^{n_x} \colon Q^{1/2} A^s x = 0, \; \forall s \in \Z_{\geq 0}\}.
  \end{equation}
  On the other hand, since $(A,Q^{1/2})$ is observable, the Cayley--Hamilton theorem ensures that
  \begin{equation}
    \{x \in \C^{n_x} \colon Q^{1/2} A^s x = 0, \; \forall s \in \Z_{\geq 0}\} = \{0\}.
  \end{equation}
  This contradicts $\xi_p \neq 0$.
  Therefore, $Q_p^{1/2} \xi_p \neq 0$ holds for all eigenvectors $\xi_p$, which completes the proof.
\end{proof}

We obtain the optimal periodic control policy under the fixed control period as follows.

\begin{lemma}
  Suppose that $(A,B)$ is controllable and $(A,Q^{1/2})$ is observable.
  Under Assumption \ref{assump:no_periodic_eigenvalue_aliasing}, the control policy that minimizes $J_c$ under the control period $p$ is given by
  \begin{equation} \label{eq:periodic_control_policy}
    \mu^{u,\mathrm{per}}_k(\Iset_k) = \begin{cases}
      \tilde{F}_p \hat{x}_k & \text{if } k \equiv 0 \pmod{p}, \\
      0 & \text{otherwise},
    \end{cases}
  \end{equation}
  where
  \begin{equation} \label{eq:periodic_lq_feedback_gain}
    \tilde{F}_p
    = - \left(B_p^{\T}\tilde{P}_pB_p + R_p\right)^{-1}
    \left(B_p^{\T}\tilde{P}_pA^p + S_p^{\T}\right),
  \end{equation}
  and $\tilde{P}_p$ is the positive-definite solution of the following algebraic Riccati equation:
  \begin{equation}
    \label{eq:algebraic_riccati_periodic}
    \tilde{P}_p = Q_p + (A^p)^{\T}\tilde{P}_p A^p - M_p,
  \end{equation}
  where $M_p \coloneqq \tilde{F}_p^{\T}(B_p^{\T}\tilde{P}_pB_p + R_p)\tilde{F}_p$.
\end{lemma}

\begin{proof}
  Under Assumption~\ref{assump:no_periodic_eigenvalue_aliasing} and the controllability of $(A,B)$, the pair $(A^p,B_p)$ is stabilizable.
  Moreover, by Lemma~\ref{lemma:observability_lifted_system}, $(A^p,Q_p^{1/2})$ is observable.
  Hence, there exists $\tilde{P}_p \succ 0$ satisfying \eqref{eq:algebraic_riccati_periodic}.

  Since $d(p)$ in \eqref{eq:periodic_average_quadratic_control_cost} is independent of $\mu^u$, minimizing $J_c$ under the fixed triggering policy \eqref{eq:periodic_triggering_policy} is equivalent to an infinite-horizon LQG problem for the system \eqref{eq:periodically_lifted_state_dynamics} with stage cost $c_p$.
  % By the separation principle, the optimal control law is obtained by replacing the state in the corresponding state-feedback law with its state estimate.
  Futhermore, since the separation principle holds, the optimal control policy is given by \eqref{eq:periodic_control_policy} with the stabilizing gain \eqref{eq:periodic_lq_feedback_gain}.
  % From \cite{bertsekas2012dynamic-ii}, this problem is characterized by the following Bellman optimality equation.
  % \begin{equation}
  %   j_p + \eta_p(\Iset_{\ell p})
  %   =
  %   \min_{u_{\ell p}}
  %   \E\left[
  %     c_p(x_{\ell p},u_{\ell p})
  %     + \eta_p(\Iset_{(\ell+1)p})
  %     \,\middle|\,
  %     \Iset_{\ell p}
  %   \right],
  %   \label{eq:average_cost_bellman_periodic}
  % \end{equation}
  % where $j_p$ is the optimal average cost associated with $c_p$ and $\eta_p(\Iset_{\ell p})$ is the corresponding relative cost given by
  % \begin{equation}
  % \label{eq:eta_p}
  %   \eta_p(\Iset_{\ell p})
  %   =
  %   \E\left[x^\T_{\ell p} \tilde{P}_p x_{\ell p} \,\middle|\, \Iset_{\ell p}\right]
  %   =
  %   \hat{x}^\T_{\ell p} \tilde{P}_p \hat{x}_{\ell p} + \tr\left(\tilde{P}_p \Sigma_{\ell p}\right).
  % \end{equation}
  % Therefore, following the results on LQ control with the average cost criterion \cite{bertsekas2012dynamic-ii}, the optimal periodic control policy is represented by the feedback law with the stabilizing gain \eqref{eq:periodic_lq_feedback_gain}.
  % Moreover, by the certainty equivalence principle, \eqref{eq:periodic_control_policy} is such an optimal policy satisfying the constraint in \eqref{prob:sparsity_promoting_regularization}.
\end{proof}

Since $\Sigma_k\to\Sigma$ as $k \to \infty$, the average control cost under the periodic policy $(\mu^{u,\mathrm{per}},\mu^{\delta,\mathrm{per}})$ is given by
\begin{align}
  &J_c(\mu^{u,\mathrm{per}},\mu^{\delta,\mathrm{per}}) \\
  &=
  \frac{1}{p}
  \left[
    \tr\left(
      \tilde{P}_p D_p \Omega_w^{(p)} D_p^\T
    \right)
    +
    \tr\left(M_p\Sigma\right)
    +
    d(p)
  \right].
  \label{eq:optimal_periodic_control_cost}
\end{align}
Using $\tilde P_p$ given by \eqref{eq:algebraic_riccati_periodic}, we set $\tilde V_{k+h}(\Iset_{k+h})$ as
\begin{equation}
  \tilde V_{k+h}(\Iset_{k+h})
  =
  \E\left[
    x_{k+h}^{\T}\tilde P_p x_{k+h}
    \,\middle|\,
    \Iset_{k+h}
  \right]
  \label{eq:periodic_quadratic_terminal_cost}
\end{equation}
and develop the rollout-based algorithm by considering the finite-horizon problem \eqref{prob:h_step_lookahead} with this terminal cost.

\begin{remark}
    Although the control period $p$ is fixed in the proposed design, one possible way to choose it is to compare periodic policies over a finite set of candidate periods.
    For each candidate $p$, the objective value in \eqref{prob:sparsity_promoting_regularization} under the optimal periodic control policy is given by $J_c(\mu^{u,\mathrm{per}},\mu^{\delta,\mathrm{per}})+\theta/p$, where $J_c$ is computed from \eqref{eq:optimal_periodic_control_cost}.
    Thus, we may select $p$ that minimizes this objective value among the candidates satisfying Assumption~\ref{assump:no_periodic_eigenvalue_aliasing}.
    The selected period is used to construct \eqref{eq:periodic_quadratic_terminal_cost}, with the lookahead horizon $h$ satisfying Assumption~\ref{assump:lookahead_horizon_multiple_of_period}.
\end{remark}

\subsection{Proposed Rollout-Based Algorithm}
\label{subsec:proposed}

We solve the problem~\eqref{prob:h_step_lookahead} by successively determining the control and triggering policies over intervals of length $h$.
Specifically, at each time $k=\ell h$, $\ell\in\mathbb{Z}_{\geq 0}$, the policies over the next $h$ time steps are determined based on the information available at that time.
By repeating this procedure for $\ell=0,1,\ldots$, we obtain the control and triggering policies over the infinite horizon.

Before proceeding to the algorithm construction, we introduce the notation used in the proposed design.
Let $\Tset$ be the set of sequences of triggering variables for which the first $h$ entries of $\{\delta_k\}_{k \in \Z_{\geq 0}}$ are free variables, whereas all subsequent entries are generated by the periodic triggering policy $\mu^{\delta, \text{per}}$.
Since $\Tset$ consists of $2^h$ admissible sequences, it can be indexed bijectively by the set $\Mset \coloneqq \{1,2,\dots,2^h\}$.
Accordingly, for each $m \in \Mset$, we denote the corresponding sequence in $\Tset$ by $\{\delta^{(m)}_k\}_{k \in \Z_{\geq 0}}$.
This sequence is specified as
\begin{equation}
  \label{eq:candidate_triggering_sequence_with_periodic_tail}
  \begin{split}
  \delta^{(m)}_k &= \rho^{(m)}_k \in \{0,1\}, \quad k \in \{0,1,\dots,h-1\}, \\
  \delta^{(m)}_k &= \mu^{\delta,\text{per}}_k(\Iset_k) =
  \begin{cases}
    1 & \text{if } k \equiv 0 \pmod{p}, \\
    0 & \text{otherwise},
  \end{cases}
  \ k \geq h.
  \end{split}
\end{equation}
% \begin{align}
%   \delta^{(m)}_k &= \rho^{(m)}_k \in \{0,1\}, \quad k \in \{0,1,\dots,h-1\}, \\
%   \delta^{(m)}_k &= \mu^{\delta,\text{per}}_k(\Iset_k) =
%   \begin{cases}
%     1 & \text{if } k \equiv 0 \pmod{p}, \\
%     0 & \text{otherwise},
%   \end{cases}
%   \ k \geq h.
%   \label{eq:candidate_triggering_sequence_with_periodic_tail}
% \end{align}
% where $\rho^{(m)}_k \in \{0,1\}$ for all $k \in \Z_{\geq 0}$.
In particular, the index $m=1$ is assigned to the sequence whose first $h$ entries also follow the periodic policy, namely,
\begin{equation}
  \rho^{(1)}_k = 
  \begin{cases}
    1 & \text{if } k \equiv 0 \pmod{p}, \\
    0 & \text{otherwise},
  \end{cases}
  \quad k \in \{0,1,\dots,h-1\}.
\end{equation}
% That is, the sequence $\{\delta^{(1)}_k\}_{k \in \Z_{\geq 0}} \in \Tset$ corresponds to the triggering sequence of the periodic policy $\mu^{\delta, \text{per}}$.

Throughout the paper, we impose the following assumption.
\begin{assumption} \label{assump:lookahead_horizon_multiple_of_period}
  $h \in \N$ and $p \in \N$ are such that $h/p \in \N$.
\end{assumption}
Assumption~\ref{assump:lookahead_horizon_multiple_of_period} ensures that the $h$-step lookahead horizon contains exactly $h/p$ actuations under the periodic policy.
This property is used in the proof of Lemma~\ref{lemma:periodic_candidate_initial_riccati_matrix}, which plays a key role in establishing the performance guarantees.
Although this assumption restricts the admissible combinations of $h$ and $p$, its practical impact is limited since the condition can be satisfied by choosing $h$ as an integer multiple of $p$.

% Assumption~\ref{assump:lookahead_horizon_multiple_of_period} ensures that the $h$-step lookahead horizon contains an integer number of periods of the periodic policy, which is useful in the performance analysis.
% Although this assumption restricts the admissible combinations of $h$ and $p$, its practical impact is limited since both quantities are design parameters and the condition can always be satisfied by choosing $h$ as an integer multiple of $p$.

We now construct the rollout-based algorithm to obtain $(\mu^u,\mu^\delta)$.
By using \eqref{eq:periodic_quadratic_terminal_cost}, the problem~\eqref{prob:h_step_lookahead} can be rewritten as
\begin{mini}|s|
  {\substack{u_k,\dots,u_{k+h-1},\\ \delta_k,\dots,\delta_{k+h-1}}}
  {
    \E\left[\sum_{i=k}^{k+h-1} c(x_i, u_i, \delta_i) + x_{k+h}^{\T}\tilde P_p x_{k+h} \,\middle|\, \Iset_k \right]
  }
  {\label{prob:h_step_lookahead_quadratic_terminal_cost}}
  {}
  \addConstraint{u_i = 0\ \text{if}\ \delta_i = 0,\quad i \in \{k,\dots,k+h-1\},}
\end{mini}
where we use the tower property to obtain
\begin{equation} \label{eq:tower_property}
  \E\left[ \E\left[x_{k+h}^{\T}\tilde{P}_p x_{k+h} \,\middle|\, \Iset_{k+h}\right] \,\middle|\, \Iset_k \right] = \E\left[x_{k+h}^{\T}\tilde{P}_p x_{k+h} \,\middle|\, \Iset_k\right].
\end{equation}
For a fixed index $m \in \Mset$, the triggering sequence is fixed as in \eqref{eq:candidate_triggering_sequence_with_periodic_tail}, and the cost in \eqref{prob:h_step_lookahead_quadratic_terminal_cost} reduces to a standard finite-horizon cost.
Thus the optimal control policy for a fixed $m \in \Mset$ is given by
\begin{equation} \label{eq:candidate_sequence_feedback_control}
  u_k = \rho^{(m)}_s F^{(m)}_s \hat{x}_k, \quad s \in \{0,1,\dots,h-1\},
\end{equation}
where
% , for any $s \in \{0,1,\dots,h-1\}$ and $m \in \Mset$, we have
\begin{equation}
  F^{(m)}_s = - \big(B^{\T}P^{(m)}_{s+1}B + R\big)^{-1} B^{\T}P^{(m)}_{s+1}A,
\end{equation}
and $P^{(m)}_s$ is computed backward from
\begin{align}
  &P^{(m)}_h = \tilde{P}_p, \\
  &P^{(m)}_s = Q + A^{\T}P^{(m)}_{s+1}A \notag \\
  &- \rho^{(m)}_s A^{\T}P^{(m)}_{s+1}B \left(B^{\T}P^{(m)}_{s+1}B + R\right)^{-1} B^{\T}P^{(m)}_{s+1}A, \notag \\
  &\qquad s \in \{0,1,\dots,h-1\}. \label{eq:candidate_sequence_riccati_recursion}
\end{align}
For the fixed sequence indexed by $m \in \Mset$ and the corresponding control input \eqref{eq:candidate_sequence_feedback_control}, the resulting value of \eqref{prob:h_step_lookahead_quadratic_terminal_cost} is given by
\begin{equation} \label{eq:candidate_sequence_h_step_cost}
  \hat{x}_k^{\T} P^{(m)}_0 \hat{x}_k + \tr\left(P^{(m)}_0 \Sigma_k\right) + \beta^{(m)}_k + \gamma^{(m)}.
\end{equation}
% where
The terms $\beta^{(m)}_k$ and $\gamma^{(m)}$ are given by
\begin{align}
  &\beta^{(m)}_k \coloneqq \sum_{\tau=0}^{h-1} \left( \tr\left(P^{(m)}_{\tau+1}\Omega_w\right) + \rho_\tau^{(m)} \tr\left(M^{(m)}_\tau \Sigma_{k+\tau}\right)\right), \\
  % &M^{(m)}_\tau \coloneqq (F^{(m)}_\tau)^{\T}\left(B^{\T}P^{(m)}_{\tau+1}B + R\right)F^{(m)}_\tau, \\
  &\gamma^{(m)} \coloneqq \theta \sum_{\tau=0}^{h-1} \rho^{(m)}_\tau
\end{align}
for each $m \in \Mset$, where $M^{(m)}_\tau \coloneqq (F^{(m)}_\tau)^{\T}(B^{\T}P^{(m)}_{\tau+1}B + R)F^{(m)}_\tau$.
These terms can be obtained following the standard LQG framework~\cite{astrom2006introduction}.
Consequently, the rollout policies are obtained by selecting an index $m \in \Mset$ that minimizes \eqref{eq:candidate_sequence_h_step_cost} and then determining the corresponding feedback law from \eqref{eq:candidate_sequence_feedback_control}.
It should be noted that, although the triggering sequence and the associated feedback gains over $h$ steps are determined at $k=\ell h$, the control input applied to the actuators is computed at each time instant using the current state estimate.
In addition, once $m \in \Mset$ is fixed, we can compute $\{P^{(m)}_s\}_{s=0}^h$, $\{F^{(m)}_s\}_{s=0}^{h-1}$, and $\{M^{(m)}_\tau\}_{\tau=0}^{h-1}$ offline.

Summarizing the above discussion, the proposed algorithm is presented below.

\emph{Algorithm~1:}
\begin{enumerate}
  \item At time $k = \ell h,\ \ell \in \Z_{\geq 0}$, compute
  \begin{align}
    \iota(\Iset_{\ell h}) \coloneqq \underset{m \in \Mset}{\argmin}\ &\hat{x}_{\ell h}^{\T}P^{(m)}_0\hat{x}_{\ell h} + \tr\left(P^{(m)}_0\Sigma_{\ell h}\right) \notag \\
    &\quad + \beta^{(m)}_{\ell h} + \gamma^{(m)}.
    \label{eq:rollout_candidate_sequence_selection_rule}
  \end{align}
  \item Set the triggering variables at $k = \ell h,\dots,(\ell+1)h-1$ as
  \begin{equation}
    \delta_k = \rho^{(\iota(\Iset_{\ell h}))}_{k - \ell h}.
  \end{equation}
  \item Compute the corresponding control input online according to
  \begin{equation} \label{eq:rollout_control_policy}
    u_k = \rho^{(\iota(\Iset_{\ell h}))}_{k - \ell h} F^{(\iota(\Iset_{\ell h}))}_{k - \ell h} \hat{x}_k.
  \end{equation}
  \item Repeat steps 1), 2), and 3).
\end{enumerate}
Therefore, Algorithm~1 determines the control policy $\mu^u_k$ and the triggering policy $\mu^{\delta}_k$ as
\begin{align}
  (\mu^u_k(\Iset_k), \mu^{\delta}_k(\Iset_k)) 
  &= \left(\rho^{(\iota(\Iset_{\ell h}))}_{k - \ell h} F^{(\iota(\Iset_{\ell h}))}_{k - \ell h} \hat{x}_k, \rho^{(\iota(\Iset_{\ell h}))}_{k - \ell h}\right), \\
  &\ell h \leq k < (\ell+1)h,\quad \ell \in \Z_{\geq 0}. \label{eq:joint_rollout_control_and_triggering_policies}
\end{align}
In the following, we denote by $(\mu^{u, \text{ro}}, \mu^{\delta, \text{ro}})$ to represent the sequence of the control and triggering policies obtained by \eqref{eq:joint_rollout_control_and_triggering_policies}.

\begin{remark}
  If \eqref{eq:rollout_candidate_sequence_selection_rule} admits multiple minimizers $m \in \Mset$, we choose the minimizer with the smallest index.
  % More precisely, for a given $x \in \R^{n_x}$, define $\zeta_x \colon \Mset \to \R$ by
  % \begin{equation}
  %   \zeta_x(m) \coloneqq x^\T P^{(m)}_0 x + \tr\left(P^{(m)}_0\Sigma_{\ell h}\right) + \beta^{(m)}_{\ell h} + \gamma^{(m)}.
  % \end{equation}
  % Then the expression
  % \begin{align}
  %   &\underset{m \in \Mset}{\argmin}\ 
  %     \hat{x}_{\ell h}^{\T}P^{(m)}_0\hat{x}_{\ell h} 
  %     + \tr\left(P^{(m)}_0\Sigma_{\ell h}\right) 
  %     + \beta^{(m)}_{\ell h} + \gamma^{(m)} \notag \\
  %   &= \min \left(\zeta_{\hat{x}_{\ell h}}^{-1}\left(\min_{m \in \Mset} \zeta_{\hat{x}_{\ell h}}(m)\right)\right)
  % \end{align}
  % is used as the rule for determining $\iota(\Iset_{\ell h})$.
\end{remark}

\begin{remark}
  Since Algorithm~1 determines $\iota(\Iset_{\ell h})$ by minimizing a function that depends only on $\hat{x}_{\ell h}$,
  % we can regard $\iota$ as a measurable function of $\hat{x}_{\ell h}$ and 
  we write $\iota(\Iset_{\ell h}) = \iota(\hat{x}_{\ell h})$ in the subsequent analysis.
\end{remark}

\section{Stability and Performance Analysis}
\label{sec:performance_and_stability_guarantees}

This section establishes theoretical stability and performance guarantees for Algorithm~1.
In Section~\ref{subsec:stability}, we prove mean-square stability of the closed-loop system induced by the rollout policies.
Section~\ref{subsec:performance_guarantee} shows that the proposed algorithm performs no worse than the periodic policy in terms of the average performance criterion.
Moreover, under extra assumptions, we establish a \emph{strict} performance improvement result in Section~\ref{subsec:strict_performance_guarantee}.

Throughout the subsequent sections, we impose the following assumptions for the analysis.
\begin{assumption}
  \label{assump:state_weight_and_stationary_kalman_filter}
  \mbox{}
  \begin{enumerate}
    % \renewcommand{\labelenumi}{(\roman{enumi})}
    % \item $\rank(C) = \rank(C^{\T}) = n_x$.
    \item $Q \succ 0$.
    \item $\Omega_{x_0}$ is a solution of the following Riccati equation:
    \begin{align}
      \Omega_{x_0} &= \Omega_w + A\Omega_{x_0}A^{\T} \\
      & - A\Omega_{x_0}C^{\T} \left(C\Omega_{x_0}C^{\T} + \Omega_v\right)^{-1} C\Omega_{x_0}A^{\T}. \label{eq:stationary_prior_covariance_riccati_equation}
    \end{align}
  \end{enumerate}
\end{assumption}
% Assumption~\ref{assump:state_weight_and_stationary_kalman_filter}(i) ensures that the steady-state Kalman gain $G$ has full row rank, since the covariance matrix $\tilde{\Sigma}$ is positive definite.
Assumption~\ref{assump:state_weight_and_stationary_kalman_filter}(i) is used for stability guarantee.
When Assumption~\ref{assump:state_weight_and_stationary_kalman_filter}(ii) holds, $\Omega_{x_0}$ is positive definite since $(A,\Omega_w^{1/2})$ is controllable.
Furthermore, this assumption implies $G_k = G$ for all $k$, i.e., the Kalman filter is stationary, which is the case of interest in this study.
% Under Assumption~\ref{assump:state_weight_and_stationary_kalman_filter}(ii), we have $\E[e_k\mid\Iset_k]=0$ and $\E[e_ke_k^\T\mid\Iset_k]=\Sigma$ for all $k\in\Z_{\geq0}$.
In the following, we write $\beta^{(m)}$ instead of $\beta^{(m)}_k$, since $\beta^{(m)}_k$ is independent of the time index $k$ under Assumption~\ref{assump:state_weight_and_stationary_kalman_filter}.

% Let us introduce the notation $(\mu^{u, \text{ro}}, \mu^{\delta, \text{ro}})$ to denote the sequence of the control and triggering policies obtained by Algorithm~1.
For a given $m \in \Mset$, we define a function $V_m$ by
\begin{equation} \label{eq:quadratic_lyapunov_function_Vm}
  V_m(x) \coloneqq x^{\T} P_0^{(m)} x,\quad \forall x \in \R^{n_x},
\end{equation}
and also define $\chi_\ell \coloneqq x_{\ell h}$ and $\hat\chi_\ell \coloneqq \hat{x}_{\ell h}$ for notational convenience.
Thus, we have $\iota(\Iset_{\ell h}) = \iota(\hat\chi_\ell)$.
We define a function $g$, which is the performance improvement achieved by the rollout policy over the periodic policy, as
\begin{align}
  g(x) &\coloneqq x^{\T} \left(P_0^{(1)} - P_0^{(\iota(x))}\right) x 
  + \tr\left(\left(P_0^{(1)} - P_0^{(\iota(x))}\right) \Sigma\right) \notag \\
  &\quad + \beta^{(1)} - \beta^{(\iota(x))} + \gamma^{(1)} - \gamma^{(\iota(x))}.
  \label{eq:rollout_improvement_function_g}
\end{align}
By Step~1 in Algorithm~1, we have $g(x) \geq 0$ for all $x \in \R^{n_x}$.

% Let us introduce the notation $(\mu^{u, \text{ro}}, \mu^{\delta, \text{ro}})$ to denote the sequence of the control and triggering policies obtained by Algorithm~1.
Before proceeding to our main results, we derive the dynamics of $\hat{\chi}_\ell$ induced by the policies $(\mu^{u, \text{ro}}, \mu^{\delta, \text{ro}})$.
From \eqref{eq:kalman_state_estimate_update}, we have
\begin{align}
  &\hat{x}_{k+1} = (A + \rho_{k-\ell h}^{(\iota(\hat{\chi}_\ell))} B F_{k-\ell h}^{(\iota(\hat{\chi}_\ell))}) \hat{x}_k \\
  &\qquad + G\left(y_{k+1} - C(A + \rho_{k-\ell h}^{(\iota(\hat{\chi}_\ell))} B F_{k-\ell h}^{(\iota(\hat{\chi}_\ell))}) \hat{x}_k\right). \label{eq:closed_loop_kalman_estimate_update}
\end{align}
We define
\begin{equation}
  \Theta_i^{(m)} \coloneqq A + \rho_i^{(m)} B F_i^{(m)}
\end{equation}
for $i \in \{0,1,\dots,h-1\}$ and $m \in \Mset$, and let
\begin{equation}
    \omega_k \coloneqq G\left(CAe_k + Cw_k + v_{k+1}\right),
\end{equation}
where $e_k = x_k - \hat x_k$.
% \begin{equation}
%     e_k = x_k - \hat x_k.
% \end{equation}
% $\omega_k \coloneqq G\left(CAe_k + Cw_k + v_{k+1}\right)$, where $e_k = x_k - \hat x_k$.
% \begin{align}
%   &\Theta_i^{(m)} \coloneqq A + \rho_i^{(m)} B F_i^{(m)}, \\
%   &\omega_k \coloneqq G\left(CAe_k + Cw_k + v_{k+1}\right).
% \end{align}
Then Equation~\eqref{eq:closed_loop_kalman_estimate_update} can be rewritten as
\begin{equation}
  \hat{x}_{k+1} = \Theta_{k-lh}^{(\iota(\hat{\chi}_\ell))} \hat{x}_k + \omega_k.
\end{equation}
By iterating this equation from $k = \ell h$ to $k = (\ell+1)h-1$, we obtain
\begin{equation}
  \label{eq:subsampled_estimate_dynamics}
  \hat{\chi}_{\ell+1} = \Phi_{\iota(\hat{\chi}_\ell)} \hat{\chi}_\ell + \Gamma_{\iota(\hat{\chi}_\ell)} \bar{\omega}_\ell, \quad \forall \ell \in \Z_{\geq 0},
\end{equation}
where
\begin{align}
  \Phi_m &\coloneqq \prod_{s=h-1}^{0} \Theta_s^{(m)}, \\
  \Gamma_m &\coloneqq \begin{bmatrix}
    \prod_{s=h-1}^{1} \Theta_s^{(m)} & \prod_{s=h-1}^{2} \Theta_s^{(m)} & \dots & \Theta_{h-1}^{(m)} & I_{n_x}
  \end{bmatrix}, \\
  \bar{\omega}_\ell &\coloneqq \begin{bmatrix}
    \omega_{\ell h}^{\T} & \omega_{\ell h+1}^{\T} & \dots & \omega_{(\ell+1)h-1}^{\T}
  \end{bmatrix}^{\T}. \label{eq:bar_omega_ell}
\end{align}
We show that the process $\{\hat\chi_\ell\}_{\ell \in \Z_{\geq 0}}$ is a Markov chain.
\begin{lemma}
  \label{lemma:subsampled_estimate_markov_property}
  % Under Assumptions~\ref{assump:no_periodic_eigenvalue_aliasing}, \ref{assump:lookahead_horizon_multiple_of_period}, and \ref{assump:state_weight_and_stationary_kalman_filter},
  The stochastic process $\{\hat{\chi}_\ell\}_{\ell \in \Z_{\geq 0}}$ driven by \eqref{eq:subsampled_estimate_dynamics} is a Markov chain.
\end{lemma}

\begin{proof}
  % Since the innovation of the Kalman filter is an independent Gaussian process and the future innovations are independent of the current information, $\bar\omega_\ell$ is Gaussian and independent of $\sigma(\Iset_{\ell h})$.
  % Moreover, since $\sigma(\hat\chi_0,\dots,\hat\chi_\ell) \subseteq \sigma(\Iset_{\ell h})$,
  % % $\bar\omega_\ell$ is also independent of $\sigma(\hat\chi_0,\dots,\hat\chi_\ell)$.
  % we have $\sigma(\bar\omega_\ell) \perp \sigma(\hat\chi_0,\dots,\hat\chi_\ell)$.
  % Furthermore, the mappings $\Phi_{\iota(\hat\chi_\ell)}$ and $\Gamma_{\iota(\hat\chi_\ell)}$ depend only on $\hat\chi_\ell$.
  The sequence $\{\omega_k\}_{k \in \Z_{\geq 0}}$ is an i.i.d. Gaussian process and independent of $\hat\chi_0$.
  Hence, by the independence of disjoint collections of independent random variables \cite[Th.~2.1.9]{durrett2019probability}, we have $\sigma(\bar\omega_\ell) \perp \sigma(\hat\chi_0,\bar\omega_0,\dots,\bar\omega_{\ell-1})$.
  Since $\sigma(\hat\chi_0,\dots,\hat\chi_\ell) \subseteq \sigma(\hat\chi_0,\bar\omega_0,\dots,\bar\omega_{\ell-1})$ holds from \eqref{eq:subsampled_estimate_dynamics}, we also have $\sigma(\bar\omega_\ell) \perp \sigma(\hat\chi_0,\dots,\hat\chi_\ell)$.
  Therefore, by \cite[Ex.~4.1.7]{durrett2019probability}, for every Borel set $\Aset \in \Borel(\R^{n_x})$, we obtain
  \begin{align}
    &\Prob\left(\hat\chi_{\ell+1} \in \Aset \mid \sigma(\hat\chi_0,\dots,\hat\chi_\ell)\right) \\
    &=
    \Prob\left(\Phi_{\iota(\hat\chi_\ell)} \hat\chi_\ell + \Gamma_{\iota(\hat\chi_\ell)} \bar\omega_\ell \in \Aset \mid \sigma(\hat\chi_0,\dots,\hat\chi_\ell)\right) \\
    &=
    \Prob\left(\hat\chi_{\ell+1} \in \Aset \mid \hat\chi_\ell\right)\quad \mathrm{a.s.},
  \end{align}
  which shows that $\{\hat\chi_\ell\}_{\Z \geq 0}$ is a Markov chain.
\end{proof}

% For later use, let $P$ denote the transition kernel of $\{\hat{\chi}_\ell\}_{\ell \in \Z_{\geq 0}}$.
% From \eqref{eq:subsampled_estimate_dynamics}, it is represented by
% \begin{equation} \label{eq:estimate_chain_gaussian_transition_kernel}
%   P(x,\Aset)
%   =
%   \int_\Aset p_{\iota(x)}(x,z)\,dz,
%   \quad
%   x \in \R^{n_x},
%   \ \Aset \in \Borel(\R^{n_x}),
% \end{equation}
% where $p_m(x,\cdot)$ is the Gaussian density with mean $\Phi_m x$ and covariance matrix $\Sigma_m$.
% The matrix $\Sigma_m$ is given by
% \begin{align}
%   \Omega_\omega
%   &\coloneqq
%   G \bigl(CA\Sigma A^\top C^\top + C\Omega_w C^\top + \Omega_v\bigr) G^\top, \\
%   \bar\Omega_\omega
%   &\coloneqq
%   I_h \otimes \Omega_\omega,\quad \Sigma_m = \Gamma_m \bar\Omega_\omega \Gamma_m^\T.
% \end{align}
% Moreover, we can represent $p(x,\cdot)$ explicitly by
% \begin{equation} \label{eq:gaussian_distribution}
%   p_m(x,z)
%   =
%   \frac{
%     \exp\left(
%       -\frac12 (z-\Phi_m x)^\top \Sigma_m^{-1}(z-\Phi_m x)
%     \right)
%   }{
%     (2\pi)^{n_x/2}\sqrt{\det(\Sigma_m)}
%   }.
% \end{equation}
% Since $p_m(x,\cdot)$ is a Gaussian density for each $m \in \Mset$ and $\Mset$ is finite, the right-hand side defines the transition kernel of $\{\hat{\chi}_\ell\}_{\ell \in \Z_{\geq 0}}$.
The Markov property of $\{\hat{\chi}_\ell\}_{\ell \in \Z_{\geq 0}}$ plays an important role in deriving the strict performance improvement result.

\subsection{Mean-Square Stability} \label{subsec:stability}

To prove mean-square stability of the closed-loop system, the following lemmas are used.

\begin{lemma}
  \label{lemma:periodic_candidate_initial_riccati_matrix}
  Consider Algorithm~1 under Assumptions~\ref{assump:no_periodic_eigenvalue_aliasing}, \ref{assump:lookahead_horizon_multiple_of_period} and \ref{assump:state_weight_and_stationary_kalman_filter}.  
  Then it holds that $P^{(1)}_0 = \tilde{P}_p$ for any $p$ satisfying Assumptions.
  % \begin{equation}
  %   J(\mu^{u, \text{per}}, \mu^{\delta, \text{per}})
  %   =
  %   \limsup_{L \to \infty} \frac1h \left(\beta^{(1)} + \gamma^{(1)}\right),
  % \end{equation}
  % where
  % \begin{align}
  %   \beta^{(1)} &= \frac{h}{p}\left( \tr\left(\tilde{P}_p D_p \Omega_w^{(p)} D_p^\T\right) + \tr\left(M_p \Sigma\right) + d(p) \right), \label{eq:periodic_candidate_quadratic_cost_constant} \\
  %   \gamma^{(1)} &= \frac{h}{p}\theta. \label{eq:periodic_candidate_actuation_penalty_constant}
  % \end{align}
\end{lemma}

\begin{proof}
  Define the following function:
  \begin{align}
    J_h(\Iset_k) &\coloneqq \E\Biggl[x_{k+h}^\T \tilde{P}_p x_{k+h} \\
    &\quad + \sum_{i=k}^{k+h-1} \bigl(x_i^\T Q x_i + u_i^\T R u_i + \theta \delta_i \bigr) \,\bigg|\, \Iset_k\Biggr].
  \end{align}
  Then we consider the following $h$-stage cost minimization problem.
  \begin{mini}|s|
    {\{u_i,\delta_i\}_{i=k}^{k+h-1}}
    {J_h(\Iset_k)}
    {\label{prob:periodic_candidate_h_stage_cost_minimization}}{}
    {}
    \addConstraint{x_{i+1}}{= A x_i + B u_i + w_i}
    \addConstraint{y_i}{= C x_i + v_i}
    % \addConstraint{\delta_i}{\in \{0,1\}}
    \addConstraint{u_i}{= 0 \quad \text{if } \delta_i = 0,\quad \delta_i \in \{0,1\}.}
  \end{mini}
  Let $H \coloneqq h/p \in \N$.
  % If the periodic triggering policy with period $p$ is optimal for \eqref{prob:periodic_candidate_h_stage_cost_minimization}, then $J_h$ can be rewritten as
  Fix $m=1$, which corresponds to the periodic triggering policy with period $p$ over the $h$-step horizon.
  Then $J_h$ can be rewritten as
  \begin{align}
    J_h(\Iset_k) &= \E\Biggl[x_{k+h}^\T \tilde{P}_p x_{k+h} + \sum_{j=0}^{H-1} \sum_{i=0}^{p-1} x_{k+jp+i}^\T Q x_{k+jp+i} \nonumber \\
    &\qquad + u_{k+jp+i}^\T R u_{k+jp+i} + \theta \delta_{k+jp+i} \,\Big|\, \Iset_k\Biggr] \\
    &= \E\Biggl[x_{k+h}^\T \tilde{P}_p x_{k+h} + \sum_{j=0}^{H-1}  x_{k+jp}^\T Q_p x_{k+jp} \nonumber \\
    &\qquad + 2x_{k+jp}^\T S_p u_{k+jp} + u_{k+jp}^\T R_p u_{k+jp} \,\Big|\, \Iset_k\Biggr] \\
    &\quad + H d(p) + H\theta.
  \end{align}
  By applying the dynamic programming principle, we obtain the following recursive equation:
  \begin{align}
    P_H &= \tilde{P}_p, \\
    P_t &= Q_p + (A^p)^\T P_{t+1} A^p - \left((A^p)^\T P_{t+1} B_p + S_p \right) \\
    &\qquad \times \left(B_p^\T P_{t+1} B_p + R_p\right)^{-1} \left(B_p^\T P_{t+1} A^p + S_p^\T \right), \nonumber \\
    &\quad t \in \{0,1,\dots,H-1\}. \label{eq:periodic_candidate_riccati_recursion}
  \end{align}
  Comparing \eqref{eq:periodic_candidate_riccati_recursion} with \eqref{eq:algebraic_riccati_periodic} shows that $P_t = \tilde{P}_p$ for all $t \in \{0,1,\dots,H-1\}$.
  % Therefore, the optimal value of \eqref{prob:periodic_candidate_h_stage_cost_minimization} corresponding to the periodic policy is given by
  Therefore, the minimum value of $J_h(\mathcal I_k)$ with respect to the control inputs, under the periodic triggering sequence $m=1$, is given by
  \begin{equation}
    \label{eq:periodic_candidate_h_step_cost}
    J_h(\Iset_k) = \hat{x}_k^\T \tilde{P}_p \hat{x}_k + \tr\left(\tilde{P}_p \Sigma\right) + \beta^{(1)} + \gamma^{(1)},
  \end{equation}
  where
  \begin{align}
    \beta^{(1)} &= \frac{h}{p}\left( \tr\left(\tilde{P}_p D_p \Omega_w^{(p)} D_p^\T\right) + \tr\left(M_p \Sigma\right) + d(p) \right), \label{eq:periodic_candidate_quadratic_cost_constant} \\
    \gamma^{(1)} &= \frac{h}{p}\theta. \label{eq:periodic_candidate_actuation_penalty_constant}
  \end{align}
  Comparing \eqref{eq:candidate_sequence_h_step_cost} and \eqref{eq:periodic_candidate_h_step_cost}, we obtain $P^{(1)}_0 = \tilde{P}_p$.
  % Furthermore, from \eqref{eq:periodic_actuation_rate} and \eqref{eq:optimal_periodic_control_cost},
  % \begin{align}
  %   &\limsup_{L \to \infty} \frac{1}{h} \left(\beta^{(1)} + \gamma^{(1)}\right) \\
  %   &= \frac{1}{p} \left(\tr\left(\tilde{P}_p D_p \Omega_w^{(p)} D_p^\T\right) + \tr\left(M_p \Sigma\right) + d(p) + \theta\right) \\
  %   &= J(\mu^{u,\text{per}}, \mu^{\delta,\text{per}}), \label{eq:periodic_average_cost_identity}
  % \end{align}
  % which completes the proof.
\end{proof}

Lemma~\ref{lemma:periodic_candidate_initial_riccati_matrix} shows that the Riccati solution of \eqref{eq:candidate_sequence_riccati_recursion} under the periodic schedule $m=1$ coincides with that of the infinite-horizon problem under \eqref{eq:periodic_triggering_policy}.

In order to prove the mean-square stability of the closed-loop system, we use
\begin{equation}
  \lambda_{\min}(Q)\E\left[\|x_{\ell h+\tau}\|^2\right]
  \leq
  \beta^{(1)} + \gamma^{(1)} + \E\left[V_1(\chi_\ell)\right],
  \ \forall\ell \in \Z_{\geq 0}
\end{equation}
for any $\tau = 0,\dots,h-1$.
Hence, we next establith the uniform boundness of $\E[V_1(\chi_\ell)]$.

To this end, we define a function $\bar c$ which is the control cost over an $h$-step interval under the control and triggering policies obtained by Algorithm~1 as follows:
\begin{align}
    &\bar{c}(\chi_\ell, \hat{\chi}_\ell, \bar{w}_\ell, \bar{\omega}_\ell) \\
    &\coloneqq \sum_{\tau=0}^{h-1} c\left(x_{\ell h+\tau},\; \rho^{(\iota(\hat{\chi}_\ell))}_\tau F^{(\iota(\hat{\chi}_\ell))}_\tau \hat{x}_{\ell h+\tau},\; \rho^{(\iota(\hat{\chi}_\ell))}_\tau\right),
    \label{eq:accumulated_h_step_stage_cost}
\end{align}
where $\bar{w}_\ell \coloneqq [w_{\ell h}^\T\ w_{\ell h+1}^\T\ \dots w_{(\ell+1)h-1}^\T]^\T$ and $\bar\omega_\ell$ be given by \eqref{eq:bar_omega_ell}.
Since Algorithm~1 minimizes the $h$-step cost in \eqref{prob:h_step_lookahead_quadratic_terminal_cost} over all $m\in\mathcal M$, for every $\ell \geq 0$, we have
\begin{align}
    &\E\Biggl[
        \sum_{\tau=0}^{h-1}
        c(x_{\ell h+\tau}, \mu^{u,\text{ro}}_{\ell h + \tau}(\Iset_{\ell h+\tau}), \mu^{\delta,\text{ro}}_{\ell h + \tau}(\Iset_{\ell h+\tau})) \\
        &\quad + x_{(\ell+1)h}^{\T}\tilde P_p x_{(\ell+1)h} \Biggm\vert \Iset_{\ell h}
    \Biggr] \\
    &=
    \E\left[
        \bar{c}(\chi_\ell, \hat{\chi}_\ell, \bar{w}_\ell, \bar{\omega}_\ell)
        + V_1(\chi_{\ell+1})\,\middle|\, \Iset_{\ell h}
    \right] \\
    &= \hat{\chi}_\ell^\T P^{(\iota(\hat{\chi}_\ell))}_0 \hat{\chi}_\ell
    +
    \tr\left(P^{(\iota(\hat{\chi}_\ell))}_0 \Sigma\right) \\
    &\quad + \beta^{(\iota(\hat{\chi}_\ell))} + \gamma^{(\iota(\hat{\chi}_\ell))},
    \label{eq:conditional_h_step_rollout_cost_identity}
\end{align}
where the first equality follows by the definitions of $\bar c$ and $V_1$ in \eqref{eq:accumulated_h_step_stage_cost} and \eqref{eq:quadratic_lyapunov_function_Vm}, respectively, and the second equality follows from \eqref{eq:rollout_candidate_sequence_selection_rule}.

The following lemma establishes the uniform boundness of $\E[V_1(\chi_\ell)]$, which will be used in the subsequent stability and performance analyses.

\begin{lemma} \label{lemma:uniform_lyapunov_moment_bound}
  Under Assumptions~\ref{assump:no_periodic_eigenvalue_aliasing}, \ref{assump:lookahead_horizon_multiple_of_period}, and \ref{assump:state_weight_and_stationary_kalman_filter}, we have
  \begin{equation} \label{eq:uniform_lyapunov_moment_bound}
    % \limsup_{L \to \infty} \E\left[V_1(\chi_L) \midd \hat{\chi}_0\right] < \infty.
    \sup_{\ell\in\mathbb{Z}_{\geq 0}}\mathbb{E}\left[V_1(\chi_\ell)\right]<\infty.
  \end{equation}
\end{lemma}

\begin{proof}
  From \eqref{eq:conditional_h_step_rollout_cost_identity}, we have
  \begin{align}
    &\E\left[
      V_1(\chi_{\ell+1}) + \bar{c}\left(\chi_\ell, \hat{\chi}_\ell, \bar{w}_\ell, \bar{\omega}_\ell\right) \midd \hat{\chi}_\ell
    \right] \\
    &= \hat{\chi}_\ell^\T P^{(\iota(\hat{\chi}_\ell))}_0 \hat{\chi}_\ell
    +
    \tr\left(P^{(\iota(\hat{\chi}_\ell))}_0 \Sigma\right) \\
    &\quad + \beta^{(\iota(\hat{\chi}_\ell))} + \gamma^{(\iota(\hat{\chi}_\ell))},
    \label{eq:conditional_rollout_cost_identity_for_stability}
  \end{align}
  where we use
  \begin{align}
    &\E\left[
      \bar{c}(\chi_\ell, \hat{\chi}_\ell, \bar{w}_\ell, \bar{\omega}_\ell)
      + V_1(\chi_{\ell+1})\,\middle|\, \Iset_{\ell h}
    \right] \\
    &=
    \E\left[
      \bar{c}(\chi_\ell, \hat{\chi}_\ell, \bar{w}_\ell, \bar{\omega}_\ell)
      + V_1(\chi_{\ell+1})\,\middle|\, \hat\chi_\ell
    \right].
  \end{align}
  Given that the function $g$ defined in \eqref{eq:rollout_improvement_function_g} satisfies $g(x) \geq 0$ for any $x \in \R^{n_x}$, we have
  \begin{align}
    &\hat\chi_\ell^\T P^{(\iota(\hat\chi_\ell))}_0 \hat\chi_\ell
    + \tr\left(P^{(\iota(\hat\chi_\ell))}_0 \Sigma\right)
    + \beta^{(\iota(\hat\chi_\ell))}
    + \gamma^{(\iota(\hat\chi_\ell))} \\
    &\leq
    \hat\chi_\ell^\T P^{(1)}_0 \hat\chi_\ell
    + \tr\left(P^{(1)}_0 \Sigma\right)
    + \beta^{(1)}
    + \gamma^{(1)}. \label{eq:rollout_cost_bounded_by_periodic_candidate}
  \end{align}
  Hence, combining \eqref{eq:conditional_rollout_cost_identity_for_stability} and \eqref{eq:rollout_cost_bounded_by_periodic_candidate} yields
  \begin{align}
    &\E\left[V_1(\chi_{\ell+1}) + \bar{c}(\chi_\ell, \hat{\chi}_\ell, \bar{w}_\ell, \bar{\omega}_\ell) \midd \hat{\chi}_\ell\right] \\
    &\leq \E\left[V_1(\chi_\ell) \midd \hat{\chi}_\ell\right] + \beta^{(1)} + \gamma^{(1)},
  \end{align}
  which results in
  \begin{align}
    \E\left[V_1(\chi_{\ell+1}) \midd \hat{\chi}_\ell\right]
    &\leq \E\left[V_1(\chi_\ell) \midd \hat{\chi}_\ell\right] + \beta^{(1)} + \gamma^{(1)} \\
    &\quad -
    \E\left[\bar{c}(\chi_\ell, \hat{\chi}_\ell, \bar{w}_\ell, \bar{\omega}_\ell) \midd \hat{\chi}_\ell\right].
    \label{eq:conditional_lyapunov_one_block_inequality}
  \end{align}
  Since $R \succ 0$ and $\theta > 0$, it follows that
  \begin{align}
    &\E\left[\bar{c}(\chi_\ell, \hat{\chi}_\ell, \bar{w}_\ell, \bar{\omega}_\ell) \midd \hat{\chi}_\ell\right] \\
    &\geq \E\left[ \sum_{\tau=0}^{h-1} x_{\ell h+\tau}^\T Q x_{\ell h+\tau} \midd \hat{\chi}_\ell\right]
    \geq \E\left[x_{\ell h}^\T Q x_{\ell h} \midd \hat{\chi}_\ell\right] \\
    &= \E\left[(\hat\chi_\ell + e_{\ell h})^\T Q (\hat\chi_\ell + e_{\ell h}) \midd \hat{\chi}_\ell\right]
    =\hat\chi_\ell^\T Q \hat\chi_\ell + \tr(Q\Sigma) \\
    &\geq \lambda_{\min}(Q) \hat\chi_\ell^\T \hat\chi_\ell \label{eq:h_step_stage_cost_quadratic_lower_bound}.
  \end{align}
  Using \eqref{eq:h_step_stage_cost_quadratic_lower_bound} to the right-hand side of \eqref{eq:conditional_lyapunov_one_block_inequality}, we have
  \begin{align}
    \E\left[V_1(\chi_{\ell+1}) \midd \hat{\chi}_\ell\right]
    &\leq \beta^{(1)} + \gamma^{(1)} + \E\left[V_1(\chi_\ell) \midd \hat{\chi}_\ell\right] \\
    &\quad - \lambda_{\min}(Q)\hat{\chi}_\ell^\T \hat{\chi}_\ell.
  \end{align}
  Furthermore, for any constant $b_1$ such that $b_1 > \lambda_{\min}(Q) > 0$ and $b_1I_{n_x} - \tilde{P}_p \succ 0$, we obtain
  % \begin{equation}
  %   \E\left[V_1(\chi_\ell) \midd \hat{\chi}_\ell\right]
  %   \leq
  %   b_1\hat{\chi}_\ell^\T \hat{\chi}_\ell + b_1\tr\left(\Sigma\right)
  % \end{equation}
  % from \eqref{eq:quadratic_lyapunov_function_Vm}.
  \begin{align}
      \E\left[V_1(\chi_\ell) \midd \hat{\chi}_\ell\right]
      &=
      \E\left[\chi_\ell^\T P^{(1)}_0 \chi_\ell \midd \hat{\chi}_\ell\right] \\
      &=
      \hat\chi_\ell^\T P^{(1)}_0 \hat\chi_\ell + \tr\left(P^{(1)}_0\Sigma\right) \\
      &\leq
      b_1\hat{\chi}_\ell^\T \hat{\chi}_\ell + b_1\tr\left(\Sigma\right)
  \end{align}
  since $P^{(1)}_0=\tilde P_p$ from Lemma~\ref{lemma:periodic_candidate_initial_riccati_matrix}.
  Then we have
  \begin{equation}
    \E\left[V_1(\chi_{\ell+1}) \midd \hat{\chi}_\ell\right]
    \leq d_1 \E\left[V_1(\chi_\ell) \midd \hat{\chi}_\ell\right] + c_1, \label{eq:conditional_true_state_lyapunov_contraction}
  \end{equation}
  where
  \begin{equation}
    d_1 = 1 - \frac{\lambda_{\min}(Q)}{b_1} < 1, \ c_1 = \beta^{(1)} + \gamma^{(1)} + \lambda_{\min}(Q) \tr\left(\Sigma\right).
  \end{equation}
  Since we have
  \begin{align}
    &\E\left[V_1(\chi_\ell) \midd \hat{\chi}_\ell\right] = V_1(\hat{\chi}_\ell) + \tr\left(P^{(1)}_0\Sigma\right), \label{eq:current_state_conditional_lyapunov_decomposition} \\
    &\E\left[V_1(\chi_{\ell+1}) \midd \hat{\chi}_\ell\right] = \E\left[V_1(\hat{\chi}_{\ell+1}) \midd \hat{\chi}_\ell\right] + \tr\left(P^{(1)}_0\Sigma\right), \label{eq:next_state_conditional_lyapunov_decomposition}
  \end{align}
  substituting \eqref{eq:current_state_conditional_lyapunov_decomposition} and \eqref{eq:next_state_conditional_lyapunov_decomposition} into \eqref{eq:conditional_true_state_lyapunov_contraction} yields
  \begin{equation}
    \label{eq:estimated_state_lyapunov_drift_bound}
    \E\left[V_1(\hat{\chi}_{\ell+1}) \midd \hat{\chi}_\ell\right]
    \leq d_1 V_1(\hat{\chi}_\ell) + c_2,
  \end{equation}
  where $c_2 \coloneqq c_1 + (d_1-1)\tr(P^{(1)}_0\Sigma)$.
  % Using the tower property and the Markov property of $\{\hat{\chi}_\ell\}_{\ell\in\Z_{\ge 0}}$, we obtain
  % \begin{align}
  %   \E\left[V_1(\hat{\chi}_{\ell+1}) \midd \hat{\chi}_0\right]
  %   &=
  %   \E\left[
  %     \E\left[V_1(\hat{\chi}_{\ell+1}) \midd \sigma(\hat{\chi}_0,\dots,\hat\chi_\ell)\right] \midd \hat{\chi}_0
  %   \right] \\
  %   &=
  %   \E\left[
  %     \E\left[V_1(\hat{\chi}_{\ell+1}) \midd \hat\chi_\ell\right] \midd \hat{\chi}_0
  %   \right] \\
  %   &\leq d_1 \E\left[V_1(\hat{\chi}_\ell) \midd \hat{\chi}_0\right] + c_2,
  % \end{align}
  % where the last inequality follows from \eqref{eq:estimated_state_lyapunov_drift_bound}.
  By taking expectations on both side, we have
  \begin{equation}
    \E\left[V_1(\hat{\chi}_{\ell+1})\right]
    \leq d_1 \E\left[V_1(\hat{\chi}_\ell)\right] + c_2.
  \end{equation}
  Iterating this inequality for $\ell=0,\dots,L-1$ gives
  % \begin{equation}
  % \label{eq:iterated_estimae_lyapunov_bound}
  %   \E\left[V_1(\hat{\chi}_L) \midd \hat{\chi}_0\right]
  %   \leq d_1^L V_1(\hat{\chi}_0) + \sum_{i=0}^{L-1} d_1^i c_2.
  % \end{equation}
  % Taking expectations on both sides yields
  \begin{equation}
    \E\left[V_1(\hat{\chi}_L)\right]
    \leq d_1^L \E\left[V_1(\hat{\chi}_0)\right] + \sum_{i=0}^{L-1} d_1^i c_2.
  \end{equation}
  Since, from \eqref{eq:current_state_conditional_lyapunov_decomposition}, it holds that
  \begin{equation}
    \E\left[V_1(\chi_L)\right]
    =
    \E\left[V_1(\hat{\chi}_L)\right]
    +
    \tr\left(P_0^{(1)}\Sigma\right),
  \end{equation}
  we obtain
  \begin{equation}
  \label{eq:expected_true_state_lyapunov_upper_bound}
    \E\left[V_1(\chi_L)\right]
    \leq d_1^L \E\left[V_1(\hat{\chi}_0)\right] + \sum_{i=0}^{L-1} d_1^i c_2 + \tr\left(P_0^{(1)}\Sigma\right).
  \end{equation}

  Since $\hat\chi_0$ follows the Gaussian distribution, we have $\E[\|\hat\chi_0\|^2] < \infty$, since Gaussian random variables are square integrable \cite[Ex.~1.6.11]{durrett2019probability}.
  This implies $\E[V_1(\hat\chi_0)]<\infty$.
  Moreover, since $c_2<\infty$ and $0<d_1<1$, the right-hand side of \eqref{eq:expected_true_state_lyapunov_upper_bound} is uniformly bounded with respect to $L \in \Z_{\geq 0}$, which proves \eqref{eq:uniform_lyapunov_moment_bound}.
\end{proof}

Building upon the above lemmas, we establish the stability guarantee of the proposed algorithm as follows.

\begin{theorem}
  \label{thm:stability_guarantee}
  Consider Algorithm~1, and suppose that Assumptions~\ref{assump:no_periodic_eigenvalue_aliasing}, \ref{assump:lookahead_horizon_multiple_of_period}, and \ref{assump:state_weight_and_stationary_kalman_filter} hold.
  Then the closed-loop system under $(\mu^{u, \text{ro}}, \mu^{\delta, \text{ro}})$ is mean-square stable, i.e.,
  \begin{equation} \label{eq:uniform_mean_square_state_bound}
    \sup_{k \in \Z_{\geq 0}} \E\left[\|x_k\|^2\right] < \infty.
  \end{equation}
\end{theorem}

\begin{proof}
  % From Lemma~\ref{lemma:uniform_lyapunov_moment_bound}, we have
  % \begin{equation}
  %   \label{eq:sup_V1}
  %   \sup_{\ell \in \Z_{\geq 0}} \E\left[V_1(\chi_\ell)\right] < \infty.
  % \end{equation}
  % Moreover, we notice that
  % \begin{align}
  %   &\E\left[
  %     V_1(\chi_{\ell+1}) + \bar{c}\left(\chi_\ell, \hat{\chi}_\ell, \bar{w}_\ell, \bar{\omega}_\ell\right) \midd \hat{\chi}_\ell
  %   \right] \\
  %   &= \hat{\chi}_\ell \left(P^{(\iota(\hat{\chi}_\ell))}_0\right) \hat{\chi}_\ell
  %   +
  %   \tr\left(\left(P^{(\iota(\hat{\chi}_\ell))}_0 \right)\Sigma\right) \\
  %   &\quad + \beta^{(\iota(\hat{\chi}_\ell))} + \gamma^{(\iota(\hat{\chi}_\ell))},
  % \end{align}
  % since $P^{(1)}_0 = \tilde{P}_p$ from Lemma~\ref{lemma:periodic_candidate_initial_riccati_matrix} and $\iota(\hat{\chi}_\ell)$ is obtained by solving \eqref{prob:h_step_lookahead_quadratic_terminal_cost} with the control policy \eqref{eq:candidate_sequence_feedback_control}.
  From \eqref{eq:conditional_h_step_rollout_cost_identity}, we have
  \begin{align}
    &\E\left[V_1(\chi_{\ell+1}) + \bar{c}\left(\chi_\ell, \hat{\chi}_\ell, \bar{w}_\ell, \bar{\omega}_\ell\right) \midd \hat{\chi}_\ell\right] - \E\left[V_1(\chi_\ell) \midd \hat{\chi}_\ell\right] \\
    &= \hat{\chi}_\ell^\T \left(P^{(\iota(\hat{\chi}_\ell))}_0 - P^{(1)}_0 \right) \hat{\chi}_\ell + \tr\left(\left(P^{(\iota(\hat{\chi}_\ell))}_0 - P^{(1)}_0\right)\Sigma\right) \\
    &\quad + \beta^{(\iota(\hat{\chi}_\ell))} + \gamma^{(\iota(\hat{\chi}_\ell))} \\
    &= \beta^{(1)} + \gamma^{(1)} - g(\hat{\chi}_\ell) \\
    &\leq \beta^{(1)} + \gamma^{(1)}, \label{eq:h_step_rollout_performance_gap_bound}
  \end{align}
  where the inequality follows from $g(x) \geq 0$ for all $x$.
  By taking expectations on both sides and using linearity of the expectation, we have
  \begin{equation}
  \label{eq:finite_expected_h_step_stage_cost}
    \E\left[\bar{c}\left(\chi_\ell, \hat{\chi}_\ell, \bar{w}_\ell, \bar{\omega}_\ell\right)\right]
    \leq
    \beta^{(1)} + \gamma^{(1)} + \E\left[V_1(\chi_\ell)\right].
  \end{equation}
  Taking the supremum in $\ell$ on both sides of \eqref{eq:finite_expected_h_step_stage_cost} yields
  \begin{align}
    &\sup_{\ell \in \Z_{\geq 0}}
    \E\left[\bar{c}\left(\chi_\ell, \hat{\chi}_\ell, \bar{w}_\ell, \bar{\omega}_\ell\right)\right] \\
    &\leq
    \beta^{(1)} + \gamma^{(1)} + \sup_{\ell \in \Z_{\geq 0}} \E\left[V_1(\chi_\ell)\right]
    <
    \infty,
    \label{eq:sup_finite_expected_h_step_stage_cost}
  \end{align}
  where we use Lemma~\ref{lemma:uniform_lyapunov_moment_bound} for the second inequality.
  Since, from \eqref{eq:h_step_stage_cost_quadratic_lower_bound}, for each $\tau=0,\dots,h-1$, it holds that
  \begin{equation}
    \E\left[\bar{c}\left(\chi_\ell, \hat{\chi}_\ell, \bar{w}_\ell, \bar{\omega}_\ell\right)\right]
    \geq
    \lambda_{\min}(Q)\E\left[\|x_{\ell h+\tau}\|^2\right]
    \quad \forall \ell \in \Z_{\geq 0},
  \end{equation}
  we obtain from \eqref{eq:sup_finite_expected_h_step_stage_cost} that
  % $\sup_{\ell \in \Z_{\geq 0}} \E[\|x_{\ell h+\tau}\|^2] < \infty$
  \begin{equation}
    \sup_{\ell \in \Z_{\geq 0}} \E\left[\|x_{\ell h+\tau}\|^2\right] < \infty
  \end{equation}
  for every $\tau=0,\dots,h-1$, which establishes \eqref{eq:uniform_mean_square_state_bound}.
\end{proof}
  
  % \begin{equation}
  % \label{eq:finite_expected_h_step_stage_cost}
  %   \E\left[\bar{c}\left(\chi_\ell, \hat{\chi}_\ell, \bar{w}_\ell, \bar{\omega}_\ell\right)\right]
  %   \leq
  %   \beta^{(1)} + \gamma^{(1)} + \E\left[V_1(\chi_\ell)\right]
  %   <
  %   \infty,
  % \end{equation}
  % where we use Lemma~\ref{lemma:uniform_lyapunov_moment_bound} for the second inequality.
  
  % Taking expectations on both sides and using \eqref{eq:uniform_lyapunov_moment_bound} implies
  % \begin{equation}
  %   \label{eq:finite_expected_h_step_stage_cost}
  %   \sup_{\ell \in \Z_{\geq 0}}
  %   \E\left[
  %     \bar{c}\left(\chi_\ell, \hat{\chi}_\ell, \bar{w}_\ell, \bar{\omega}_\ell\right)
  %   \right]
  %   < \infty.
  % \end{equation}
  % From \eqref{eq:h_step_stage_cost_quadratic_lower_bound}, for each $\tau=0,\dots,h-1$, it holds that
  % \begin{equation}
  %   \E\left[\bar{c}\left(\chi_\ell, \hat{\chi}_\ell, \bar{w}_\ell, \bar{\omega}_\ell\right)\right]
  %   \geq
  %   \lambda_{\min}(Q)\E\left[\|x_{\ell h+\tau}\|^2\right]
  %   \quad \forall \ell \in \Z_{\geq 0}.
  %   % \leq
  %   % \E\left[\bar{c}\left(\chi_\ell, \hat{\chi}_\ell, \bar{w}_\ell, \bar{\omega}_\ell\right)\right],
  %   % \quad \forall \ell \in \Z_{\geq 0}.
  % \end{equation}
  % Taking the supremum in $\ell$ and using \eqref{eq:finite_expected_h_step_stage_cost} yields
  % \begin{equation}
  %   \sup_{\ell \in \Z_{\geq 0}} \E\left[\|x_{\ell h+\tau}\|^2\right] < \infty
  % \end{equation}
  % for $\tau=0,\dots,h-1$, which establishes \eqref{eq:uniform_mean_square_state_bound}.

\subsection{Performance Guarantee} \label{subsec:performance_guarantee}

This section establishes the \emph{nonstrict} performance improvement result in \eqref{eq:target_nonstrict_periodic_performance_bound}, which means that the rollout policy $(\mu^{u, \text{ro}}, \mu^{\delta, \text{ro}})$ performs no worse than the periodic policy $(\mu^{u, \text{per}}, \mu^{\delta, \text{per}})$ for the average cost criterion $J$ in \eqref{prob:sparsity_promoting_regularization}.
Obtaining performance guarantees for multi-step rollout algorithms is generally hard \cite{bertsekas2017dynamic-i}, and hence the following theorem is one of the main results of this paper.

\begin{theorem}
  \label{thm:performance_guarantee}
  Consider Algorithm~1, and suppose that Assumptions~\ref{assump:no_periodic_eigenvalue_aliasing}, \ref{assump:lookahead_horizon_multiple_of_period}, and \ref{assump:state_weight_and_stationary_kalman_filter} hold.
  Then the following inequality holds.
  \begin{equation} \label{eq:performance_guarantee}
    J(\mu^{u, \text{ro}}, \mu^{\delta, \text{ro}}) 
    \leq
    J(\mu^{u, \text{per}}, \mu^{\delta, \text{per}}).
  \end{equation}
\end{theorem}

\begin{proof}
  The proof builds upon the arguments in Theorem~\ref{thm:stability_guarantee}.
  We first rewrite the objective function $J$ under the rollout policies $(\mu^{u, \text{ro}}, \mu^{\delta, \text{ro}})$.
  % Let
  % \begin{align}
  % S_N &\coloneqq
  % \E\Bigg[
  %   \sum_{k=0}^{N-1}
  %   x_k^\T Qx_k
  %   +(\mu^{u,\mathrm{ro}}_k(\Iset_k))^\T R \mu^{u,\mathrm{ro}}_k(\Iset_k) \\
  %   &\quad +\theta\mu^{\delta,\mathrm{ro}}_k(\Iset_k)
  % \Bigg]. \label{eq:def_SN}
  % \end{align}
  % Since $\{S_{hL}/(hL)\}_{L\geq1}$ is a subsequence of $\{S_N/N\}_{N\geq1}$, we have
  % \begin{equation}
  % \label{eq:SN_limsup}
  % \limsup_{L\to\infty}\frac{S_{hL}}{hL}
  % \leq
  % \limsup_{N\to\infty}\frac{S_N}{N}.
  % \end{equation}
  % Conversely, writing $N=hL+r$ with $0\leq r<h$ and using the monotonicity of $S_N$, we obtain
  % \begin{equation}
  % \frac{S_N}{N}
  % \leq
  % \frac{h(L+1)}{N}
  % \frac{S_{h(L+1)}}{h(L+1)}.
  % \end{equation}
  % Since $h(L+1)/N\to1$ as $N\to\infty$, the reverse inequality of \eqref{SN_limsup} follows.
  % Hence, we have
  % \begin{equation}
  % \label{eq:inequality_SN}
  % \limsup_{N\to\infty}\frac{S_N}{N}
  % =
  % \limsup_{L\to\infty}\frac{S_{hL}}{hL}.
  % \end{equation}
  % For the policies $(\mu^{u, \text{ro}}, \mu^{\delta, \text{ro}})$,
  we have
  \begin{align}
    &J(\mu^{u, \text{ro}}, \mu^{\delta, \text{ro}}) \\
    &= \limsup_{N \to \infty}
    \frac{1}{N} \E\Biggl[
      \sum_{k=0}^{N-1}
      x_k^\T Q x_k
      + (\mu^{u,\text{ro}}_k(\Iset_k))^\T R \mu^{u,\text{ro}}_k(\Iset_k) \\
      &\quad + \theta \mu^{\delta,\text{ro}}_k(\Iset_k)
    \Biggr] \\
    &= \limsup_{L \to \infty} \frac{1}{hL}
    \E\Biggl[
      \sum_{\ell=0}^{L-1}
      \sum_{\tau=0}^{h-1}
      x_{\ell h+\tau}^\T Q x_{\ell h+\tau} \\
      &\quad + (\mu^{u,\text{ro}}_{\ell h+\tau}(\Iset_{\ell h+\tau}))^\T R \mu^{u,\text{ro}}_{\ell h+\tau}(\Iset_{\ell h+\tau})
      + \theta \mu^{\delta,\text{ro}}_{\ell h+\tau}(\Iset_{\ell h+\tau})
    \Biggr] \\
    &= \limsup_{L \to \infty} \frac{1}{hL}
    \E\left[
      \sum_{\ell=0}^{L-1}
      \bar{c}\left(\chi_\ell, \hat{\chi}_\ell, \bar{w}_\ell, \bar{\omega}_\ell\right)
    \right], \label{eq:average_cost_as_h_step_stage_cost}
  \end{align}
  where $\bar c$ is defined in \eqref{eq:accumulated_h_step_stage_cost}.
  % The second equality follows from and \eqref{eq:inequality_SN}.

  We recall that \eqref{eq:h_step_rollout_performance_gap_bound} holds for $(\mu^{u, \text{ro}}, \mu^{\delta, \text{ro}})$.
  Hence, summing both sides of \eqref{eq:h_step_rollout_performance_gap_bound} over $\ell=0,\dots,L-1$ yields
  \begin{align}
    &\sum_{\ell=0}^{L-1} \E\left[V_1(\chi_{\ell+1}) + \bar{c}\left(\chi_\ell, \hat{\chi}_\ell, \bar{w}_\ell, \bar{\omega}_\ell\right) \midd \hat{\chi}_\ell\right] \\
    &\quad - \sum_{\ell=0}^{L-1} \E\left[V_1(\chi_\ell) \midd \hat{\chi}_\ell\right] \\
    &\leq L\left(\beta^{(1)} + \gamma^{(1)}\right). \label{eq:summed_h_step_performance_inequality}
  \end{align}
  % Taking expectations of both sides, we have
  % \begin{align}
  %   &\E\left[\sum_{\ell=0}^{L-1} \bar{c}\left(\chi_\ell, \hat{\chi}_\ell, \bar{w}_\ell, \bar{\omega}_\ell\right)\right] \\
  %   &\leq
  %   L\left(\beta^{(1)} + \gamma^{(1)}\right)
  %   + \E\left[V_1(\chi_0)\right]
  %   - \E\left[V_1(\chi_L)\right] \\
  %   &\leq
  %   L\left(\beta^{(1)} + \gamma^{(1)}\right)
  %   + \E\left[V_1(\chi_0)\right]. \label{eq:finite_horizon_performance_bound}
  % \end{align}
  Taking expectations of both sides and using linearity of the expectation, we have
  \begin{align}
    &\E\left[\sum_{\ell=0}^{L-1} \bar{c}\left(\chi_\ell, \hat{\chi}_\ell, \bar{w}_\ell, \bar{\omega}_\ell\right)\right] \\
    &\leq
    L\left(\beta^{(1)} + \gamma^{(1)}\right)
    +
    \E\left[\sum_{\ell=0}^{L-1} V_1(\chi_\ell) - V_1(\chi_{\ell+1})\right] \\
    &=
    L\left(\beta^{(1)} + \gamma^{(1)}\right)
    + \E\left[V_1(\chi_0)\right]
    - \E\left[V_1(\chi_L)\right] \\
    &\leq
    L\left(\beta^{(1)} + \gamma^{(1)}\right)
    + \E\left[V_1(\chi_0)\right].
    \label{eq:finite_horizon_performance_bound}
  \end{align}
  Dividing both sides by $hL$ and taking the limit superior as $L \to \infty$, the left-hand side of \eqref{eq:finite_horizon_performance_bound} is equal to $J(\mu^{u, \text{ro}}, \mu^{\delta, \text{ro}})$ from \eqref{eq:average_cost_as_h_step_stage_cost}.
  % is expressed as
  % \begin{equation}
  %   \limsup_{L \to \infty} \frac{1}{hL} \E\left[\sum_{\ell=0}^{L-1} \bar{c}\left(\chi_\ell, \hat{\chi}_\ell, \bar{w}_\ell, \bar{\omega}_\ell\right)\right]
  %   = J(\mu^{u,\text{ro}}, \mu^{\delta,\text{ro}}) \label{eq:left}
  % \end{equation}
  % from \eqref{eq:average_cost_as_h_step_stage_cost}.
  On the right-hand side of \eqref{eq:finite_horizon_performance_bound}, we obtain
  \begin{align}
    &\limsup_{L \to \infty} \frac{1}{h} \left(\beta^{(1)} + \gamma^{(1)}\right) \\
    &= \frac{1}{p} \left(\tr\left(\tilde{P}_p D_p \Omega_w^{(p)} D_p^\T\right) + \tr\left(M_p \Sigma\right) + d(p) + \theta\right) \\
    &= J(\mu^{u,\text{per}}, \mu^{\delta,\text{per}}), \label{eq:periodic_average_cost_identity}
  \end{align}
  where the second equality follows from \eqref{eq:optimal_periodic_control_cost} and \eqref{eq:periodic_actuation_rate}.
  In addition, since $\E[V_1(\chi_0)] < \infty$ by Lemma~\ref{lemma:uniform_lyapunov_moment_bound}, we have
  \begin{equation}
    \limsup_{L \to \infty} \frac{1}{hL} \E\left[V_1(\chi_0)\right] = 0. \label{eq:vanishing_initial_lyapunov_average}
  \end{equation}
  Combining \eqref{eq:average_cost_as_h_step_stage_cost}, \eqref{eq:periodic_average_cost_identity}, and \eqref{eq:vanishing_initial_lyapunov_average} gives \eqref{eq:performance_guarantee}.
\end{proof}

% \begin{remark}
%   For each $\theta$ and $h$, the period $p$ satisfying $h/p \in \N$ can be selected so that the corresponding periodic control policy minimizes the cost $J_c(\mu^{u,\mathrm{per}},\mu^{\delta,\mathrm{per}})
%   + \theta/p$.
%   This policy can be viewed as the best periodic control policy, and we can compute $\tilde{P}_p$ using this policy.
%   Consequently, Theorem~\ref{thm:performance_guarantee} guarantees that, for the given $\theta$ and $h$, we can construct a rollout-based algorithm whose performance is no worse than that of the best periodic control policy.
% \end{remark}

\subsection{Strict Performance Guarantee} \label{subsec:strict_performance_guarantee}

This section establishes the \emph{strict} performance improvement result in \eqref{eq:target_strict_periodic_performance_improvement}.
Under the additional assumptions, we guarantee that the rollout policy strictly outperforms the periodic policy.
To this end, preliminary results on Markov chains are described in Appendix~\ref{append:preliminary}, which are used in the following performance analysis.

In order to derive the strict performance improvement, we impose the following assumptions.
\begin{assumption}
  \label{assump:strict_performance_improvement_conditions}
  \mbox{}
  \begin{enumerate}
    \item $\rank(C) = n_x$.
    \item There exists $m \in \Mset\setminus\{1\}$ such that $\lambda_{\max}(P^{(1)}_0-P^{(m)}_0) > 0$ or $\tr((P^{(1)}_0-P^{(m)}_0)\Sigma)+\beta^{(1)}-\beta^{(m)}+\gamma^{(1)}-\gamma^{(m)} > 0$.
  \end{enumerate}
\end{assumption}
Assumption~\ref{assump:strict_performance_improvement_conditions}(i) ensures that the steady-state Kalman gain $G$ has full row rank, since the covariance matrix $\tilde{\Sigma}$ is positive definite and we have $\rank G = \rank C$.
The conditions in Assumption~\ref{assump:strict_performance_improvement_conditions}(ii) are both checkable offline under Assumptions~\ref{assump:no_periodic_eigenvalue_aliasing}--\ref{assump:state_weight_and_stationary_kalman_filter}.

Let $f \coloneqq g+1$, where $g$ is defined in \eqref{eq:rollout_improvement_function_g}.
Since $g$ is nonnegative and measurable, $f$ is also nonnegative and measurable.
% Then we can state the following lemma.
Then we establish the $f$-ergodicity of the Markov chain $\{\hat\chi_\ell\}_{\ell \in \Z_{\geq 0}}$.
% that $\{\hat\chi_\ell\}_{\ell \in \Z_{\geq 0}}$ is $f$-ergodic.

\begin{lemma}
  \label{lemma:f_ergodicity_of_estimate_chain}
  Suppose that Assumptions~\ref{assump:no_periodic_eigenvalue_aliasing}, \ref{assump:lookahead_horizon_multiple_of_period}, \ref{assump:state_weight_and_stationary_kalman_filter}, and \ref{assump:strict_performance_improvement_conditions}(i) hold.
  Then $\{\hat\chi_\ell\}_{\ell \in \Z_{\geq 0}}$ is an $f$-ergodic chain.
\end{lemma}

\begin{proof}
  See Appendix~\ref{append:key_lemmas} and \ref{append:ergodicity}.
\end{proof}

Moreover, under Assumption~\ref{assump:strict_performance_improvement_conditions}(ii), the following lemma guarantees that, for at least one state, the choice of $m \in \Mset$ according to Algorithm~1 differs from the periodic one.

\begin{lemma} \label{lemma:pointwise_strict_candidate_advantage}
  Suppose that Assumption~\ref{assump:strict_performance_improvement_conditions}(ii) holds.
  Then there exist $x \in \R^{n_x}$ and $m \in \Mset\setminus\{1\}$ such that
  \begin{align}
    x^{\T} P^{(m)}_0 x
    + \tr\left(P^{(m)}_0\Sigma\right) 
    + \beta^{(m)} + \gamma^{(m)} \\
    <
    x^{\T} P^{(1)}_0 x
    + \tr\left(P^{(1)}_0\Sigma\right) 
    + \beta^{(1)} + \gamma^{(1)}. \label{eq:strict_candidate_cost_advantage}
  \end{align}
\end{lemma}

\begin{proof}
  Let us define $D_m \coloneqq P^{(1)}_0-P^{(m)}_0$ and $d_m \coloneqq \tr((P^{(1)}_0-P^{(m)}_0)\Sigma)+\beta^{(1)}-\beta^{(m)}+\gamma^{(1)}-\gamma^{(m)}$.
  If $d_m>0$ holds for some $m\neq1$, the choice $x=0$ satisfies \eqref{eq:strict_candidate_cost_advantage}.
  
  Next, suppose that $d_m \leq 0$ for every $m \in \Mset$.
  From Assumption~\ref{assump:strict_performance_improvement_conditions}(ii), there exists $m \neq 1$ such that $\lambda_{\max}(D_m)>0$.
  % Let $v$ be an eigenvector satisfying $D_m v = \lambda_{\max}(D_m)v$ and $\|v\|^2=1$.
  % For $a \in \R$, we take $x=av$.
  Let $x$ be an eigenvector satisfying $D_m x = \lambda_{\max}(D_m)x$ with sufficiently large $\|x\|$.
  Then we have
  % Since $\lambda_{\max}(D_m) > 0$, we have
  \begin{equation}
    x^\T D_m x + d_m
    % = a^2 v^\T D_m v + d_m
    % = a^2 \lambda_{\max}(D_m) + d_m.
    = \lambda_{\max}(D_m)\|x\|^2 + d_m
    > 0,
  \end{equation}
  since $\lambda_{\max}(D_m)>0$.
  % Since $\lambda_{\max}(D_m)>0$,  it holds that $a^2 \lambda_{\max}(D_m) + d_m > 0$ for sufficiently large $a$.
  Therefore, if $\lambda_{\max}(D_m)>0$ or $d_m>0$ holds for some $m \in \Mset \setminus \{1\}$, there exists $x$ such that \eqref{eq:strict_candidate_cost_advantage} holds.
  % which is positive for sufficiently large $\|x\|$.
\end{proof}

Building upon the above lemmas, we establish the following strict performance improvement result.

\begin{theorem}
  \label{thm:strict_performance_guarantee}
  Consider Algorithm~1, and suppose that Assumptions~\ref{assump:no_periodic_eigenvalue_aliasing}, \ref{assump:lookahead_horizon_multiple_of_period}, \ref{assump:state_weight_and_stationary_kalman_filter}, and \ref{assump:strict_performance_improvement_conditions} hold.
  % Then the following inequality holds.
  Then we have
  \begin{equation}
    \label{eq:exact_average_cost_improvement}
    J(\mu^{u, \text{ro}}, \mu^{\delta, \text{ro}})
    =
    J(\mu^{u, \text{per}}, \mu^{\delta, \text{per}})
    -
    \frac{1}{h} \int_{\R^{n_x}} g\, d\pi_{\hat\chi},
  \end{equation}
  where $\pi_{\hat\chi}$ is the invariant probability measure of $\{\hat\chi_\ell\}_{\ell \in \Z_{\geq 0}}$.
  Moreover, it holds that
  \begin{equation}
  \label{eq:positive_stationary_improvement}
    \int_{\R^{n_x}} g\, d\pi_{\hat\chi} > 0
  \end{equation}
  and
  \begin{equation} \label{eq:strict_performance_guarantee}
    J(\mu^{u, \text{ro}}, \mu^{\delta, \text{ro}}) 
    <
    J(\mu^{u, \text{per}}, \mu^{\delta, \text{per}}).
  \end{equation}
\end{theorem}

\begin{proof}
  First, we recall that
  \begin{align}
    &\E\left[V_1(\chi_{\ell+1}) + \bar{c}\left(\chi_\ell, \hat{\chi}_\ell, \bar{w}_\ell, \bar{\omega}_\ell\right) \midd \hat{\chi}_\ell\right] - \E\left[V_1(\chi_\ell) \midd \hat{\chi}_\ell\right] \\
    &= \beta^{(1)} + \gamma^{(1)} - g(\hat{\chi}_\ell)
    \label{eq:exact_h_step_rollout_improvement_identity}
  \end{align}
  from \eqref{eq:h_step_rollout_performance_gap_bound}.
  % By following the same argument in the proof of Theorem~\ref{thm:performance_guarantee}, we have
  Summing both sides over $\ell=0,\dots,L-1$, dividing by $hL$, and taking expectations yields
  \begin{align}
    &\frac{1}{hL}\E\left[\sum_{\ell=0}^{L-1} \bar{c}\left(\chi_\ell, \hat{\chi}_\ell, \bar{w}_\ell, \bar{\omega}_\ell\right)\right] \\
    &=
    \frac{1}{h} \left(\beta^{(1)} + \gamma^{(1)}\right)
    - \frac{1}{hL} \E\left[\sum_{\ell=0}^{L-1} g(\hat\chi_\ell)\right] \\
    &\quad + \frac{1}{hL}\left(\E\left[V_1(\chi_0)\right] - \E\left[V_1(\chi_L)\right]\right) \label{eq:finite_horizon_strict_improvement_decomposition}
  \end{align}
  From \eqref{eq:uniform_lyapunov_moment_bound} and \eqref{eq:vanishing_initial_lyapunov_average}, it holds that
  \begin{equation}
    \label{eq:vanishing_lyapunov_boundary_average}
    \lim_{L \to \infty}
    \frac{1}{hL}\left(\E\left[V_1(\chi_0)\right] - \E\left[V_1(\chi_L)\right]\right)
    =
    0.
  \end{equation}
  Moreover, since $\{\hat\chi_\ell\}_{\ell \in \Z_{\geq 0}}$ is $f$-ergodic and $\E[V_1(\hat\chi_0)]<\infty$, Proposition~\ref{prop:f_ergodic_expected_time_average} in Appendix~\ref{append:preliminary} yields
  \begin{equation}
    \label{eq:ergodic_average_of_improvement}
    \lim_{L \to \infty}
    \frac{1}{hL} \E\left[\sum_{\ell=0}^{L-1} g(\hat\chi_\ell)\right]
    =
    \frac{1}{h} \int_{\R^{n_x}} g\, d\pi_{\hat\chi}.
  \end{equation}
  % where $\pi_{\hat\chi}$ is an invariant measure of the chain $\{\hat\chi_\ell\}_{\ell \in \Z_{\geq 0}}$.
  Hence, by \eqref{eq:average_cost_as_h_step_stage_cost}, \eqref{eq:periodic_average_cost_identity}, \eqref{eq:vanishing_lyapunov_boundary_average} and \eqref{eq:ergodic_average_of_improvement}, taking the limit supremum in \eqref{eq:finite_horizon_strict_improvement_decomposition} gives
  \begin{equation}
    J(\mu^{u, \text{ro}}, \mu^{\delta, \text{ro}})
    =
    J(\mu^{u, \text{per}}, \mu^{\delta, \text{per}})
    -
    \frac{1}{h} \int_{\R^{n_x}} g\, d\pi_{\hat\chi},
  \end{equation}
  which establish \eqref{eq:exact_average_cost_improvement}.
  % Therefore, it suffices to show that $\int_{\R^{n_x}} g\, d\pi_{\hat\chi} > 0$.

  Next, we show \eqref{eq:positive_stationary_improvement} and \eqref{eq:strict_performance_guarantee}.
  From Lemma~\ref{lemma:pointwise_strict_candidate_advantage}, there exists $x_\star \in \R^{n_x}$ such that $g(x_\star) > 0$.
  We define a constant $\varepsilon \coloneqq g(x_\star)/2 > 0$ and a set
  \begin{equation}
    \Oset_\varepsilon
    \coloneqq
    g^{-1}((\varepsilon,\infty))
    =
    \{x \in \R^{n_x} \colon g(x) > \varepsilon\},
  \end{equation}
  where $g^{-1}((\varepsilon,\infty))$ denotes the inverse image of $(\varepsilon,\infty)$ under $g$.
  Note that $\Oset_\varepsilon$ is nonempty since $x_\star \in \Oset_\varepsilon$.
  By the definition, it holds that
  \begin{equation}
    \label{eq:positive_improvement_on_open_set}
    g(x) \geq \varepsilon,\quad \forall x \in \Oset_\varepsilon.
  \end{equation}
  Since the minimum of finitely many continuous functions is continuous, $g$ is continuous.
  Combining this property with the fact that $(\varepsilon,\infty)$ is an open set, $\Oset_\varepsilon$ is also an open set \cite[Th.~4.8]{rudin1976principles}.
  Hence, we have $\pi_{\hat\chi}(\Oset_\varepsilon)>0$ since $\{\hat\chi_\ell\}_{\ell \in \Z_{\geq 0}}$ is open set irreducible from Lemma~\ref{lemma:open_set_irreducible_aperiodic_t_chain}.
  Consequently, we obtain
  \begin{equation}
    \int_{\R^{n_x}} g\, d\pi_{\hat\chi}
    \geq
    \int_{\Oset_\varepsilon} g\, d\pi_{\hat\chi}
    \geq
    \varepsilon \pi_{\hat\chi}(\Oset_\varepsilon)
    >
    0,
  \end{equation}
  which proves \eqref{eq:positive_stationary_improvement}.
  Conbining \eqref{eq:exact_average_cost_improvement} and \eqref{eq:positive_stationary_improvement} implies \eqref{eq:strict_performance_guarantee}.
\end{proof}

For a given $x$, $g(x)$ represents the performance improvement achieved by Algorithm~1 over a single $h$-step window.
Accordingly, \eqref{eq:exact_average_cost_improvement} characterizes the performance improvement over the infinite horizon, while $\frac{1}{h}\int_{\R^{n_x}} g\,d\pi_{\hat\chi}$ quantifies the corresponding expected reduction in the average cost compared to the periodic policy.

\begin{remark}
The proof of Theorem~\ref{thm:strict_performance_guarantee} follows an approach similar to that in \cite{antunes2014rollout}, which investigates the ZOH setting with full-state observation.
However, the analysis in \cite{antunes2014rollout} cannot be directly applied to our setting.
In \cite{antunes2014rollout}, the Markov process includes both the state and the previously applied control input due to the ZOH mechanism.
In contrast, no control input is retained between actuations in our framework, and the triggering variables are determined based on state estimates rather than the actual states.
Hence, our Markov process is driven by the state estimate $\hat\chi_\ell$, whose transition kernel differs from that in \cite{antunes2014rollout}.
In particular, under full-state observation, the irreducibility of the Markov chain can be guaranteed by the positive definiteness of the noise covariance matrix.
However, in our setting, the conditional covariance matrix of $\hat\chi_\ell$ given by \eqref{eq:covariance_matrix_hatchi_ell} is not necessarily positive definite, even when $\Omega_w$ and $\Omega_v$ are positive definite.
This makes it difficult to establish the irreducibility of the chain $\{\hat\chi_\ell\}_{\ell \in \Z_{\geq 0}}$, thereby affecting the subsequent analysis.
To overcome this difficulty, we use Assumption~\ref{assump:strict_performance_improvement_conditions}(i) to ensure that the conditional covariance matrix of $\hat\chi_\ell$ is positive definite.
Consequently, the one-step transition kernel of the chain $\{\hat\chi_\ell\}_{\ell \in \Z_{\geq 0}}$ has an everywhere positive Gaussian density, which allows us to establish open set irreducibility.
% Therefore, the $f$-ergodicity of this process must be established separately.
% We explicitly establish its open-set irreducibility, T-chain property, and strong aperiodicity, and verify a drift condition for $f=g+1$ to prove $f$-ergodicity.
% Together with the condition $\E[V_1(\hat\chi_0)]<\infty$, we use $f$-ergodicity to justify the convergence in \eqref{eq:ergodic_average_of_improvement} for the generally unbounded function $g$.
\end{remark}
  
% \begin{remark}
%   For a given state $x$, $g(x)$ represents the performance improvement achieved by Algorithm~1 over a single $h$-step window.
%   Accordingly, \eqref{eq:exact_average_cost_improvement} characterizes the performance improvement over the infinite horizon, while $\frac{1}{h}\int_{\R^{n_x}} g\,d\pi_{\hat\chi}$ quantifies the corresponding expected reduction in the average cost compared to the periodic policy.
%   As in \cite{antunes2014rollout}, evaluating the exact value of $\frac{1}{h}\int_{\R^{n_x}} g\,d\pi_{\hat\chi}$ is generally hard since the invariant distribution $\pi_{\hat\chi}$ is not available explicitly.
%   Nevertheless, it can be approximated numerically using Monte Carlo simulation \cite{athreya2003estimating}.
% \end{remark}

\section{Event-Based Control With a Hard Constraint}
\label{sec:hard_constraint}

In the previous sections, we investigated event-based control via sparsity-promoting regularization.
In this section, we consider an alternative formulation in which the average actuation rate is explicitly constrained.
Our objective is to design control and triggering policies that guarantee performance and stability while satisfying a prescribed constraint on the actuation rate.

Based on the setup in Section~\ref{sec:problem_formulation}, we consider the following optimal control problem:
\begin{mini}|s|
  {\mu^u,\mu^{\delta}}
  {J_c(\mu^u,\mu^{\delta})}
  {\label{prob:fixed_average_actuation_rate_control}}
  {}
  \addConstraint{J_r(\mu^{\delta}) = \bar{r}}
  \addConstraint{u_k = 0\quad\text{if}\ \delta_k = 0,}
\end{mini}
where $\bar{r}\in(0,1]$ denotes the prescribed actuation rate.

The problem~\eqref{prob:fixed_average_actuation_rate_control} differs from the problem considered in the previous sections in how the trade-off between control performance and the actuation rate is specified.
In the sparsity-promoting formulation~\eqref{prob:sparsity_promoting_regularization}, the actuation rate is penalized through the weighting parameter $\theta$.
However in this formulation, for a given value of $\theta$, the actuation rate achieved by the resulting policy is generally not known in advance.
In contrast, \eqref{prob:fixed_average_actuation_rate_control} explicitly prescribes the actuation rate by restricting the admissible policies to those satisfying $J_r(\mu^\delta)=\bar r$.

Although the problems \eqref{prob:fixed_average_actuation_rate_control} and \eqref{prob:sparsity_promoting_regularization} share the same system model, information structure, and actuation mechanism, Algorithm~1 cannot be directly applied to \eqref{prob:fixed_average_actuation_rate_control}.
This is because, in Algorithm~1, all actuation sequences over the lookahead horizon are considered, and the sequence minimizing the finite-horizon cost is selected.
Consequently, the number of active instants at which $\delta_k=1$ in the selected sequence is not fixed in advance, and the resulting triggering policy is not guaranteed to satisfy the rate constraint.

To address this issue, we slightly modify Algorithm~1 by restricting the candidate actuation sequences in the $h$-step window.
Specifically, we consider only those actuation sequences over the $h$-step lookahead horizon that yield the prescribed actuation rate.
To this end, we assume that $\bar r=1/p$ and define
\begin{equation}
  \Sset
  \coloneqq
  \left\{
    (\rho_0,\dots,\rho_{h-1})\in\{0,1\}^h
    \colon
    \sum_{j=0}^{h-1}\rho_j = h\bar r
  \right\}.
  \label{eq:fixed_rate_admissible_triggering_sequences}
\end{equation}
Then the set $\Sset$ contains $\binom{h}{h\bar r}$ actuation sequences.
Let $\Mset_{\bar r}$ be an index set for the elements of $\Sset$, and write
\begin{equation}
  \Sset
  =
  \left\{
    (\rho_0^{(m)},\dots,\rho_{h-1}^{(m)})
    \colon
    m\in\Mset_{\bar r}
  \right\}.
\end{equation}
Since each sequence in $\Sset$ contains exactly $h\bar r$ actuation instants in every $h$-step window, the resulting triggering policy satisfies $J_r(\mu^\delta)=\bar r$.
We assign $m=1$ to the periodic actuation sequence, which belongs to
$\Sset$ since $\bar r=1/p$ and $h/p\in\N$.

For each $m\in\Mset_{\bar r}$, the corresponding control policy is obtained in the same manner as in Section~\ref{subsec:proposed}.
Thus, the control cost associated with the $m$th admissible actuation sequence is given by
\begin{equation}
  \label{eq:fixed_rate_candidate_quadratic_control_cost}
  \hat{x}_k^{\T}P_0^{(m)}\hat{x}_k
  +
  \tr\left(P_0^{(m)}\Sigma_k\right)
  +
  \beta_k^{(m)}.
\end{equation}
The actuation sequence is then selected by minimizing \eqref{eq:fixed_rate_candidate_quadratic_control_cost} over $m\in\Mset_{\bar r}$.

Therefore, the rollout-based algorithm for \eqref{prob:fixed_average_actuation_rate_control} is summarized below.

\emph{Algorithm~2:}
\begin{enumerate}
  \item At time $k=\ell h$, $\ell\in\Z_{\geq0}$, compute
  \begin{equation}
    \label{eq:fixed_rate_rollout_sequence_selection_rule}
    \iota(\Iset_{\ell h})
    \coloneqq
    \underset{m\in\Mset_{\bar r}}{\argmin}\,
      \hat{x}_{\ell h}^{\T}P_0^{(m)}\hat{x}_{\ell h}
      +
      \tr\left(P_0^{(m)}\Sigma_{\ell h}\right)
      +
      \beta_{\ell h}^{(m)}.
  \end{equation}
  \item Set the triggering variables for $k=\ell h,\dots,(\ell+1)h-1$ as
  \begin{equation}
    \delta_k
    =
    \rho_{k-\ell h}^{(\iota(\Iset_{\ell h}))}.
  \end{equation}
  \item Compute the corresponding control input according to
  \begin{equation}
    \label{eq:fixed_rate_rollout_feedback_control}
    u_k
    =
    \rho_{k-\ell h}^{(\iota(\Iset_{\ell h}))}
    F_{k-\ell h}^{(\iota(\Iset_{\ell h}))}
    \hat{x}_k.
  \end{equation}
  \item Repeat steps 1), 2), and 3) at $k=(\ell+1)h$.
\end{enumerate}

Regarding the stability and performance analyses for Algorithm~2, the same frameworks developed in the previous sections remain applicable, since Algorithm~2 changes only the set of candidate actuation sequences.
For each fixed admissible actuation sequence, the control policy and the corresponding control cost are obtained in the same manner as in Algorithm~1.
Moreover, the periodic actuation sequence belongs to $\Sset$, and hence the $h$-step cost selected by Algorithm~2 is no larger than that associated with the periodic policy.
Therefore, the same comparison arguments can be applied directly, with the regularized cost replaced by the control cost.

More precisely, the following results hold for the rollout policy obtained by Algorithm~2.

\begin{corollary}
  Consider Algorithm~2, and suppose that Assumptions~\ref{assump:no_periodic_eigenvalue_aliasing}, \ref{assump:lookahead_horizon_multiple_of_period}, and \ref{assump:state_weight_and_stationary_kalman_filter} hold.
  Then the closed-loop system under $(\mu^{u,\text{ro}},\mu^{\delta,\text{ro}})$ is mean-square stable.
  Moreover, we have $J_r(\mu^{\delta,\text{ro}}) = J_r(\mu^{\delta,\text{per}}) = \bar r$ and
  \begin{equation}
    J_c(\mu^{u,\text{ro}},\mu^{\delta,\text{ro}})
    \leq
    J_c(\mu^{u,\text{per}},\mu^{\delta,\text{per}}).
  \end{equation}
\end{corollary}

\begin{corollary}
  Consider Algorithm~2, and suppose that Assumptions~\ref{assump:no_periodic_eigenvalue_aliasing}, \ref{assump:lookahead_horizon_multiple_of_period}, \ref{assump:state_weight_and_stationary_kalman_filter}, and \ref{assump:strict_performance_improvement_conditions}(i) hold.
  Suppose further that there exists $m\in\Mset_{\bar r}\setminus\{1\}$ such that $\lambda_{\max}(P^{(1)}_0-P^{(m)}_0)>0$ or $\tr((P^{(1)}_0-P^{(m)}_0)\Sigma) + \beta^{(1)}-\beta^{(m)} > 0$.
  Then we have $J_r(\mu^{\delta,\text{ro}}) = J_r(\mu^{\delta,\text{per}}) = \bar r$ and
  \begin{equation}
    J_c(\mu^{u,\text{ro}},\mu^{\delta,\text{ro}})
    <
    J_c(\mu^{u,\text{per}},\mu^{\delta,\text{per}}).
  \end{equation}
\end{corollary}

\section{Illustrative Example}
\label{sec:simulation_example}

In this section, we evaluate the effectiveness of the rollout-based algorithm for the sparsity-promoting regularization problem, which is proposed in Section~\ref{subsec:proposed}.

We consider a two-mass spring system studied in \cite{antunes2014rollout}.
The system consists of two masses $m_1$ and $m_2$ connected by a spring with stiffness $\kappa_s$.
The parameters are chosen as $m_1=m_2=1$ and $\kappa_s=2\pi^2$.
Both the control input and the process noise act only on mass $m_1$.
The continuous-time system is modeled by
\begin{align}
  dx(t) &= \left(A_C x(t) + B_C u(t)\right) dt + D_C dw(t), \label{eq:simulation_continuous_state_dynamics} \\
  dy(t) &= C_C x(t) dt + dv(t), \label{eq:simulation_continuous_output_dynamics}
\end{align}
where
\begin{align}
  A_C &= \begin{bmatrix}
    0 & 0 & 1 & 0\\
    0 & 0 & 0 & 1\\
    -\frac{\kappa_s}{m_1} & \frac{\kappa_s}{m_1} & 0 & 0\\
    \frac{\kappa_s}{m_2} & -\frac{\kappa_s}{m_2} & 0 & 0
  \end{bmatrix},\quad
  B_C = \begin{bmatrix}
    0\\
    0\\
    \frac{1}{m_1}\\
    0
  \end{bmatrix}, \\
  D_C &= \begin{bmatrix}
    0\\
    0\\
    0.4\\
    0
  \end{bmatrix},\quad
  C_C = \begin{bmatrix}
    1 & 0 & 0 & 0\\
    0 & 1 & 0 & 0
  \end{bmatrix}. \label{eq:simulation_diffusion_and_output_matrices}
\end{align}
The state vector is given by $x(t) = [x^1(t)\ x^2(t)\ \dot{x}^1(t)\ \dot{x}^2(t)]^{\T} \in \R^4$, where $x^1(t)$ and $x^2(t)$ represent the displacements of masses $m_1$ and $m_2$, respectively.
The noise processes $w(t)$ and $v(t)$ are assumed to be independent Wiener processes of dimensions one and $n_y$, respectively, with incremental covariance matrices $dt$ and $10^{-5}I_{n_y}dt$.
Discretizing \eqref{eq:simulation_continuous_state_dynamics} and \eqref{eq:simulation_continuous_output_dynamics} with a sampling period of $t_s=0.1$ yields the discrete-time system \eqref{eq:plant} and \eqref{eq:measurement}.
The resulting noise sequences $\{w_k\}$ and $\{v_k\}$ are zero-mean, i.i.d. Gaussian sequences with covariance matrices
\begin{equation}
  \Omega_w = \int_0^{t_s} e^{A_C \tau} D_C D_C^{\T} e^{A_C^{\T} \tau}\, d\tau, \quad
  \Omega_v = \frac{10^{-5}}{t_s} I_{n_y}.
\end{equation}
For the initial state, we set $\bar{x}_0=[1\ -1\ 0\ 0]^{\T}$ and choose $\Omega_{x_0}$ to satisfy \eqref{eq:stationary_prior_covariance_riccati_equation}.
The control horizon is set to 600 time steps, and the weighting matrices are chosen as
\begin{align}
  Q &= \begin{bmatrix}
    0.1336 & -0.0936 & -0.0327 & 0.0347\\
    -0.0936 & 0.1336 & 0.0347 & -0.0327\\
    -0.0327 & 0.0347 & 0.0377 & 0.0024\\
    0.0347 & -0.0327 & 0.0024 & 0.0377
  \end{bmatrix},\\
  R &= 0.1.
\end{align}

\begin{remark}
  The matrix $C_C$ in \eqref{eq:simulation_diffusion_and_output_matrices} does not satisfy Assumption~\ref{assump:strict_performance_improvement_conditions}(i).
  Hence, the strict performance improvement established in Theorem~\ref{thm:strict_performance_guarantee} is not theoretically guaranteed for this example.
  Nevertheless, we use this example to investigate the empirical performance of the proposed method in a setting where the sufficient conditions for the strict performance guarantee are not fulfilled.
\end{remark}

% We evaluate three methods for the problem~\eqref{prob:sparsity_promoting_regularization}.
% The first one is Algorithm~1 presented in Section~\ref{subsec:proposed}, with $h=6$.
% At $k=6\ell$ for $\ell \in \Z_{\geq 0}$, the triggering variables and the corresponding control inputs are computed according to \eqref{eq:joint_rollout_control_and_triggering_policies}.
% The second method is the periodic control described in Section~\ref{subsec:opt_periodic}, with $p \in \{1,2,3,6\}$.
% The third method is an $\ell_1$-relaxation approach, where $\delta_k$ in \eqref{eq:average_actuation_rate} is replaced by $\|u_k\|_{\ell_2/\ell_1}$, known as the $\ell_2/\ell_1$ norm \cite{eldar2010block}.
% To ensure robustness against disturbances, we employ an MPC scheme with a prediction horizon of 30 and solve the $\ell_1$-relaxed problem at each time step using CVX \cite{grant2014cvx}.
% This method is hereafter referred to as $\ell_1$-relax.+MPC.
% We conduct Monte Carlo simulations for each method using different values of the weighting parameter $\theta \in \{0, 0.02, \dots, 2\}$, and compare the means of the control cost and the actuation rate across trials for each $\theta$.
We compare three strategies for problem~\eqref{prob:sparsity_promoting_regularization}.
The first is the proposed rollout-based method, Algorithm~1 in Section~\ref{subsec:proposed}, with a lookahead horizon of $h=6$.
At each time $k=6\ell$, $\ell\in\Z_{\geq 0}$, the triggering sequence is determined for the subsequent six-step interval, and the control inputs are calculated according to \eqref{eq:joint_rollout_control_and_triggering_policies}.
The second strategy is the periodic policy introduced in Section~\ref{subsec:opt_periodic}, for which the control period is selected from $p\in\{1,2,3,6\}$.
This period is chosen to minimize the average cost $J_c(\mu^{u,\text{per}}, \mu^{\delta,\text{per}})+\theta/p$ for each $\theta$.
As the third strategy, we consider an $\ell_1$-relaxation of the binary variable $\delta_k$ in \eqref{eq:average_actuation_rate} with $\|u_k\|_{\ell_2/\ell_1}$, which is known as the combined $\ell_2/\ell_1$ norm \cite{eldar2010block}.
To account for the affect of disturbances, we solve the $\ell_1$-relaxed problem in an MPC fashion with a prediction horizon of 30.
% The resulting optimization problem is solved at every time step using CVX \cite{grant2014cvx}.
We refer to this strategy as $\ell_1$-relax.+MPC.
For each method, we conducted Monte Carlo simulations over 100 trials for $\theta\in\{0,0.02,\ldots,3\}$.
The empirical mean control cost and mean actuation rate are evaluated for each value of $\theta$.

% Fig.~\ref{fig:Jr_vs_Jc} shows the average control cost versus the average actuation rate, and Fig.~\ref{fig:theta_curves} presents each cost as a function of $\theta$.
% In Fig.~\ref{fig:theta_curves}, error bars indicate one standard deviation around the mean over 50 trials.
% For the periodic control, $p \in \{1,2,3,6\}$ is chosen so that the average cost associated with the periodic policy \eqref{eq:optimal_periodic_control_cost} is minimized for each $\theta$.
% Compared with periodic control, the proposed algorithm achieves lower control costs for all $\theta$, while maintaining nearly the same average actuation rate.
% Although $\ell_1$-relax.+MPC yields better control performance than the proposed method, its average actuation rate is relatively high.
% Thus, our method provides a superior trade-off compared with the other two approaches, highlighting the effectiveness of the rollout-based algorithm we propose.

\begin{figure}[t]
  \centering
  \includegraphics[width=0.98\columnwidth]{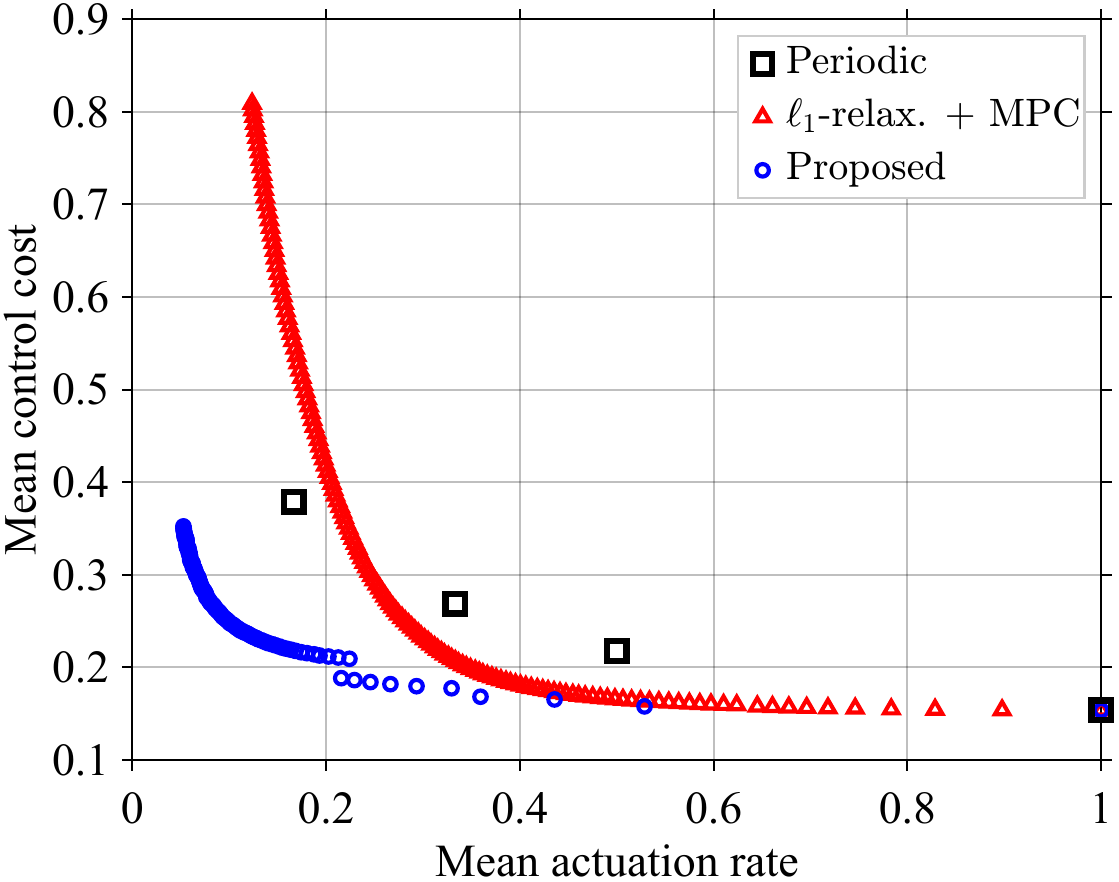}
  \caption{Trade-off curve between the mean control cost and mean actuation rate for the three methods with $\theta\in\{0,0.02,\dots,3\}$. The means are computed over 100 trials.}
  \label{fig:Jr_vs_Jc}
\end{figure}

Fig.~\ref{fig:Jr_vs_Jc} illustrates the trade-off between the control cost and the actuation rate for the three methods.
For comparable actuation rates, the proposed method achieves lower control costs than periodic control and $\ell_1$-relax.+MPC.
These results indicate that the rollout-based algorithm provides a favorable trade-off between control performance and actuation rate.

% Fig.~\ref{fig:theta_curves} shows the mean control cost and the mean actuation rate as functions of $\theta$.
% The error bars in Fig.~\ref{fig:theta_curves} represent one standard deviation across the 100 trials.
% The proposed method achieves a lower mean control cost than periodic control for every $\theta$.
% Furthermore, as $\theta$ increases, its mean actuation rate decreases and becomes lower than that of periodic control after $\theta=0.4$.
% Although $\ell_1$-relax.+MPC achieves a lower control cost untill about $\theta=0.7$, it requires a higher actuation rate than the proposed method.
% For larger values of $\theta$, the control cost of $\ell_1$-relax.+MPC increases substantially, whereas the proposed method maintains a low control cost with fewer control actions.
% Overall, these simulation results demonstrate the effectiveness of the proposed rollout-based method in balancing control performance and actuation rate, even when the conditions for the strict performance guarantee are not satisfied.

\begin{figure}[t]
  \centering
  \includegraphics[width=0.98\columnwidth]{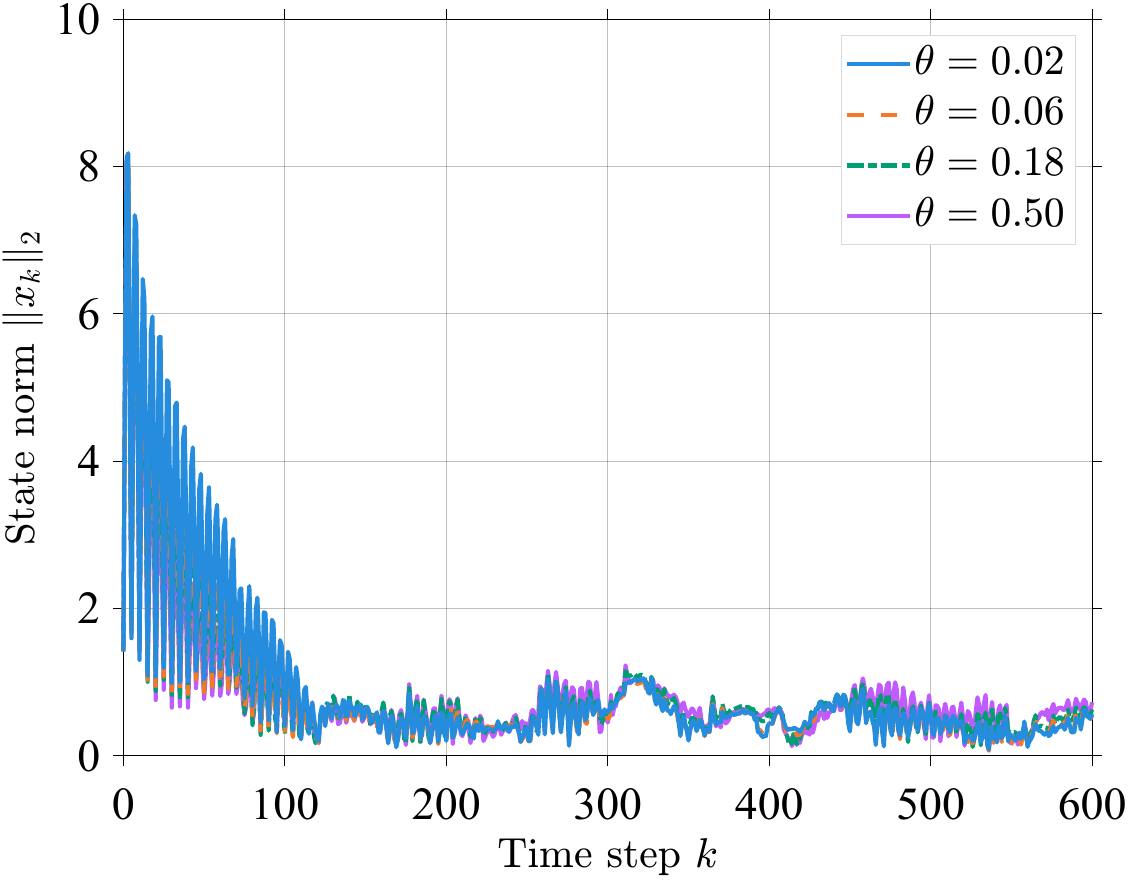}
  \caption{Trajectories of the state norm under the rollout-based algorithm. These trajectories are obtained using the same initial state and disturbance sequence across all values of $\theta$.}
  \label{fig:state_norm}
\end{figure}

\begin{figure}[t]
  \centering
  \includegraphics[width=0.98\columnwidth]{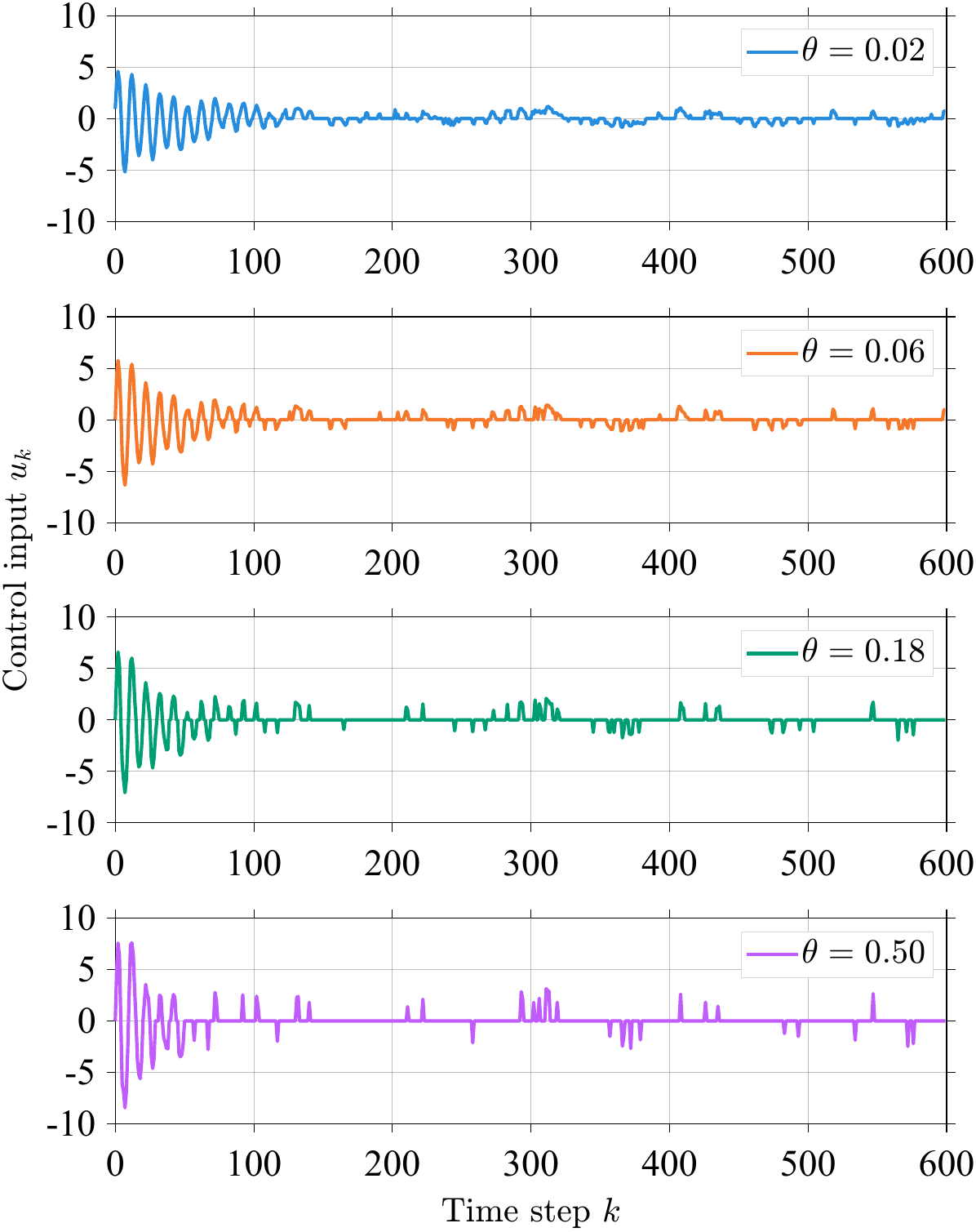}
  \caption{Control inputs corresponding to the state trajectories in Fig.~\ref{fig:state_norm}.}
  \label{fig:control_input}
\end{figure}

Fig.~\ref{fig:state_norm} and Fig.~\ref{fig:control_input} show representative trajectories of the state norms and the corresponding control inputs under the rollout-based algorithm for $\theta\in\{0.02,0.06,0.18,0.5\}$.
These trajectories are obtained using the same initial state and disturbance sequence.
For these values of $\theta$, the optimal periodic control periods are $p=1,2,3,6$, respectively.
In all cases, the state norms decrease substantially within the first 100 time steps and remain small thereafter.
Correspondingly, control actions occur frequently during the first 100 time steps, whereas they become sparse after $k=100$.
Furthermore, as shown in Fig.~\ref{fig:control_input}, larger values of $\theta$ impose a greater penalty on actuation, resulting in sparser control inputs.
% These results illustrate that the rollout-based method applies control inputs frequently to improve transient response, while substantially reducing the number of control actions once the response becomes steady.
These results highlight the effectiveness of the rollout-based algorithm in balancing the trade-off between control performance and actuation rate.

% \begin{figure}[t]
%   \centering
%   % \includegraphics[width=\linewidth]{figs/theta_curves.pdf}
%   \includegraphics[width=0.98\columnwidth]{figs/comparison_tradeoff_vs_theta.pdf}
%   \caption{Mean control cost (top) and mean actuation rate (bottom) as functions of $\theta$, with error bars indicating the standard deviation from the mean.}
%   \label{fig:theta_curves}
% \end{figure}

\section{Conclusion}
\label{sec:conclusion}

We have studied linear quadratic control to balance quadratic performance and actuation rate.
For the sparsity-promoting regularization problem, we developed a rollout-based algorithm and showed that it ensures mean-square stability and achieves an average cost no greater than that of periodic control.
We derived an explicit expression for the performance improvement over periodic control and showed that this improvement is strictly positive under additional conditions.
In addition, we considered the hard-constraint problem and developed a corresponding rollout-based design that satisfies a prescribed average actuation rate while providing stability and performance guarantees.
A numerical example demonstrated that the proposed method improves the trade-off between control performance and actuation rate.
% Future work includes relaxing the conditions for strict performance improvement and characterizing the relationship between the sparsity-promoting regularization and hard-constraint formulations.

Future work includes two directions.
The first is to relax the conditions for strict performance improvement.
This would extend the theoretical guarantees to a broader class of systems.
The second is to incorporate practical actuator constraints, such as input saturation and input rate constraints, into the proposed framework.
This extension would allow us to investigate the trade-off between LQ control performance and actuation resource usage under more realistic operating conditions.

\section*{Acknowledgement}
% The authors acknowledge the use of OpenAI ChatGPT for grammar correction, rephrasing of the manuscript text, and assistance in coding for the numerical simulation in Section~\ref{sec:simulation_example}.
% The AI assistance involved code implementation of the $\ell_1$-relaxed MPC and troubleshooting.
% The authors reviewed the generated code and verified its consistency with the mathematical formulations presented in this paper.
The authors acknowledge the use of OpenAI's ChatGPT for grammar correction and rephrasing of the manuscript, as well as for assistance with the code implementation of the $\ell_1$-relaxation-based MPC method in Section~\ref{sec:simulation_example}.
The authors carefully reviewed the edited text, verified the consistency of the implemented code with the mathematical formulation, and take full responsibility for the content of the manuscript.
% The authors take full responsibility for the content of the manuscript.

\appendix
% \section*{Appendix}
\subsection{Preliminaries on Markov Chains}
\label{append:preliminary}

We describe several definitions and properties for Markov chains on a Euclidean space, based on \cite{meyn2012markov}.

Let $\bm{X} = \{X_k\}_{k \in \Z_{\geq 0}}$ be a stochastic process with $\R^d$-valued random variables.
A mapping $P \colon \R^d\times\Borel(\R^d) \to [0,1]$ is a transition kernel if $P(\cdot,\Aset)$ is measurable for each $\Aset\in\Borel(\R^d)$ and $P(x,\cdot)$ is a probability measure for each $x\in\R^d$.
We say $\bm{X}$ is a Markov chain with transition kernel $P$ if
\begin{equation}
  \Prob(X_{k+1} \in \Aset \mid \sigma(X_0,\dots,X_k))
  =
  P(X_k,\Aset)\quad \mathrm{a.s.}
\end{equation}
for all $\Aset \in \Borel(\R^d)$ and all $k \in \Z_{\geq0}$.
Let $P^n$ denote the $n$-step transition kernel, defined by
\begin{align}
  &P^0(x,\Aset)
  \coloneqq
  \begin{cases}
    1 & x\in\Aset,\\
    0 & x\notin\Aset,
  \end{cases}\\
  &P^n(x,\Aset)
  \coloneqq
  \int_{\R^d} P(x,dy)P^{n-1}(y,\Aset),\quad n \geq 1.
\end{align}

\begin{definition}
  \label{def:phi_irreducible}
  Let $\varphi$ be a nontrivial $\sigma$-finite measure on $\Borel(\R^d)$.
  Then the chain $\bm{X}$ is called $\varphi$-irreducible if, for any $x \in \R^d$ and any $\Aset \in \Borel(\R^d)$ satisfying $\varphi(\Aset)>0$, we have
  \begin{equation}
    \sum_{n\geq1} P^n(x,\Aset) > 0.
  \end{equation}
  % \begin{equation}
  %   \Prob(\tau_\Aset < \infty \mid X_0=x) > 0,\quad
  %   \tau_\Aset \coloneqq \inf\{k\geq 1 \colon X_k \in \Aset\},
  % \end{equation}
  % where $\inf\emptyset=\infty$.
\end{definition}

\begin{definition}
  \label{def:open_set_irreducible}
  We call the chain $\bm{X}$ open set irreducible if, for any $x \in \R^d$ and every nonempty open set $\Oset \subset \R^d$, we have
  \begin{equation}
    \sum_{n\geq1} P^n(x,\Oset) > 0.
  \end{equation}
\end{definition}

\begin{definition}
  \label{def:strong_aperiodicity}
  A $\varphi$-irreducible chain is called strongly aperiodic if there exist a set $\mathcal{K}$ and a nontrivial measure $\nu$ such that $\nu(\mathcal{K})>0$ and, for all $x \in \mathcal{K}$ and $\Aset \in \Borel(\R^d)$, we have
  \begin{equation}
    P(x,\Aset) \geq \nu(\Aset).
  \end{equation}
\end{definition}
% Note that strong aperiodicity is a sufficient condition for aperiodicity \cite[Sec.~5.4]{meyn2012markov}.

For a distribution $a=\{a(n)\}_{n\in\Z_{\geq 0}}$ on $\Z_{\geq 0}$, define
\begin{equation}
  K_a(x,\Aset)
  \coloneqq
  \sum_{n=0}^{\infty} P^n(x,\Aset)a(n),\quad x \in \R^d,\ \Aset \in \Borel(\R^d).
\end{equation}

\begin{definition}
  \label{def:t_chain}
  The chain $\bm{X}$ is called a T-chain if there exist a distribution $a$ on $\Z_{\geq 0}$ and a function $T \colon \R^d\times\Borel(\R^d) \to [0,1]$ such that  $T(x,\cdot)$ is a measure on $\Borel(\R^d)$ for each $x\in\R^d$, $T(\cdot,\Aset)$ is lower semicontinuous for all $\Aset \in \Borel(\R^d)$, and for all $x \in \R^d$ and $\Aset \in \Borel(\R^d)$, we have
  \begin{equation}
    \label{eq:t_chain_minorization_condition}
    K_a(x,\Aset) \geq T(x,\Aset),\quad 0 < T(x,\R^d) \leq 1.
  \end{equation}
\end{definition}
In particular, if we can choose $a(1)=1$ and $a(n) = 0$ for $n\neq 1$, the condition \eqref{eq:t_chain_minorization_condition} reduces to
\begin{equation}
  P(x,\Aset) \geq T(x,\Aset),\quad 0 < T(x,\R^d) \leq 1.
\end{equation}
Moreover, every open set irreducible T-chain admits an irreducibility measure
$\varphi$ \cite[Prop.~6.2.2]{meyn2012markov}.

Let $\pi$ be a $\sigma$-finite measure on $\Borel(\R^d)$.
We call $\pi$ an invariant measure if
\begin{equation}
  \pi(\Aset) = \int_{\R^d} \pi(dx)P(x,\Aset),\quad \Aset \in \Borel(\R^d).
\end{equation}
Furthermore, we say the invariant measure $\pi$ is finite if $\pi(\R^d) < \infty$.
In this case, we can normalize the measure to construct a probability measure on the same space.

For $\Aset \in \Borel(\R^d)$, define the return time to $\Aset$ by $\tau_A \coloneqq \inf\{k\geq 1 \colon X_k \in \Aset\}$,
% \begin{equation}
%   \tau_A \coloneqq \inf\{k\geq 1 \colon X_k \in \Aset\},
% \end{equation}
where $\inf\emptyset=\infty$.
Let $L(x,\Aset)$ denote the probability of the event $\{\tau_A<\infty\}$ for the chain with the initial distribution $\delta_x$.
\begin{definition}
  \label{def:harris_recurrence}
  % For $\Aset \in \Borel(\R^d)$, let $L(x,\Aset)$ denote the return time probability with initial distribution $\delta_x$.
  A $\varphi$-irreducible chain is called Harris recurrent if, for any $x \in \R^d$ and any $\Aset \in \Borel(\R^d)$ satisfying $\varphi(\Aset)>0$, we have $L(x,\Aset)=1$.
  % \begin{equation}
  %   \Prob(\tau_\Aset < \infty \mid X_0=x) = 1.
  % \end{equation}
  In addition, if the Harris recurrent chain has an invariant probability measure $\pi$, it is called positive Harris recurrent.
\end{definition}

% The existence and finiteness of an invariant measure are certified as follows.
% \begin{proposition}[{\cite[Th.~10.0.1]{meyn2012markov}}]
%   \label{prop:invariant_measure}
%   A Harris recurrent T-chain admits an invariant measure $\pi$ that is unique up to constant multiples.
%   Moreover, this invariant measure is finite if there exists a compact set $\mathcal{C}\subset\R^d$ such that $\sup_{x \in \mathcal{C}} \E[\tau_\mathcal{C} \mid X_0=x]< \infty$.
% \end{proposition}

% The following notion plays a crucial role to analyze stationary behavior of Markov chains.
% \begin{definition}
%   A function $V$ is called coercive if $V(x) \to \infty$ as $\|x\| \to \infty$.
%   Equivalently, the sublevel sets $\{x\colon V(x)\leq r\}$ are compact for all $r>0$.
% \end{definition}

% The following statement connects $\varphi$-irreducible T-chains and Harris recurrent chains.
% \begin{proposition}[{\cite[Th.~9.2.2, Th.~9.4.1]{meyn2012markov}}]
%   \label{prop:recurrence}
%   Let $\bm X$ be a $\varphi$-irreducible T-chain.
%   Suppose that there exist a compact set $\mathcal C\subset\R^d$ and a coercive
%   function $V$ such that
%   \begin{equation}
%     \int_{\R^d} P(x,dy)V(y) - V(x) \leq -1,
%     \quad \forall x\in\R^d\setminus\mathcal C.
%   \end{equation}
%   Then $\bm X$ is Harris recurrent.
%   % If, in addition, $\bm X$ admits an invariant probability measure $\pi$, then $\bm X$ is positive Harris recurrent.
% \end{proposition}

\begin{definition}
  \label{def:f_ergodicity}
  Let $f \colon \R^d \to [1,\infty)$ be a measurable function.
  Then the chain $\bm{X}$ is called $f$-ergodic if all of the following conditions are satisfied.
  \begin{enumerate}
    \item $\bm{X}$ is a positive Harris recurrent chain with an invariant probability measure $\pi$.
    \item $\int_{\R^d} \pi(dx)f(x) < \infty$.
    \item For any $x \in \R^d$, it holds that
    \begin{equation}
      \lim_{n\to\infty}
      \sup_{|g|\leq f}
      \left|
        \int_{\R^d} P^n(x,dy)g(y) - \int_{\R^d} \pi(dy)g(y)
      \right|
      =
      0.
    \end{equation}
  \end{enumerate}
\end{definition}

We can verify the $f$-ergodicity of the chain by the following drift criterion.
\begin{proposition}[{\cite[Th.~14.0.1]{meyn2012markov}}]
  \label{prop:f_ergodicity_drift}
  Let $f\geq 1$ be a measurable function.
  Suppose that $\bm X$ is a $\varphi$-irreducible and aperiodic T-chain.
  If there exist a compact set $\mathcal C\subset\R^d$, a constant $b<\infty$, and a nonnegative measurable function $V\colon \R^d \to [0,\infty)$ such that
  % $V(x_0)<\infty$ for some $x_0 \in \R^d$ and
  \begin{equation}
    \label{eq:f_ergodicity_drift_criterion}
    \int_{\R^d} P(x,dy)V(y)-V(x)
    \leq
    -f(x)+b\mathbb{I}_{\mathcal C}(x),
    \quad \forall x\in\R^d,
  \end{equation}
  then $\bm X$ is $f$-ergodic.
\end{proposition}

Note that the drift condition \eqref{eq:f_ergodicity_drift_criterion} can also be expressed as
\begin{equation}
  \E\left[V(X_{k+1}) \midd X_k\right] - V(X_k)
  \leq
  -f(X_k)+b\mathbb{I}_{\mathcal C}(X_k),
  \ \forall X_k\in\R^d.
\end{equation}

For an $f$-ergodic chain, if a measurable function $g$ satisfies $|g|\leq f$, then it holds that
\begin{equation}
  \label{eq:f_ergodic_pointwise_expectation_limit}
  \lim_{n \to \infty}
  \int_{\R^d} P^n(x,dy) g(y)
  =
  \int_{\R^d} \pi(dy)g(y)
\end{equation}
for every  $x \in \R^d$.
Moreover, if the initial distribution $\lambda$ satisfies
\begin{equation}
  \int_{\R^d} \lambda(dx) V(x)
  = \E\left[V(X_0)\right]
  < \infty,
\end{equation}
where $V$ is such that \eqref{eq:f_ergodicity_drift_criterion} holds, we have
\begin{align}
  &\lim_{n \to \infty}
  \int_{\R^d} \lambda(dz)
  \left|\int_{\R^d} P^n(z,dy) g(y) - \int_{\R^d} P^n(x,dy) g(y)\right| \\
  &=
  0 \label{eq:integrated_transition_expectation_convergence}
\end{align}
for each $x$ \cite[Th.~14.3.4]{meyn2012markov}.
As a consequence, we obtain the following statement for $f$-ergodic chains.
\begin{proposition}
  \label{prop:f_ergodic_expected_time_average}
  Let $f \geq 1$ be a measurable function.
  Suppose that the chain $\bm X$ is $f$-ergodic with invariant probability measure $\pi$.
  Let $g\colon\R^d\to\R$ be a measurable function satisfying $|g|\leq f$ and $V$ be a nonnegative measurable function such that \eqref{eq:f_ergodicity_drift_criterion} holds and $\E[V(X_0)]<\infty$.
  Then we have
  \begin{equation}
    \lim_{N\to\infty}
    \frac1N
    \E\left[
    \sum_{k=0}^{N-1}g(X_k)
    \right]
    =
    \int_{\R^d} g\, d\pi.
  \end{equation}
\end{proposition}

\begin{proof}
  Fix any $x\in\R^d$.
  Then we have
  \begin{align}
    &\left|
      \E\left[g(X_k)\right]
      -
      \int_{\R^d}P^k(x,dy)g(y)
    \right| \\
    &=
    \left|
      \int_{\R^d}\lambda(dz)
      \left[
        \int_{\R^d}P^k(z,dy)g(y)
        -
        \int_{\R^d}P^k(x,dy)g(y)
      \right]
    \right| \\
    &\leq
    \int_{\R^d}\lambda(dz)
    \left|
      \int_{\R^d}P^k(z,dy)g(y)
      -
      \int_{\R^d}P^k(x,dy)g(y)
    \right|.
  \end{align}
  Hence, by \eqref{eq:integrated_transition_expectation_convergence}, we obtain
  \begin{equation}
    \label{eq:initial_distribution_expectation_convergence}
    \lim_{k\to\infty}
    \left|
      \E\left[g(X_k)\right]
      -
      \int_{\R^d}P^k(x,dy)g(y)
    \right|
    =
    0.
  \end{equation}
  Since \eqref{eq:f_ergodic_pointwise_expectation_limit} holds, \eqref{eq:initial_distribution_expectation_convergence} yields
  \begin{equation}
    \lim_{k\to\infty}
    \E\left[g(X_k)\right]
    =
    \int_{\R^d}g\, d\pi.
  \end{equation}
  The claim follows by taking the Ces\`{a}ro mean.
\end{proof}

\subsection{Two Key Lemmas for Ergodicity}
\label{append:key_lemmas}

To prove Lemma~\ref{lemma:f_ergodicity_of_estimate_chain}, we establish the following lemmas.

% \begin{lemma}
%   \label{lemma:coerciveness}
%   The function $V_1$ defined in \eqref{eq:quadratic_lyapunov_function_Vm} is coercive.
% \end{lemma}
% \begin{proof}
%   To establish the coerciveness of $V_1$, it is sufficient to show that the sublevel sets $\Lset_r(x) \coloneqq \{x \in \R^{n_x} \colon V_1(x) \leq r\}$ are precompact, i.e., their closures are compact, for each $r > 0$.
%   Since $P^{(1)}_0$ is positive definite, we have
%   \begin{equation}
%     \Lset_r(x) \subseteq \left\{ x \in \R^{n_x} \colon \|x\|^2 \leq \frac{r}{\lambda_{\min}(P^{(1)}_0)} \right\}.
%   \end{equation}
%   The set on the right-hand side is bounded for every $r > 0$, and hence $\Lset_r(x)$ is also bounded.
%   Moreover, the continuity of $V_1$ implies that the inverse image $\Lset_r(x) = V_1^{-1}([0,r])$ is closed \cite[Th.~4.8]{rudin1976principles}.
%   By the Heine--Borel theorem \cite[Th.~2.41]{rudin1976principles}, $\Lset_r(x)$ is compact.
%   Thus its closure is also compact for each $r > 0$ \cite[Th.~2.27]{rudin1976principles}, which establishes the claim.
% \end{proof}

\begin{lemma}
  \label{lemma:open_set_irreducible_aperiodic_t_chain}
  Under Assumptions~\ref{assump:no_periodic_eigenvalue_aliasing}, \ref{assump:lookahead_horizon_multiple_of_period}, \ref{assump:state_weight_and_stationary_kalman_filter}, and \ref{assump:strict_performance_improvement_conditions}(i), $\{\hat{\chi}_\ell\}_{\ell \in \Z_{\geq 0}}$ is an open set irreducible and strongly aperiodic T-chain.
\end{lemma}

\begin{lemma}
  \label{lemma:estimate_chain_f_drift_condition}
  The function $V_1$ satisfies the following drift condition:
  \begin{equation}
    \label{eq:estimate_chain_foster_lyapunov_drift}
    \E\left[V_1(\hat{\chi}_{\ell+1}) \midd \hat{\chi}_\ell\right] - V_1(\hat{\chi}_\ell)
    \leq -f(\hat{\chi}_\ell) + b \mathbb{I}_\Cset(\hat{\chi}_\ell),
    \ \forall \hat{\chi}_\ell \in \R^{n_x},
  \end{equation}
  where $b \coloneqq \beta^{(1)} + \gamma^{(1)} + 1$
  % \begin{equation}
  %   \label{eq:def_const_for_drift_cond}
  %   b \coloneqq \beta^{(1)} + \gamma^{(1)} + 1
  % \end{equation}
  and
  \begin{equation}
    \label{eq:compact_set_for_estimate_chain_drift}
    \Cset \coloneqq \left\{\chi \in \R^{n_x} \colon \lambda_{\min}(Q) \chi^\T \chi \leq \beta^{(1)} + \gamma^{(1)} + 1\right\}.
  \end{equation}
\end{lemma}

% \begin{proof}
\par\indent\textit{Proof of Lemma~\ref{lemma:open_set_irreducible_aperiodic_t_chain}:}
  It suffices to prove the following three statements:
  \begin{enumerate}
      \item $\{\hat{\chi}_\ell\}_{\ell \in \Z_{\geq 0}}$ is open set irreducible.
      \item $\{\hat{\chi}_\ell\}_{\ell \in \Z_{\geq 0}}$ is a T-chain.
      \item $\{\hat{\chi}_\ell\}_{\ell \in \Z_{\geq 0}}$ is strongly aperiodic.
  \end{enumerate}
  % (i) $\{\hat{\chi}_\ell\}_{\ell \in \Z_{\geq 0}}$ is open set irreducible;
  % (ii) $\{\hat{\chi}_\ell\}_{\ell \in \Z_{\geq 0}}$ is a T-chain; and
  % (iii) $\{\hat{\chi}_\ell\}_{\ell \in \Z_{\geq 0}}$ is strongly aperiodic.

  % \emph{Proof of (i):}
  \subsubsection*{Proof of 1)}
  Let us define matrices by
  \begin{align}
    \Omega_\omega
    &\coloneqq
    G \bigl(CA\Sigma A^\top C^\top + C\Omega_w C^\top + \Omega_v\bigr) G^\top, \\
    \bar\Omega_\omega
    &\coloneqq
    I_h \otimes \Omega_\omega,\quad \Sigma_m = \Gamma_m \bar\Omega_\omega \Gamma_m^\T.
    \label{eq:covariance_matrix_hatchi_ell}
  \end{align}
  Since $\Omega_v \succ 0$ and Assumption~\ref{assump:strict_performance_improvement_conditions}(i) holds, we have $\Omega_\omega \succ 0$ and $\bar\Omega_\omega \succ 0$.
  Moreover, $\Gamma_m$ has full row rank for every $m \in \Mset$ by construction.
  Hence, we have $\Sigma_m \succ 0$ for all $m \in \Mset$, and the transition kernel of $\{\hat{\chi}_\ell\}_{\ell \in \Z_{\geq 0}}$ is represented by
  \begin{equation} \label{eq:estimate_chain_gaussian_transition_kernel}
    P(x,\Aset)
    =
    \int_\Aset p_{\iota(x)}(x,z)\,dz,
    \quad
    x \in \R^{n_x},
    \ \Aset \in \Borel(\R^{n_x}),
  \end{equation}
  where $p_m(x,\cdot)$ is the Gaussian density with mean $\Phi_m x$ and covariance matrix $\Sigma_m$.
  Since $\Sigma_m \succ 0$, the density $p_m(x,\cdot)$ is strictly positive on $\R^{n_x}$ for each $x \in \R^{n_x}$.
  % and we can represent $p_m$ explicitly by
  % \begin{equation} \label{eq-append:gaussian_distribution}
  %   p_m(x,z)
  %   =
  %   \frac{
  %     \exp\left(
  %       -\frac12 (z-\Phi_m x)^\top \Sigma_m^{-1}(z-\Phi_m x)
  %     \right)
  %   }{
  %     (2\pi)^{n_x/2}\sqrt{\det(\Sigma_m)}
  %   }.
  % \end{equation}
  % Since $\Sigma_m \succ 0$, the mapping $z \mapsto p_m(x,z)$ is continuous and strictly positive on $\R^{n_x}$ \cite[Sec.~4.4.3]{meyn2012markov}.
  Then, for any nonempty open set $\Oset \subset \R^{n_x}$, we have
  \begin{equation}
    P(x,\Oset)=\int_{\Oset} p_{\iota(x)}(x,z)\,dz > 0,
    \quad \forall x \in \R^{n_x}.
  \end{equation}
  % since $p_{\iota(x)}(x,z)>0$ for all $z\in\R^{n_x}$.
  Therefore, the chain $\{\hat{\chi}_\ell\}_{\ell \in \Z_{\geq 0}}$ is open set irreducible.

  % \emph{Proof of (ii):}
  \subsubsection*{Proof of 2)}
  For $x\in\R^{n_x}$ and $\Aset\in\Borel(\R^{n_x})$, we define
  \begin{equation}
    T(x,\Aset)
    \coloneqq
    \int_{\Aset} \underline p(x,z)\,dz, \quad
    \underline p(x,z)
    \coloneqq
    \min_{m\in\Mset} p_m(x,z). \label{eq:common_density_minorization_kernel}
  \end{equation}
  % Since $p_m$ is continuous, $\underline p$ is also continuous, and hence measurable~\cite[Sec.~1.11]{rudin1987real}.
  % Since $\Mset$ is finite and $p_m$ is continuous and nonnegative, $\underline p$ is measurable and nonnegative.
  Since $\Mset$ is finite and each $p_m$ is continuous and
  nonnegative on $\R^{n_x}\times\R^{n_x}$,
  the function $\underline p$ is also continuous and nonnegative on $\R^{n_x}\times\R^{n_x}$.
  Since continuous functions are measurable, for each fixed $x\in\R^{n_x}$, the mapping $z\mapsto\underline p(x,z)$ is Borel measurable~\cite[Sec.~1.11]{rudin1987real}.
  Moreover, $\underline p(x,\cdot)$ is integrable since
  \begin{equation}
    \label{eq:minorization_kernel_mass_upper_bound}
    \int_{\R^{n_x}} \underline p(x,z)\,dz
    \leq
    \int_{\R^{n_x}} p_m(x,z)\,dz
    = 1
  \end{equation}
  holds for every $m \in \Mset$.
  Thus, $T(x,\cdot)$ is a measure on $\Borel(\R^{n_x})$~\cite[Exer.~1.5.4]{durrett2019probability}.

  Next, we show that $T(\cdot,\Aset)$ is lower semicontinuous.
  Specifically, for every $x\in\R^{n_x}$ and every sequence $\{x_n\}$ in $\R^{n_x}$ satisfying $x_n\to x$, we show that
  \begin{equation}
    T(x,\Aset)
    \leq
    \liminf_{n\to\infty}T(x_n,\Aset),
    \quad
    \forall \Aset\in\Borel(\R^{n_x}).
  \end{equation}
  % Let $\{x_n\}$ be a sequence that converges to some point $x$.
  Since $\Mset$ is finite and $\underline p$ is continuous on $\R^{n_x} \times \R^{n_x}$, the mapping $x \mapsto \underline p(x,z)$ is also continuous for each $z$.
  Hence, we have
  \begin{equation}
    \underline p(x,z)
    =
    \lim_{n\to\infty}\underline p(x_n,z),
    \quad
    \forall z\in\R^{n_x}
    \label{eq:underline_p_pointwise_convergence}
  \end{equation}
  for any sequence $\{x_n\}$ that converges to $x$ \cite[Th.~4.2]{rudin1976principles}.
  Since $\mathbb I_{\Aset}(z)$ does not depend on $n$, \eqref{eq:underline_p_pointwise_convergence} yields
  \begin{equation}
    \mathbb I_{\Aset}(z)\underline p(x,z)
    =
    \liminf_{n\to\infty}
    \mathbb I_{\Aset}(z)\underline p(x_n,z),
    \quad
    \forall z\in\R^{n_x}.
  \end{equation}
  Hence, we obtain
  \begin{equation}
    T(x,\Aset)
    =
    \int_{\Aset} \underline p(x,z)\,dz
    =
    \int_{\R^{n_x}} \liminf_{n\to\infty} \mathbb{I}_\Aset(z) \underline p(x_n,z)\,dz.
  \end{equation}
  % From \eqref{eq:underline_p_pointwise_convergence}, we have
  % \begin{equation}
  %   T(x,\Aset)
  %   =
  %   \int_{\Aset} \underline p(x,z)\,dz
  %   =
  %   \int_{\R^{n_x}} \liminf_{n\to\infty} \mathbb{I}_\Aset(z) \underline p(x_n,z)\,dz.
  % \end{equation}
  Note that the indicator function $\mathbb{I}_\Aset$ is Borel measureable for all $\Aset \in \Borel(\R^{n_x})$~\cite[Sec.~1.9(d)]{rudin1987real}, and hence the function $z \mapsto \mathbb{I}_\Aset(z) \underline p(x_n,z)$ is nonnegative and measurable \cite[Sec.~1.9(c)]{rudin1987real}.
  % Since $\mathbb I_{\Aset}(z)$ does not depend on $n$, \eqref{eq:underline_p_pointwise_convergence} yields
  % \begin{equation}
  %   \mathbb I_{\Aset}(z)\underline p(x,z)
  %   =
  %   \liminf_{n\to\infty}
  %   \mathbb I_{\Aset}(z)\underline p(x_n,z),
  %   \quad
  %   \forall z\in\R^{n_x}.
  % \end{equation}
  % Hence, we obtain
  % \begin{equation}
  %   T(x,\Aset)
  %   =
  %   \int_{\R^{n_x}}
  %   \liminf_{n\to\infty}
  %   \mathbb I_{\Aset}(z)\underline p(x_n,z)\,dz.
  % \end{equation}
  Consequently, by Fatou's lemma \cite[Lemma~1.28]{rudin1987real}, we have
  \begin{align}
    &\int_{\R^{n_x}} \liminf_{n\to\infty} \mathbb{I}_\Aset(z) \underline p(x_n,z)\,dz \\
    &\leq
    \liminf_{n\to\infty} \int_{\R^{n_x}} \mathbb{I}_\Aset(z) \underline p(x_n,z)\,dz
    =
    \liminf_{n\to\infty}T(x_n,\Aset),
    \label{eq:fatou_bound_for_kernel_lower_semicontinuity}
  \end{align}
  which shows that $T(\cdot,\Aset)$ is lower semicontinuous.

  By the definition of $P$, we have
  \begin{align}
    P(x,\Aset)
    &= \int_{\Aset} p_{\iota(x)}(x,z)\,dz \\
    &\geq \int_{\Aset} \min_{m\in\Mset} p_m(x,z)\,dz
    = T(x,\Aset).
  \end{align}
  From \eqref{eq:minorization_kernel_mass_upper_bound}, we have $T(x,\R^{n_x}) \leq 1$ for any $x \in \R^{n_x}$.
  % Moreover, since $\underline p(x,z) \leq p_m(x,z)$ for all $z$, it holds that
  % \begin{equation}
  %   \label{eq:T-chain_highbound}
  %   T(x,\R^{n_x})
  %   =
  %   \int_{\R^{n_x}} \underline p(x,z)\,dz
  %   \leq \int_{\R^{n_x}} p_m(x,z)\,dz
  %   = 1.
  % \end{equation}
  % Moreover, since each $p_m$ is strictly positive and $\Mset$ is finite, $\underline p(x,z)>0$ for any $(x,z) \in \R^{n_x} \times \R^{n_x}$.
  % Moreover, since $\underline p(x,z)>0$ for every $x \in \R^{n_x}$, by \cite[Ex.~1.4.1]{durrett2019probability}, we have
  Moreover, since $\underline p(x,z) > 0$ for every $(x,z) \in \R^{n_x}\times\R^{n_x}$, by \cite[Exer.~1.4.1]{durrett2019probability}, we have
  \begin{equation}
    \label{eq:positive_minorization_kernel_mass}
    T(x,\R^{n_x}) = \int_{\R^{n_x}} \underline p(x,z)\, dz >0
  \end{equation}
  % since $\underline p$ is strictly positive.
  Therefore, we obtain $0 < T(x,\R^{n_x}) \leq 1$ for any $x \in \R^{n_x}$.

  % \emph{Proof of (iii):}
  \subsubsection*{Proof of 3)}
  Fix a compact set $\Kset\subset\R^{n_x}$ with nonempty interior.
  From the proof of (i), $\Sigma_m\succ0$ for all $m\in\Mset$.
  Hence, for each $m\in\Mset$, the density $p_m$ is continuous and strictly positive on $\Kset \times \Kset$.
  Since $\Kset \times \Kset$ is compact, there exists $\delta>0$ such that
  \begin{equation}
    \label{eq:uniform_transition_density_lower_bound}
    p_m(x,z)\ge \delta,
    \quad
    \forall (x,z)\in \Kset \times \Kset,\ \forall m\in\Mset.
  \end{equation}
  We introduce a measure $\nu$ on $\Borel(\R^{n_x})$ as
  \begin{equation}
    \label{eq:minorizing_measure_definition}
    \nu(\Bset)=\delta\mu^{\text{Leb}}(\Bset \cap \Kset),
    \quad \Bset\in\Borel(\R^{n_x}),
  \end{equation}
  where $\mu^{\text{Leb}}$ denotes the Lebesgue measure on $\Borel(\R^{n_x})$.
  Note that $\nu(\Kset)>0$ since $\nu(\Kset) = \delta\mu^{\text{Leb}}(\Kset) > 0$.
  Moreover, for any $x\in \Kset$ and $\Bset\in\Borel(\R^{n_x})$, we have
  \begin{align}
    P(x,\Bset)
    &=
    \int_\Bset p_{\iota(x)}(x,z)\,dz
    \geq
    \int_{\Bset \cap \Kset} p_{\iota(x)}(x,z)\,dz \\
    &\geq
    \delta\mu^{\text{Leb}}(\Bset \cap \Kset)
    =
    \nu(\Bset) \label{eq:strong_aperiodicity_minorization}
  \end{align}
  from \eqref{eq:uniform_transition_density_lower_bound} and \eqref{eq:minorizing_measure_definition}.
  By parts 1) and 2), $\{\hat\chi_\ell\}_{\ell\in\mathbb Z_{\geq0}}$ is an open set irreducible T-chain and hence is $\varphi$-irreducible for some irreducibility measure $\varphi$~\cite[Prop.~6.2.2]{meyn2012markov}.
  Since $\nu(\Kset)>0$, \eqref{eq:strong_aperiodicity_minorization} implies that the chain is strongly aperiodic.

  Combining 1) -- 3) proves the claim.
  \hfill\QED
% \end{proof}

% \begin{lemma}
%   \label{lemma:estimate_chain_f_drift_condition}
%   The function $V_1$ satisfies the following drift condition.
%   \begin{equation}
%     \label{eq:estimate_chain_foster_lyapunov_drift}
%     \E\left[V_1(\hat{\chi}_{\ell+1}) \midd \hat{\chi}_\ell\right] - V_1(\hat{\chi}_\ell)
%     \leq -f(\hat{\chi}_\ell) + b \mathbb{I}_\Cset(\hat{\chi}_\ell),
%     \ \forall \hat{\chi}_\ell \in \R^{n_x},
%   \end{equation}
%   where $b \coloneqq \beta^{(1)} + \gamma^{(1)} + 1$ and
%   \begin{equation}
%     \label{eq:compact_set_for_estimate_chain_drift}
%     \Cset \coloneqq \left\{\chi \in \R^{n_x} \colon \lambda_{\min}(Q) \chi^\T \chi \leq \beta^{(1)} + \gamma^{(1)} + 1\right\}.
%   \end{equation}
% \end{lemma}

% \begin{proof}
\par\indent\textit{Proof of Lemma~\ref{lemma:estimate_chain_f_drift_condition}:}
  % Since $\hat{x}_k$ is a sufficient statistics with respect to $\Iset_k$, and
  % By the orthogonal property of the Kalman filter and the sufficient statistic property of $\hat{x}_k$ with respect to $\Iset_k$, for any $k \in \Z_{\geq 0}$, it follows that
  % \begin{align}
  %   \E\left[e_k^{\T} \hat{x}_k \midd \hat{x}_k\right] &= \E\left[(x_k - \hat{x}_k)^{\T} \hat{x}_k \midd \hat{x}_k\right] \\
  %   &= \left(\E\left[x_k\midd \hat{x}_k\right] -\hat{x}_k\right)^{\T} \hat{x}_k
  %   = 0.
  % \end{align}
  % Hence, we obtain
  % \begin{align}
  %   \E\left[V_1(\chi_\ell) \midd \hat{\chi}_\ell\right] &= \E\left[(\hat{\chi}_\ell + e_{\ell h})^{\T} P^{(1)}_0 (\hat{\chi}_\ell + e_{\ell h}) \midd \hat{\chi}_\ell\right] \\
  %   &= \E\left[V_1(\hat{\chi}_\ell) \midd \hat{\chi}_\ell\right] + \tr\left(P^{(1)}_0 \Sigma\right),
  % \end{align}
  % and
  % \begin{align}
  %   &\E\left[V_1(\chi_{\ell+1}) \midd \hat{\chi}_\ell\right] \\
  %   &= \E\left[(\hat{\chi}_{\ell+1} + e_{(\ell+1)h})^{\T} P^{(1)}_0 (\hat{\chi}_{\ell+1} + e_{(\ell+1)h}) \midd \hat{\chi}_\ell\right] \\
  %   &= \E\left[V_1(\hat{\chi}_{\ell+1}) \midd \hat{\chi}_\ell\right] + \tr\left(P^{(1)}_0 \Sigma\right).
  % \end{align}
  We recall that it holds that
  \begin{align}
    &\E\left[V_1(\chi_\ell) \midd \hat{\chi}_\ell\right] = V_1(\hat{\chi}_\ell) + \tr\left(P^{(1)}_0\Sigma\right), \\
    &\E\left[V_1(\chi_{\ell+1}) \midd \hat{\chi}_\ell\right] = \E\left[V_1(\hat{\chi}_{\ell+1}) \midd \hat{\chi}_\ell\right] + \tr\left(P^{(1)}_0\Sigma\right)
  \end{align}
  from \eqref{eq:current_state_conditional_lyapunov_decomposition} and \eqref{eq:next_state_conditional_lyapunov_decomposition}.
  Hence, we have
  % From \eqref{eq:h_step_stage_cost_quadratic_lower_bound} and \eqref{eq:h_step_rollout_performance_gap_bound}, we obtain
  \begin{align}
    &\E\left[V_1(\hat{\chi}_{\ell+1}) \midd \hat{\chi}_\ell\right]
    -
    V_1(\hat{\chi}_\ell) \\
    &= 
    \E\left[V_1(\chi_{\ell+1}) \midd \hat{\chi}_\ell\right]
    -
    \E\left[V_1(\chi_\ell) \midd \hat{\chi}_\ell\right] \\
    &=
    \beta^{(1)} + \gamma^{(1)} - g(\hat{\chi}_\ell)
    - \E\left[\bar{c}\left(\chi_\ell, \hat{\chi}_\ell, \bar{w}_\ell, \bar{\omega}_\ell\right) \midd \hat{\chi}_\ell\right] \\
    &\leq
    \beta^{(1)} + \gamma^{(1)} - \lambda_{\min}(Q) \hat{\chi}_\ell^{\T}\hat{\chi}_\ell - g(\hat{\chi}_\ell),
  \end{align}
  where the second equality and the last inequality follow from \eqref{eq:h_step_rollout_performance_gap_bound} and \eqref{eq:h_step_stage_cost_quadratic_lower_bound}, respectively.
  % \begin{align}
  %   &\E\left[V_1(\chi_{\ell+1}) + \bar{c}\left(\chi_\ell, \hat{\chi}_\ell, \bar{w}_\ell, \bar{\omega}_\ell\right) \midd \hat{\chi}_\ell\right] - \E\left[V_1(\chi_\ell) \midd \hat{\chi}_\ell\right] \\
  %   &\quad + \beta^{(\iota(\hat{\chi}_\ell))} + \gamma^{(\iota(\hat{\chi}_\ell))} \\
  %   &= \beta^{(1)} + \gamma^{(1)} - g(\hat{\chi}_\ell)
  % \end{align}
  % and the last inequality follows by \eqref{eq:h_step_stage_cost_quadratic_lower_bound}.
  By the definition of the set $\Cset$ in \eqref{eq:compact_set_for_estimate_chain_drift}, for any $\hat{\chi}_\ell \in \R^{n_x}\setminus\Cset$, we obtain
  \begin{equation}
    \E\left[V_1(\hat{\chi}_{\ell+1}) \midd \hat{\chi}_\ell\right] - V_1(\hat{\chi}_\ell)
    \leq
    -(g(\hat{\chi}_\ell) + 1)
    =
    -f(\hat{\chi}_\ell). \label{eq:negative_drift_outside_compact_set}
  \end{equation}
  When $\hat{\chi}_\ell \in \Cset$, we have
  \begin{align}
    &\E\left[V_1(\hat{\chi}_{\ell+1}) \midd \hat{\chi}_\ell\right]
    -
    V_1(\hat{\chi}_\ell) \\
    &\leq
    \beta^{(1)} + \gamma^{(1)} - g(\hat{\chi}_\ell)
    % = -f(\hat{\chi}_\ell) + \beta^{(1)} + \gamma^{(1)} + 1.
    = -f(\hat\chi_\ell) + b,
    \label{eq:bounded_drift_inside_compact_set}
  \end{align}
  % where $b$ is defined in \eqref{eq:def_const_for_drift_cond}.
  where $b \coloneqq \beta^{(1)} + \gamma^{(1)} + 1$.
  Combining \eqref{eq:negative_drift_outside_compact_set} and \eqref{eq:bounded_drift_inside_compact_set} gives \eqref{eq:estimate_chain_foster_lyapunov_drift}.
  \hfill\QED
% \end{proof}

\subsection{Proof of Lemma~\ref{lemma:f_ergodicity_of_estimate_chain}}
\label{append:ergodicity}

Building upon Lemmas~\ref{lemma:open_set_irreducible_aperiodic_t_chain} and \ref{lemma:estimate_chain_f_drift_condition}, we prove that $\{\hat{\chi}_\ell\}_{\ell \in \Z_{\geq 0}}$ is an $f$-ergodic chain.

From Lemma~\ref{lemma:open_set_irreducible_aperiodic_t_chain}, $\{\hat{\chi}_\ell\}_{\ell \in \Z_{\geq 0}}$ is a $\varphi$-irreducible and aperiodic T-chain.
% Since $g$ is measurable and $g \geq 0$, $f \geq 1$ is also measurable.
% The continuity of $V_1$ implies that it is also measurable \cite[Chap.~1]{rudin1987real}, and it is nonnegative since $P^{(1)}_0 \succ 0$.
Since $V_1$ is continuous and $P^{(1)}_0 \succ 0$, $V_1$ is nonnegative and measurable.
Moreover, since $\Cset$ is closed and bounded, it is compact by the Heine--Borel theorem \cite[Th.~2.41]{rudin1976principles}.
Therefore, from Proposition~\ref{prop:f_ergodicity_drift} in Appendix~\ref{append:preliminary}, we conclude that $\{\hat{\chi}_\ell\}_{\ell \in \Z_{\geq 0}}$ is $f$-ergodic.
\hfill\QED

% \balance
\bibliographystyle{IEEEtran}
\bibliography{myrefs}

\begin{IEEEbiography}[
{\includegraphics[width=1in,height=1.25in,clip,keepaspectratio]{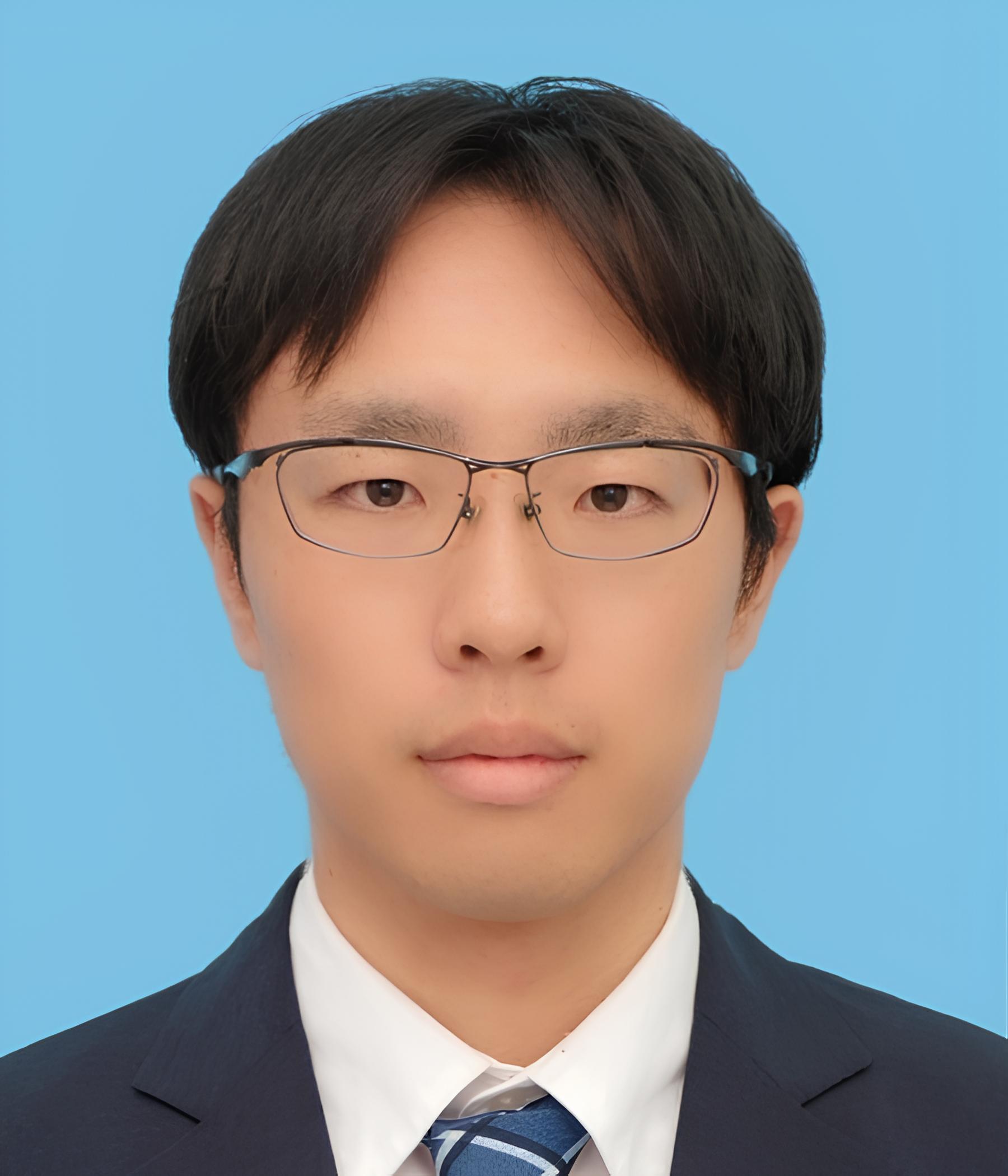}}
]{Shumpei Nishida} (Student Member, IEEE)
received the B.Eng. and M.Eng. degrees from Ritsumeikan University, Kusatsu, Japan, in 2023 and 2025, respectively.
He is currently pursuing the Ph.D. degree at Ritsumeikan University, Japan.

Since September 2026, he has been a visiting Ph.D. student at Eindhoven University of Technology, Eindhoven, the Netherlands.
His research interests include networked control systems and stochastic control.

Mr.~Nishida received the SICE International Young Authors Award for CDC 2024 from the Society of Instrument and Control Engineers.
He was a finalist for the Best Student Paper Award at the 2026 American Control Conference.
\end{IEEEbiography}

\begin{IEEEbiography}[{\includegraphics[width=1in,height=1.25in,clip,keepaspectratio]
{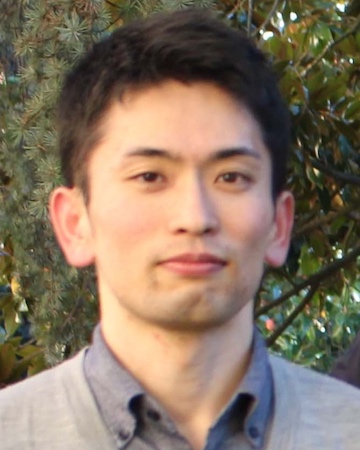}}]{Kunihisa Okano}  (Member, IEEE)
received the B.Eng.\ degree in systems science from Osaka University,
Toyonaka, Japan, in 2006, the M.I.Sc.T.\ degree in information physics and
computing from the University of Tokyo, Tokyo, Japan, in 2008, and the D.Eng.\
degree in computational intelligence and systems science from Tokyo Institute
of Technology, Yokohama, Japan, in 2013.

He was with NTT DATA Corp., Tokyo, Japan,
from 2008 to 2010 and a visiting scholar at the University of California,
Santa Barbara, CA, USA, from 2013 to 2016.
From 2016 to 2021, he was an Assistant Professor at
Tokyo University of Science, Tokyo, Japan, and then at Okayama University, Okayama, Japan.
Since 2021, he has been an Associate Professor with the College of Science
and Engineering, Ritsumeikan University, Japan.
His research interests are in networked control systems.
\end{IEEEbiography}

\end{document}